\documentclass{emulateapj}
\usepackage{color}
\usepackage{graphicx}

\newcommand{\mscred}{{\sc mscred}}
\newcommand{\wfpred}{{\sc wfpred}}
\newcommand{\allframe}{{\sc allframe}}

\slugcomment{Revised version 3.0}

\shorttitle{SX Phe stars in the Fornax galaxy}
\shortauthors{Poretti et al.}

\begin{document}

\title{Variable stars in the Fornax dSph Galaxy.\\ 
II. Pulsating stars below the horizontal branch  
\altaffilmark{1}}
\author{Ennio~Poretti\altaffilmark{2}, Gisella~Clementini\altaffilmark{3},
Enrico~V.~Held\altaffilmark{4}, Claudia~Greco\altaffilmark{3,5},
Mario~Mateo\altaffilmark{6},
Luca~Dell'Arciprete\altaffilmark{2}, 
Luca~Rizzi\altaffilmark{7}, Marco~Gullieuszik\altaffilmark{4}, 
and Marcella~Maio\altaffilmark{3}}
\altaffiltext{1}{Based on observations collected at ESO,
La Silla, Chile, Proposal 68.A-0281}
\altaffiltext{2}{INAF-Osservatorio Astronomico di Brera, Via E. Bianchi 46, I-23807 Merate (LC), Italy}
\altaffiltext{3}{INAF-Osservatorio Astronomico di Bologna, Via Ranzani 1, I-40127 Bologna, Italy}
\altaffiltext{4}{INAF-Osservatorio Astronomico di Padova, Vicolo dell'Osservatorio 5, I-35122 Padova, Italy}
\altaffiltext{5}{Current address: Observatoire de Gen\`eve, 51 ch. des Maillettes, CH-1290 Versoix,
 Switzerland}
\altaffiltext{6}{Department of Astronomy, University of Michigan, 821 Dennison Building, Ann Arbor, MI 48109-1090}
\altaffiltext{7}{Joint Astronomy Centre, 660 N. A'ohoku Place, University Park, Hilo, HI 96720, USA}

\begin{abstract}
                                                                                                    
We have carried out an intensive survey of the northern region of the Fornax
dwarf spheroidal galaxy 
with the aim of detecting the galaxy's short--period pulsating
stars ($P<0.25$~days). Observations collected over three consecutive nights with the Wide Field Imager
of the 2.2m MPI telescope at ESO allowed us to detect 85 high--amplitude
(0.20--1.00 mag in $B$-light) variable stars
with periods in the range from 0.046 to 0.126~days, 
similar to SX Phoenicis stars in Galactic metal-poor stellar populations.
The plots of the observed periods vs. the $B$ and $V$ magnitudes  show a dispersion largely exceeding
the observational errors.  To disentangle the matter, we separated the first-overtone
from the fundamental-mode pulsators and tentatively identified a group
of subluminous variables, about 0.35~mag fainter than the others. Their
nature as either 
metal-poor intermediate-age stars or stars 
formed by the merging of close binary systems is discussed.
The rich sample of the Fornax variables also led us to reconstruct the
Period--Luminosity
relation for short--period pulsating stars. An excellent linear fit, 
$M_V=-1.83(\pm0.08)-3.65(\pm0.07)\times\log P_F$, was obtained
using 153 $\delta$ Scuti and SX Phoenicis stars in a number of different stellar systems.
\end{abstract}

\keywords{galaxies: distance and redshifts  -- galaxies: individual (Fornax) --  $\delta$~Scuti
-- stars: variables: others   -- blue stragglers -- techniques: photometric}

\section{Introduction}
In globular clusters, the systematic detection of 
the short--period ($P < 0.25$~days) pulsating stars
located below the horizontal branch (namely, SX Phe and
$\delta$ Sct stars) started about twenty years ago (e.g., 
\citealt{baas,ngc5053}). 
At that time, Nemec, Nemec \& Lutz (1994) 
listed 14 SX Phe stars in NGC~5053, $\omega$~Cen and NGC~5466.
The use of these variable stars
as distance indicators has made some progress in the last years
(e.g., \citealt{pych,mazur,olech}).
However, due to their intrinsic faintness, not many of these pulsating
stars have been discovered so far in stellar systems outside the
Milky Way:
\cite{matcarina} and \cite{carina} describe the first results obtained
for SX Phe stars in the Carina dwarf spheroidal (dSph) galaxy;
McNamara, Clementini \& Marconi (2007)
have used a sample of $\delta$ Sct stars
to estimate the distance to the Large Magellanic Cloud. To exploit in a
more complete way the potential of these variable stars as distance
indicators and stellar population tracers we have carried out 
a search for pulsating stars below the horizontal branch 
as starting point for an extensive project on the Fornax dSph.
This galaxy is an ideal target since: (i) it is known to host
a mix of old and intermediate--age stars with different metal abundances
\citep[see, e.g.,][and refs. therein]{battaglia06, coleman}; 
(ii) its size and distance make the
observation of large fractions of the galaxy, down to a limiting  magnitude of about
3.0 mag below the horizontal branch, 
practical using a medium--class telescope
equipped with a wide--field imager, using 
exposure times fully adequate to reveal the galaxy's short--period
pulsators.  Our project also allowed a
more comprehensive study of several classes of variable stars in the 
Fornax dSph galaxy, and it was extended over the course of the years
to cover 
the galaxy's field and its system of five globular clusters, through
the acquisitions of new observations  
with other telescopes
(\citealt{clementini06,for4}). 

The nomenclature of the short--period pulsating stars below the horizontal branch
is confusing.
In the Milky Way there is a physical distinction between $\delta$ Sct
and SX Phe stars: the former are Population~I stars (Pop.~I), 
the latter are Population~II (Pop.~II) objects.
Low--amplitude, nonradial modes are typical of the $\delta$ Sct stars,
but intensive and accurate surveys in globular clusters have provided
observational evidences that they  are excited also in the SX Phe stars
(\citealt{pych,mazur,olech}).  Therefore, the amplitude can vary from a few 0.001~mag
to several 0.1~mag both in the $\delta$ Sct and in the SX Phe variables.
The $\delta$ Sct stars showing amplitude larger than 0.20~mag are generally referred to as
High--Amplitude $\delta$ Sct (HADS) stars. As in the case of high--amplitude
SX Phe stars, the main pulsation period is a radial mode.
The pulsation period can provide a rough
separation between HADS and SX Phe stars, since periods of HADS vary from 0.07 to
0.25~days \citep{ogle},
while most of the SX Phe stars in globular clusters have  $P<0.10$~days.
However, since there is some overlap in the period
range spanned by the two types of variables, 
disentangling HADS from SX Phe stars may not 
be an easy task if no details about the metallicity are known
\footnote{In the past, it was 
quite common to use the term ``Dwarf Cepheids" 
to identify both HADS and SX Phe stars, since their light curves are reminding
those of the Classical Cepheids. This term has an unclear meaning in an astrophysical context,
since it groups  stars belonging to different populations, and has not been used here.}.
 In this respect, it is appropriate to use the SX Phe term
to identify the short--period Pop.~II variables found in globular clusters.
We will adopt this name also in the case of Fornax, 
since the range of metal abundance
observed in Fornax is $-2.2<$[Fe/H]$<-0.7$ 
\citep{saviane,pont04,battaglia06,gullieus07,coleman}, so that 
all the short--period pulsating stars in this galaxy 
are more similar to the SX Phe stars according to 
the nomenclature scheme described above.
However, we note 
that in a different environment such as the Fornax dSph galaxy
the criteria used in the Milky Way may not be adequate to describe
the mixture of stars having such different ages and metallicities.

  \section{Observations and Data Reductions}
                                                                                                    
\subsection {Observations}
                                                                                                    
Observations of one field north of the center of Fornax dSph, centered
at $\alpha = 2^h 39^m 59^s$, $\delta = -34\degr 10\arcmin 00\arcsec$
(J2000.0), were obtained using the Wide-Field Imager (WFI)
at the 2.2~m ESO/MPI telescope. The mosaic is composed of 8 EEV ``{\sc
ccd~44}'' type CCDs, each having $2048 \times 4096$ pixels. The pixel
scale is $0.238\arcsec$ pixel$^{-1}$ yielding a field-of-view of $34 \times
33$ arcmin$^2$.

The Fornax field was observed on three consecutive nights (Nov. 8--11, 2001)
for 6.2, 7.5, and 7.7~hours, respectively.
To maximize the
stability of differential photometry of variable stars, no dithering was
applied.
Our goals were to obtain data in a two--color system and at least one
very dense time series. We noted that the passband of the ESO842 $B$--filter is larger 
than that of the ESO 843 $V$--filter;
hence, the former filter 
was more suitable to reach the desired signal--to--noise ratio in a short time, an
observational constraint posed by the expected very short periods. Moreover, the
amplitudes of the SX Phe stars are expected to be slightly larger in $B$--light
than in $V$--light. 
Hence, 3--4~consecutive images were
taken in $B$--light (700~s each), followed by a single exposure in $V$
(1000~s). We collected 61 images in $B$ and 16 in $V$ across the 3~nights.
This strategy ensured the dense $B$ time series necessary to perform a
reliable frequency analysis, and allowed us to obtain the mean
brightness and amplitude values in a two-color system and reliably
place the variable stars on the Fornax dSph color-magnitude diagram
(CMD).
                                                                                                    
The sky conditions were generally stable and photometric, especially on
the third night used as our reference for absolute photometric
calibration.  The standard fields Ru\,149, Ru\,152, and T\,Phe 
\citep[see][]{landolt92} were observed to this purpose in all CCDs, with
typical exposure times 300--350~s in $B$ and 450--500~s in $V$.  
The seeing conditions varied from $1.0\arcsec$ to $1.8\arcsec$.

\subsection{Data Reduction}


\begin{deluxetable}{l l l l l}
\tablewidth{0pt}
\tablecaption{Calibration coefficients for the WFI}
\tablehead{
\multicolumn{1}{c}{CCD} &
\multicolumn{1}{c}{$zp_B$} &
\multicolumn{1}{c}{$c_B$} &
\multicolumn{1}{c}{$zp_V$} &
\multicolumn{1}{c}{$c_V$}
}
\startdata
1    & 24.755 & 0.263 & 24.292 & $-$0.086 \\
2    & 24.694 & 0.281 & 24.238 & $-$0.074 \\
3    & 24.692 & 0.254 & 24.227 & $-$0.082 \\
4    & 24.729 & 0.264 & 24.280 & $-$0.073 \\
5    & 24.734 & 0.256 & 24.282 & $-$0.075 \\
6    & 24.659 & 0.289 & 24.222 & $-$0.090 \\
7    & 24.674 & 0.257 & 24.223 & $-$0.088 \\
8    & 24.722 & 0.266 & 24.284 & $-$0.090 \\
\enddata
\label{tab:calib}
\end{deluxetable}

                                                                                                    
Image reduction of the CCD mosaic data was accomplished in
IRAF\footnote{The Image Reduction and Analysis Facility (IRAF) software
is provided by the National Optical Astronomy Observatories (NOAO),
which is operated by the Association of Universities for Research in
Astronomy (AURA), Inc., under contract to the National Science
Foundation.}
using standard procedures. All multi-extension images were
bias-subtracted and divided by twilight flat-fields using
the mosaic reduction package \mscred\ \citep{vald98}.  A bad-pixel mask
was created, and CCD blemishes and bad columns interpolated over.  
This step was necessary since CCD defects significantly increase 
the number of false detections by the
image subtraction software we used to identify variable sources 
(ISIS~2.1, \citealt{alard}).
                                                                                                    
The images were astrometrically calibrated using a polynomial solution
computed on astrometric fields from \citet{ston+99} to remove distortion
and the photometric effects of the variable pixel area.  The images were
then registered onto a common distortion-free coordinate grid using
stars in the USNO-A2.0 Catalogue \citep{usnoa2}.  The package \mscred\
was used to this purpose, along with the pipeline script package
\wfpred\ developed by two of us (LR and EVH) at the Padua Astronomical
Observatory. The multi-extension images were then split and the
individual CCD images analyzed separately.

                                                                                                    
Crowded field stellar photometry was performed with the \allframe\
package \citep{allframe}, with spatially varying point-spread functions
(PSFs) independently computed for each CCD and each filter.  
Aperture corrections were used to match the PSF photometry to
large-aperture photometry on selected reference images taken on the best
photometric night, using growth-curve analysis of bright isolated stars.
                                                                                                    
The absolute calibration was derived from standard star observations
on the best night.  A set of linear calibration relations was computed
for every CCD:
                                                                                                    
\begin{eqnarray}
B  &=&  b  +  c_B (B-V)  +  zp_B \\
V  &=&  v  +  c_V (B-V)  +  zp_V
\end{eqnarray}
                                                                                                    
\noindent
where $b$ and $v$ are the instrumental magnitudes normalized to 1~s
exposure and corrected for atmospheric extinction.
The color terms and zero-points of the calibration are given in
Table~\ref{tab:calib}.  These relations were
used to calibrate, independently for each CCD, the mean instrumental
PSF-fitting magnitudes, after appropriate aperture correction, and
the individual data points in the light curves.
The zero-point uncertainty on the calibrated magnitudes is of the
order 0.04~mag in $B$ and $V$, including the uncertainties on the
calibration, the aperture corrections, and the residual zero-point
variations within each CCD.

\subsection{Variable star search}
Variable stars were
identified by applying the Image Subtraction Technique. 
We used the package
ISIS~2.1 \citep{alard}
which was run on the $B$ and $V$ image sequences independently.
This package identifies variable sources by direct comparison 
of the time-sequence of images, and is specifically designed to allow the detection of faint,
small amplitude variables in crowded fields. 
Each CCD of the WFI mosaic was analyzed individually.
The $B$ and $V$ catalogs of the candidate variables were cross-identified 
and  conservative selection criteria were applied to the objects displaying variability flags.
  On average we detected about 300--400 candidate
variables in each of the 8 CCDs of the WFI mosaic.
The reference frames produced by the Image Subtraction 
were matched to the astrometric reference frames 
during the photometric reduction process.
Instrumental $b$ and $v$ time series of the candidate variables 
were obtained by matching
the ISIS catalogs and the \allframe\,   photometric catalogs.
In each chip we then selected a number of comparison stars (up to 12),
chosen amongst the most stable objects, and built differential light curves
that were subsequently examined using period-search user--interactive
tools (see \citealt{for4} and Clementini et al. 2008, in preparation, for details) in order
to confirm the variability, and derive periods and type of variability of
the confirmed variables.
Finally, the instrumental \allframe\, photometry of the confirmed variables
was transformed to the standard Johnson-Cousins photometric system using the
equations derived in the calibration process.

\begin{figure}
\plotone{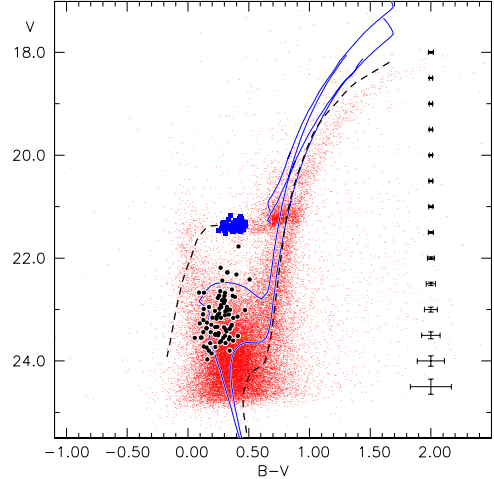}
\caption{
Color-magnitude diagram of the Fornax field (only data in one of the
CCDs of the WFI mosaic, CCD\#6, with \allframe\, errors less than 0.06~mag
 have been plotted for clarity), with superimposed our total sample of
SX~Phe stars (85 objects, {\it black dots}) together with a sub-sample
of 80 RR~Lyr stars (marked by {\it blue squares}).  The error bars
represent $\pm 3\sigma$ typical uncertainties of the mean magnitudes and
colors at different magnitude levels. 
The solid lines are the isochrones from \citet{gira+2000}, with $Z=0.001$ 
and ages 2.2 and 7.1 Gyr,  plotted 
to show the positions of the turnoff points at different ages.
The dashed black line is the mean ridge line of the Galactic globular cluster M3 \citep{m3}.
See the electronic edition of the journal for a color version
of this figure.}
\label{cmd}
\end{figure}

We have identified an extraordinary large number of bona fide
 variable stars
of different types in our collection of time series of candidate variables 
(about 2500 in total). They include: several hundreds RR Lyr stars,
85 SX Phe stars, many eclipsing binaries, and a number of variables with periods longer than
about 1.0~days (likely anomalous and Population II Cepheids).
Figure~\ref{cmd} shows the CMD of the Fornax dSph obtained from the present data,
with the
SX Phe stars and a sub-sample of the RR Lyr variables
shown as circles and squares, respectively.
The isochrones from \citet[solid lines]{gira+2000}
bracket the bulk of the Fornax SX Phe stars. The mean ridge line  of the Galactic
globular cluster M3 \citep[dashed line]{m3} provides a good fit of the old ($t>10$ Gyr) stellar
component producing the Fornax RR Lyrae stars.
We have adopted the following parameters for M3: [Fe/H]=$-1.66\pm0.06$ \citep{zinn}, 
$E(B-V)=0.01$ \citep{harris03}, and a dereddened distance modulus of 
$(m-M)_0$=15.04~mag, as inferred 
from the average magnitude of the cluster RR Lyrae stars \citep{cc}.
Preliminary results on the RR Lyr stars were presented
in \cite{greco05}. In the present paper we will
focus on the SX Phe stars.
A more detailed analysis of the Fornax RR Lyr stars and of the
other types of variables 
will be presented in the following papers of this series.

\begin{deluxetable*}{l r r rcrrc l}
\tablewidth{0pt}
\tabletypesize{\scriptsize}
\tablecaption{Identification and properties of the Fornax dSph SX Phe stars
\label{basic}}
\tablehead{
\multicolumn{1}{c}{Name\tablenotemark{a}}  & \multicolumn{1}{c}{$\alpha$ (2000.0)} & \multicolumn{1}{c}{$\delta$ (2000.0)} &
\multicolumn{1}{c}{Period} &
\multicolumn{1}{c}{Epoch (T$_{\rm max}$)} &
\multicolumn{1}{c}{$\langle B \rangle$} & A$_B$ &  $\langle B \rangle$ - $\langle V \rangle$ & \multicolumn{1}{c}{Type\tablenotemark{b}}\\
&\multicolumn{1}{c}{} & \multicolumn{1}{c}{}   & \multicolumn{1}{c}{(days)} &  \multicolumn{1}{c}{(HJD-2452222)} & & &
}
\startdata
8\_V9899   &  2 41 14.2  & --34  23    02.7  &   0.04619  &  0.266  &  23.98  &   0.53  &   -      &  F    \\
7\_V8618   &  2 40 20.2  & --34  23    58.6  &   0.04917  &  0.170  &  24.14  &   0.99  &   0.34  &  SL   \\
2\_V3796   &  2 40 12.1  & --34  09    07.3  &   0.05137  &  0.253  &  23.96  &   0.77  &   0.23  &  F    \\
5\_V6170   &  2 38 51.9  & --34  24    46.7  &   0.05326  &  0.289  &  23.27  &   0.23  &   0.32  &  FO   \\
5\_V15474  &  2 39 02.6  & --34  20    26.4  &   0.05338  &  0.238  &  23.66  &   0.51  &   0.36  &  F    \\
5\_V16343  &  2 38 42.1  & --34  20    00.3  &   0.05339  &  0.256  &  23.92  &   0.88  &   0.41  &  F    \\
7\_V5215   &  2 40 12.2  & --34  25    23.7  &   0.05413  &  0.262  &  23.43  &   0.49  &   0.27  &  FO   \\
2\_V16907  &  2 40 18.5  & --34  01    19.6  &   0.05490  &  0.309  &  23.88  &   0.44  &   0.34  &  F    \\
\enddata
%
\tablecomments{Units of right ascension are hours, minutes, and seconds,
and units of declination are degrees, arcminutes, and arcseconds.
Table~\ref{basic} is published in its entirety in the electronic edition
of the {\it Astrophysical Journal}. A portion is shown here for guidance
regarding its form and content.}
\tablenotetext{a}{Variable stars are ordered by increasing period, the
first digit of the ID gives the number of the CCD in the WFI mosaic, the
second part is the DAOPHOT identifier.}
\tablenotetext{b}{F: fundamental radial-mode pulsator; FO: first overtone
radial-mode pulsator; SL: subluminous variable.}
\end{deluxetable*}
\normalsize

\begin{deluxetable}{cccc}
\tablewidth{0pt}
\tablecaption{$B, V$ photometry of the Fornax dSph SX Phe stars}
\tablehead{
\multicolumn{4}{c}{Star 8$\_$V9899 - {\rm P=0.04619}} \\
\hline
\\[0.5mm]
{\rm HJD} & {V}  & {\rm HJD } & {B}\\
{\rm ($-$2452222)} &   & {\rm ($-$2452222) }
}
\startdata
  0.578 &  23.89 & 0.704 &  23.76 \\
  0.591 &  23.97 & 0.795 &  23.79 \\
  0.634 &  23.52 & 1.629 &  23.75 \\
  0.644 &  23.77 & 1.714 &  23.70 \\
  0.669 &  24.07 & 1.800 &  23.38 \\
  0.678 &  23.93 & 2.556 &  23.91 \\
  0.689 &  23.82 & 2.643 &  23.76 \\
\enddata
\label{data}
\tablecomments{Table~\ref{data} is published in its entirety in the
electronic edition of the {\it Astrophysical Journal}. A portion is
shown here for guidance regarding its form and content.}
\end{deluxetable}

\begin{figure}
\plotone{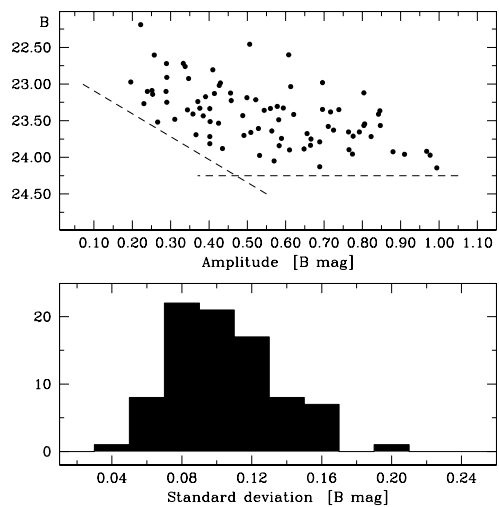}
\caption{
Precision and depth of our survey at the SX Phe stars magnitude
level. {\it Upper panel}: $B$ magnitudes of the detected
SX Phe stars $vs.$ the amplitude of their light variation.
{\it Lower panel}: Distribution of
the standard deviation showing the measurement precision for variable
stars in the 22.5--24.0 mag range.
}
\label{soglia}
\end{figure}

\begin{figure}
\plotone{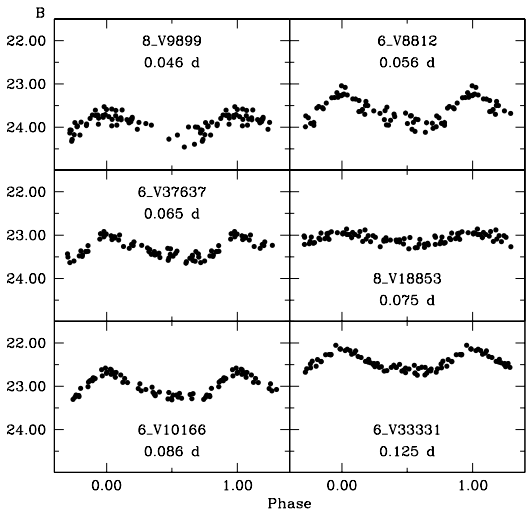}
\caption{
Examples of $B$ light-curves of SX Phe stars in Fornax. Note how the
mean brightness increases with increasing pulsation period. Light curves
of all variable stars are available in the electronic edition of the
journal.
}
\label{lc}
\end{figure}

\section[]{The Fornax SX Phoenicis star sample}
The analysis of the time series described in the previous section allowed us
to identify 85 SX Phe variables according to the criteria we describe
in this section.
It should be pointed out that the identification of short--period variables
is, in general, more difficult than, for instance, the detection of
RR Lyr stars.   Firstly, they are intrinsically much fainter than the 
RR Lyr stars
(indeed, only the use of the image subtraction permitted us to be successful
in Fornax), and secondly,
while the about half-a-day periodicity makes the RR Lyr variables
clearly stand out just looking at the intranight
light curves, this is not the case for the SX Phe stars, whose shorter
periods, smaller amplitudes and the fewer number of points per cycle make the
regular variability hardly discernible by simply plotting the points sequence.
Therefore, we performed the frequency analysis of the data by applying a least--squares
iterative sine--wave method \citep{vani} both to the whole 3-nights
time series and to the measurements of each individual night separately.
The comparison of the power spectra obtained in the two different ways allowed us to reject a number of
spurious candidates, i.e., stars for
which the scatter in just one night (or in a part of it) mimicked an apparent
variability. 
The fit of such an apparent variability can produce a false peak in the power spectrum owing to the
limited time coverage of our data.
To ensure the best quality of our sample, we thus accepted as bona--fide SX Phe 
stars only candidates whose power spectra showed the presence
of a peak at the same frequency value in all the three nights (see \citealt{luca}
for a detailed description of the full procedure).
                                                                                                                                 
Figure~\ref{soglia} shows the depth of our survey of the Fornax field at the
magnitude level of the SX Phe stars. The faintest
SX Phe stars in our sample have $\langle B\rangle $=24.0~mag. At this brigthness level
the smallest detectable amplitudes
are of about 0.5~mag, but they become smaller with increasing the star's brightness and
a 0.25~mag variability could be detected in stars with $\langle B\rangle >$23.5~mag.
In particular, small amplitude ($< 0.10$ mag) pulsators would be detectable
only if they have $B<$23.0, whilst 
almost all the SX Phe variables in Fornax are fainter than $B$=23.0.
The  700~s exposure time also makes very difficult  
the detection of small amplitude, very--short period pulsators.

Figure~\ref{lc} shows examples of the $B,V$ light curves of the Fornax SX Phe stars
we obtained at the end of the whole procedure.
The full atlas of light curves is published in the electronic
edition of the journal.
Identification (IDs and coordinates) and parameters of the light curves
(period, time of maximum light, mean $\langle B\rangle$ magnitude and
$\langle B\rangle - \langle V\rangle $ color, and $B$-amplitude $-A_B-$)
are provided in Table~\ref{basic},  which is published in its entirety
in the electronic edition of the journal.
They were calculated using cosine-series truncated at the
last significant term (usually the first harmonic, in exceptional cases
the second or the third) and  a least--squares fit.
Error bars on the $\langle B \rangle$ and $\langle V \rangle$ magnitudes are of
0.01 and 0.04 mag, respectively. Errors in the periods range from 
2$\times 10^{-5}$ to 7$\times 10^{-5}$~days.
The last column of  Table~\ref{basic} gives the pulsation characteristics. They were
 identified following the procedures described in Section 4.
The $B$ and $V$ photometry of all the SX Phe stars we
identified in Fornax is provided in  Table~\ref{data}, available
in electronic form in the electronic edition of the journal.

The lower panel of Fig.~\ref{soglia}
shows the distribution of the standard deviations of the $B$ light curve best fit models,
indicating a median precision of 0.10 mag. We also note that in seven cases
only (8\_V28121, 3\_V6718, 6\_V39151, 8\_V1625, 8\_V9899, 6\_V14592, and  8\_V20859)
the fit of the $V$--data was unsuccessful due to the incomplete phase coverage.
As usual for single--site observations, the frequency detections can be  affected
by the $\pm1$~days$^{-1}$ alias ambiguity. However, this has negligible
effects on the $\log P$ values for $P\leq0.10$~days;
for $P\geq0.11$~days
 the alias ambiguity  can shift the $\log P$ values  by --0.05 or +0.05 only.
                                                                                                                                 
The high amplitudes ($>$0.20~mag) and  the short
periods ($P\le$0.08~days in 83\% of the cases) of the stars in the Fornax sample
rule out  a significant
contamination  from rotational variables and binaries, which are expected to show smaller
amplitudes and longer periods. 
We stress that the 85 SX Phe variables make the Fornax dSph galaxy unique in richness of such
high--amplitude, short--period  pulsators. The globular clusters $\omega$ Cen
\citep{olech} and M55 \citep{pych} are also rich in SX Phe stars, but most of them
have amplitudes smaller than 0.10~mag and seem to pulsate in
nonradial modes. 

In CCD\,6 the globular cluster Fornax~3 (For~3) is resolved into stars, at least
in its outer regions, where several RR Lyr stars were identified.
Two SX Phe stars were found in the outskirts of For~3, namely 6\_V40765 and 6\_V38403.
Their vicinity to For~3 suggests that they could be cluster members.
On the other hand, star 7\_V22094 is at least
1.0 mag brighter than other SX Phe stars of similar period.
Assuming they also have same absolute magnitude, the 1.0 mag difference implies a ratio
of 0.6 between the distances. Therefore, 7\_V22094 possibly 
does not belong to the Fornax galaxy and it could be a star in the halo
of the Milky Way.

\begin{figure*}
\plottwo{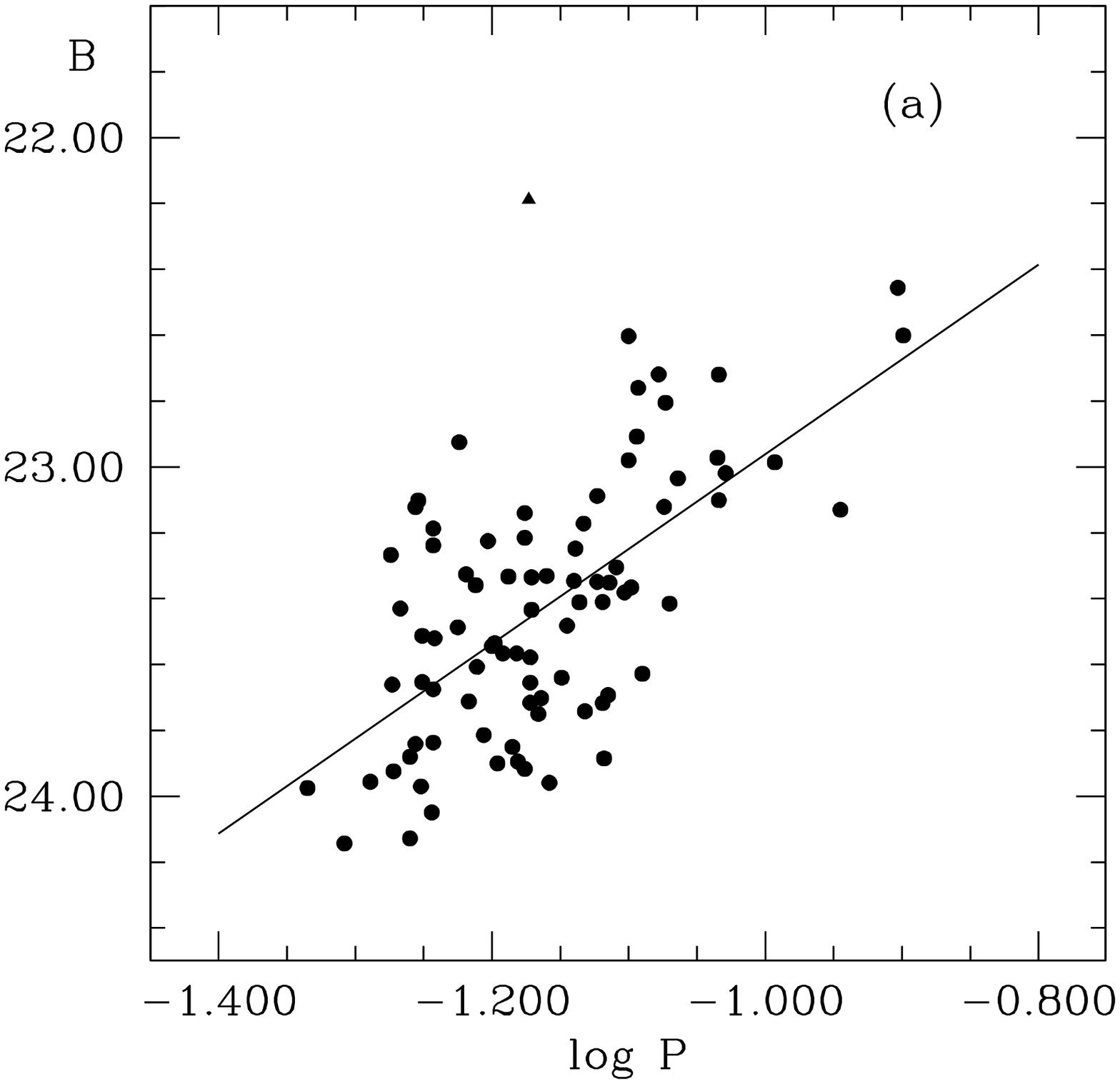}{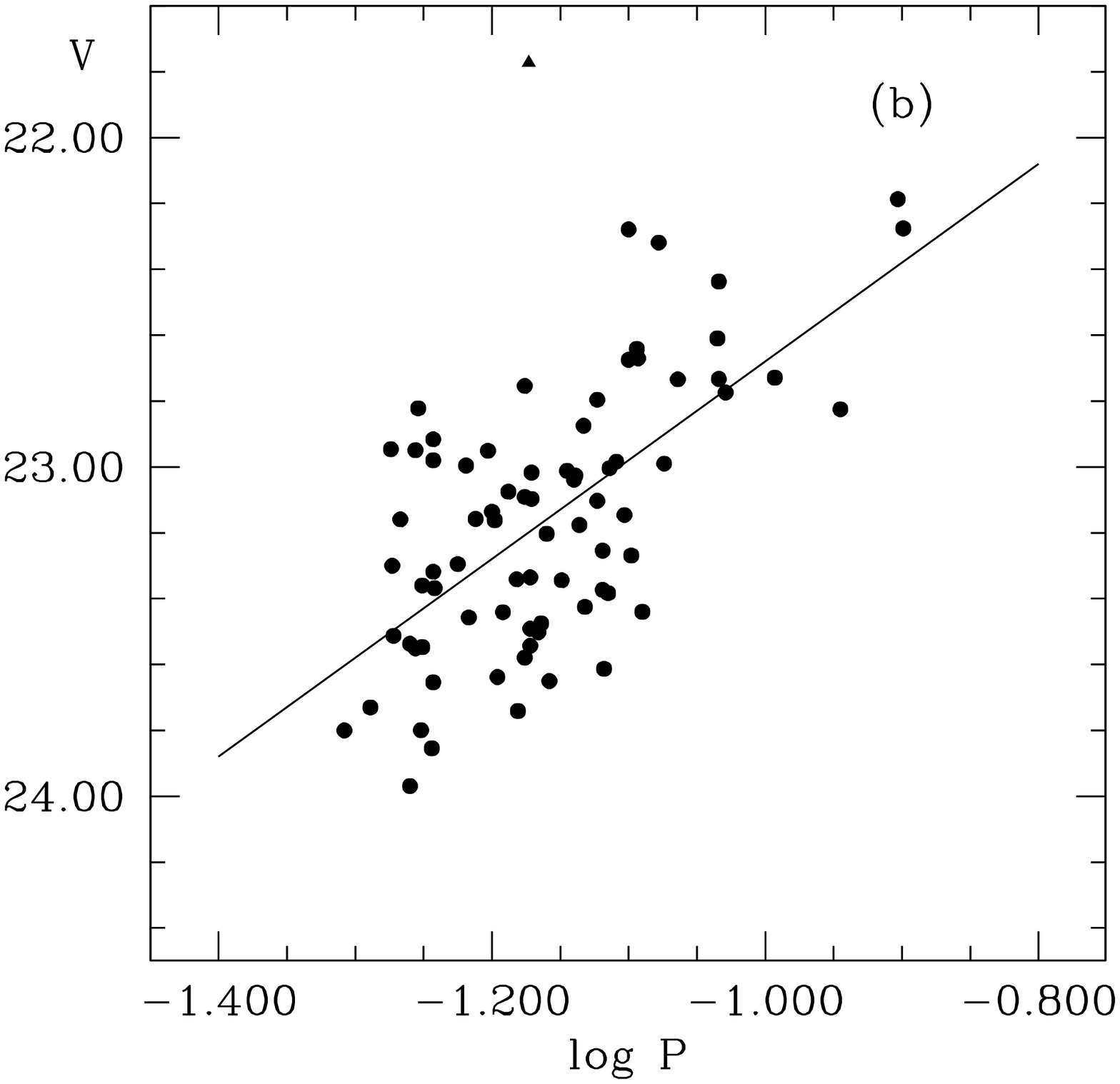}
\plottwo{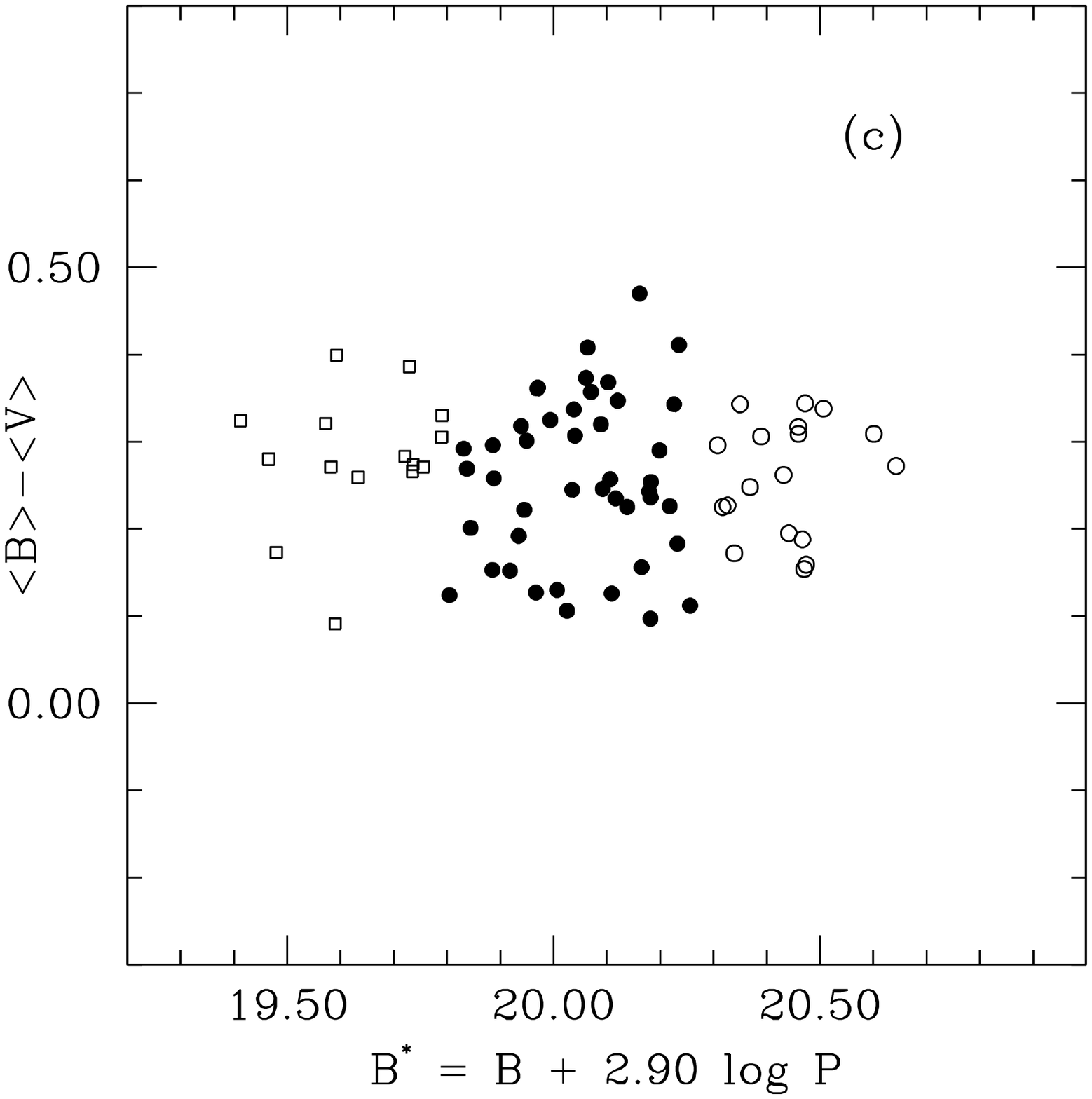}{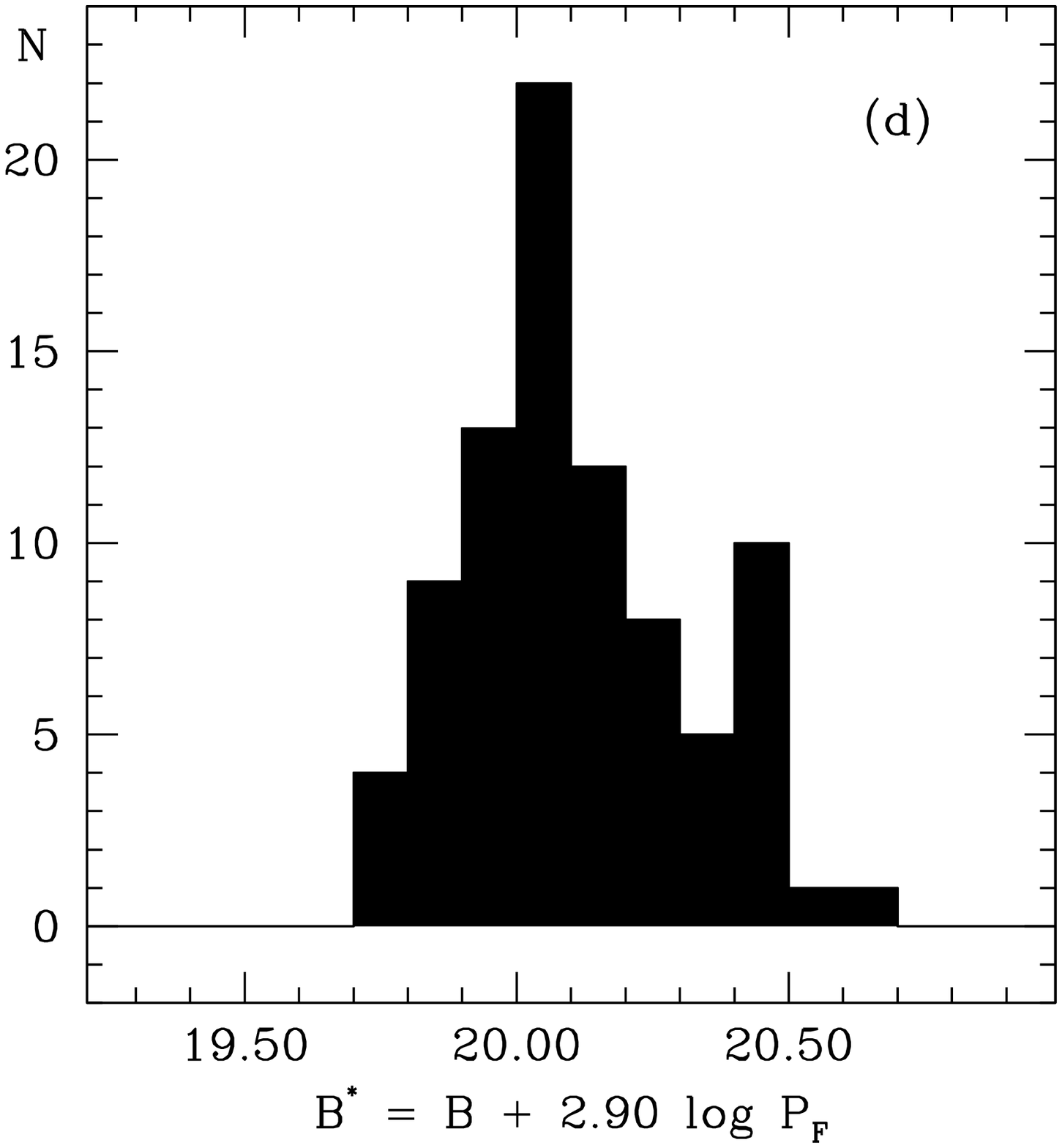}
\caption{Properties of the Fornax SX Phe variables. {\it Panel a: }
Mean $\langle B\rangle$
magnitudes plotted against observed periods. The
solid line is the least--squares fit calculated from these values
excluding star 7$\_$V22094 (filled triangle) which does
not belong to the Fornax dSph; {\it Panel b: }
Mean$\langle V\rangle$ magnitudes plotted against observed periods.
The solid line is the $P-L$
relation by \citet{nz} which was superposed to the data
assuming for Fornax a reddened
distance modulus of 20.78 mag; {\it Panel c: }  $B^\ast$--magnitudes
plotted against color indices (see text).
Filled circles are
stars pulsating in the fundamental radial mode; open squares are stars
pulsating in the first overtone; open circles indicate the group of
subluminous stars; {\it Panel d: }
Distribution of the $B^\ast$--magnitudes. 
}
\label{groups}
\end{figure*}

\section {Characteristics of the Fornax SX Phe stars}
\label{sec:charsxp}

As shown in Fig.~\ref{cmd}, 
most of the SX Phe stars lie at $22.4 <V< 23.6$ mag, hence from 1.0 to 2.2 mag fainter than
the well--defined central part of the galaxy horizontal branch, which is filled by the RR Lyr stars.
Their $B-V$ indices span about 0.40 mag, but
84\% of the SX Phe stars have $B-V$ colors in the range from 0.10 to 0.35 mag.
Taking into account the uncertainties in the $B-V$ values from the least--squares fit
($\pm0.04$~mag) and from the calibration procedures ($\pm$0.04~mag), the Fornax instability
strip appears to be slightly
larger than that of the  Milky Way. We note that the Galactic HADS and
SX Phe stars are located in the $(b-y)_0 = 0.13$--0.25 mag interval \citep{vienna},
which translates into a $B-V$ range narrower than 0.20 mag.
                                                                                                                                 
Due to the larger number of datapoints, our
$\langle B\rangle$ magnitudes are more accurate than the $\langle V\rangle$ ones and therefore
more suitable to study the characteristics of
the Fornax SX Phe stars. 
The properties of SX Phe stars in Fornax are summarized in Fig.~\ref{groups}.
Panel {\it a} shows the
mean $\langle B\rangle$ apparent magnitudes of the Fornax SX Phe stars
plotted versus  observed periods.
The least--squares linear fit over 84 stars (excluding star 7\_V22094)
is shown by the solid line; its standard deviation is 0.29~mag. 
This dispersion is  too large to be accounted for by geometrical effect.
If we consider the tidal radius of Fornax,
the difference in apparent magnitude between stars with same absolute magnitude
located at opposite sides of the galaxy does not exceed 0.10~mag.
On the other hand, since the RR Lyr stars are all grouped in the expected narrow range of the
horizontal branch, we can also exclude the existence of regions with different absorption
in Fornax. 
Regarding possible observational errors or variability misidentification, 
the SX Phe stars define very clearly the
instability strip below the horizontal branch. 

The spread of the $B$ magnitudes is 1.4~mag in the narrow $\log P$ interval
from --1.15 to --1.05 
(Fig.~\ref{groups}, panel {\it a}). 
Such spread is as large
as the total range spanned by $B$ magnitudes all along the observed periods.
 It follows that the dispersion of the points must be intrinsic to the SX Phe stars.
The same conclusion can be drawn when comparing 
the mean $\langle V\rangle$ magnitudes and 
observed periods of our SX Phe sample to the
Period--Luminosity ($P-L$) relations of \cite{nz}. We adopt $E_{B-V}$=+0.02 mag and 
$m-M$=20.72 mag
(dereddened value, \citealt{greco05,rizzi}) 
and [Fe/H]~$\approx -1$ dex (average metallicity of the Fornax stars,
\citealt{saviane,battaglia06}).
This comparison is shown in panel {\it b} of Fig.~\ref{groups}, where
the solid line is the $P-L$ relation of \cite{nz} which is valid for [Fe/H]$>-1.5$ dex.
If we adopt instead [Fe/H]=$-$2.0~dex,
and use the second $P-L$ relation given by \cite{nz}, which is valid for
[Fe/H]$<-1.5$ dex, we obtain a parallel line
shifted by only 0.06~mag toward fainter magnitudes.
                                                                                                                                 
Notwithstanding the large dispersion along the $y$--axis in panels {\it a} and
{\it b} of Fig.~\ref{groups}, it is quite evident that the Fornax SX Phe variables follow
a $P-L$ relation, since stars with longer periods are also brighter
(see also Fig.~\ref{lc}).
In the following subsections we will investigate the physical reasons that may cause the
observed  scatter of  the $B$ magnitudes at a given period.
                     
\subsection{Fundamental-mode and first-overtone pulsators}
When examining panels {\it a} and {\it b} of Fig.~\ref{groups},
it appears that several  stars are much brighter than expected from the $P-L$ relations
and  even more  are much fainter.
To provide a quantitative analysis  of the different groups of SX Phe stars that may be present in
our Fornax sample,
we calculated the $B^\ast$ values defined as
 $B^\ast$=$B+2.90~\log~P_F$, where $P_F$ is the period of the fundamental radial mode.
Here, the slope is that given by \cite{nz}. By using the observed periods,
we precisely evidenced 17 stars being 0.35~mag (or more) brighter than the 
value predicted by the $P-L$ relations.
We note that the aliasing effect ($\pm$1~days$^{-1}$) will produce a shift of only 0.10~mag
at $\log P=-1.115$~days, unable to account for the bright magnitudes observed for these 17
objects.

The presence of such  brighter stars in our SX Phe sample
can be explained by a different mode of pulsation.
First-overtone pulsation has been attributed to stars too bright for
the observed periods both in other galaxies (Carina, \citealt{matcarina} and 
\citealt{carina}; LMC, \citealt{lmc})
and in globular clusters ($\omega$ Cen, \citealt{olech};  M55, \citealt{pych}).
Indeed,  the fundamental period of a SX Phe star  pulsating 
in the first-overtone (FO) mode will be 1.290~times longer 
than the observed one,
according to the period ratio typical of the SX Phe pulsators
 ($P_{FO}/P_F=0.775$, \citealt{hadsdm}).
Therefore, the FO--star has to be shifted rightward by +0.111 in the
the $\log~P-B,V$  planes, in which the fundamental periods
have to be plotted on the abscissae. 
Assuming a slope value of the $P-L$ relations
in the range from --2.90 to --3.70, the period shift caused by the different
pulsation mode corresponds to a brightening of 0.32--0.40 mag.

Similarly to what usually done in other stellar systems,
we assume that the 17 stars identified by means of the procedure described above
(see panel {\it c} of Fig.~\ref{groups} for the 15 stars
having reliable color indices)
are FO--pulsators and  calculate their 
F--mode periods as $\log P_F$=$\log P$+0.111. 
This identification, though arbitrary, allows us to improve the quality
of the fit, by obtaining 
a well--mixed group
composed of ``fundamentalized" first--overtone and fundamental-mode pulsators.

\subsection{The subluminous variables}
The identification of the brighter stars as first--overtone
 pulsators   strongly reduces
the scatter above the $P-L$ relations. However,
the most remarkable feature in the $\log~P-B,V$ planes still remains the presence of a large
number of subluminous stars well below the $P-L$ relations toward the shortest periods,
both when the $P-L$
relation is calculated from the data as in panel {\it a} of Fig.~\ref{groups}, or when
it is assumed (Fig.~\ref{groups},  panel {\it b}).
To explain the stars below the $P-L$ lines by a different pulsation mode
we should admit that this mode has a period longer than that of the fundamental radial
mode. In such a case, these stars have to be shifted leftward in the $\log~P-B,V$ planes.
However, periods longer than the fundamental radial one could be identified only as
nonradial modes driven by gravity,
but these modes  are unable to produce the observed large amplitudes.

The stars below the $P-L$ lines can be better pointed out after removing the $\log P$ dependence.
We calculated the
distribution of the $B^\ast$ magnitudes within classes with an  amplitude bin of 0.10~mag
(Fig.~\ref{groups}, panel {\it d}).
We note that most of the subluminous stars have $\log P \simeq -1.20$; 
for these values the $\pm1$~days$^{-1}$
ambiguity can affect the distribution in a marginal way.

The histogram shows an asymmetric  distribution: 
there are 26 stars to the left of the central peak and 36 ones to the right. 
This fact cannot be ascribed
to an observational bias, which should work in the opposite sense (i.e., in reducing the
number of fainter stars, which are closer to the detection threshold; see Fig.~\ref{soglia}).
By using the \cite{ashman} test,
we considered the 21 variables  having $B^\ast>20.27$ as belonging to a group of subluminous stars
(Fig.~\ref{groups}, panel {\it c}).
The introduction of this group of subluminous stars  
accounts for the dispersions in the magnitudes of the SX Phe variables
(Fig.~\ref{groups}, panels {\it a} and {\it b}) and for the asymmetry in
their distribution (Fig.~\ref{groups}, panel {\it d}).

\begin{figure}
\plotone{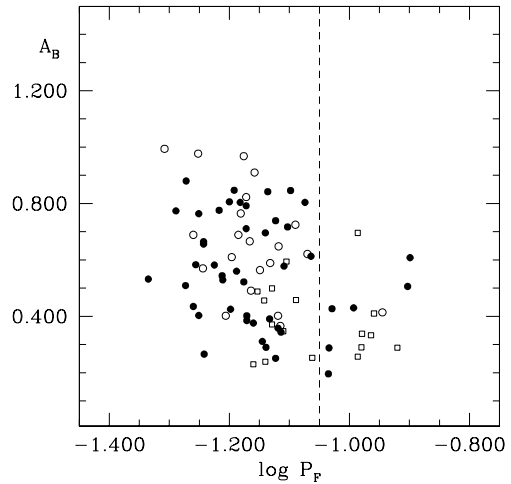}
\caption{
 Period--amplitude diagram of the Fornax SX Phe stars: a decrease in the
amplitude is observed for $\log P > -1.05 $.
Filled circles
are F pulsators, open squares are FO pulsators, open circles are subluminous variables.}
\label{ampli}
\end{figure}

Figure~\ref{groups}, panel {\it c} shows that there
 are no differences in the average $B-V$ color of subluminous
and other stars (0.26~mag for both groups), but 
that the subluminous variables span a narrower range
in $B-V$ than the other stars (0.15--0.34 against 0.09--0.47 mag).
The $\log~P-A_B$ plot shows an abrupt decrease in the amplitudes at $\log~P=-1.05$,
i.e., $P = 0.09$~days
(Fig.~\ref{ampli}) regardless  of the luminosity of the stars.
The change in light curve properties around this period value was already observed
in Galactic HADS and SX Phe stars (\citealt{ogle} and references therein) and in the
$\omega$ Cen SX Phe stars \citep{omega}.
  

\begin{figure}
\plotone{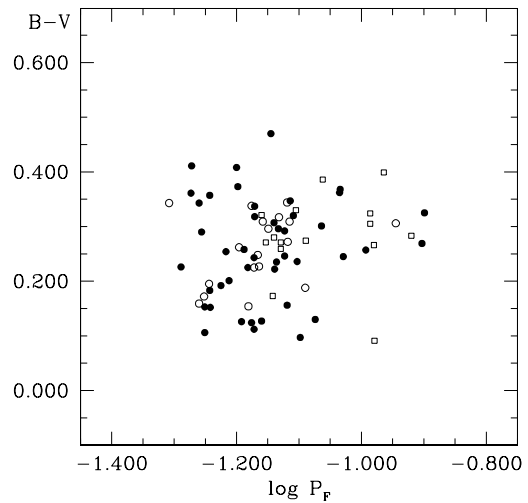}
\caption{Color versus period distribution of the SX Phe stars in Fornax.
Same symbols as in Fig.~\ref{ampli}.}
\label{bv}
\end{figure}

\section {$P-L$ relations for the SX Phe stars}\label{v21}
\begin{figure*}
\plottwo{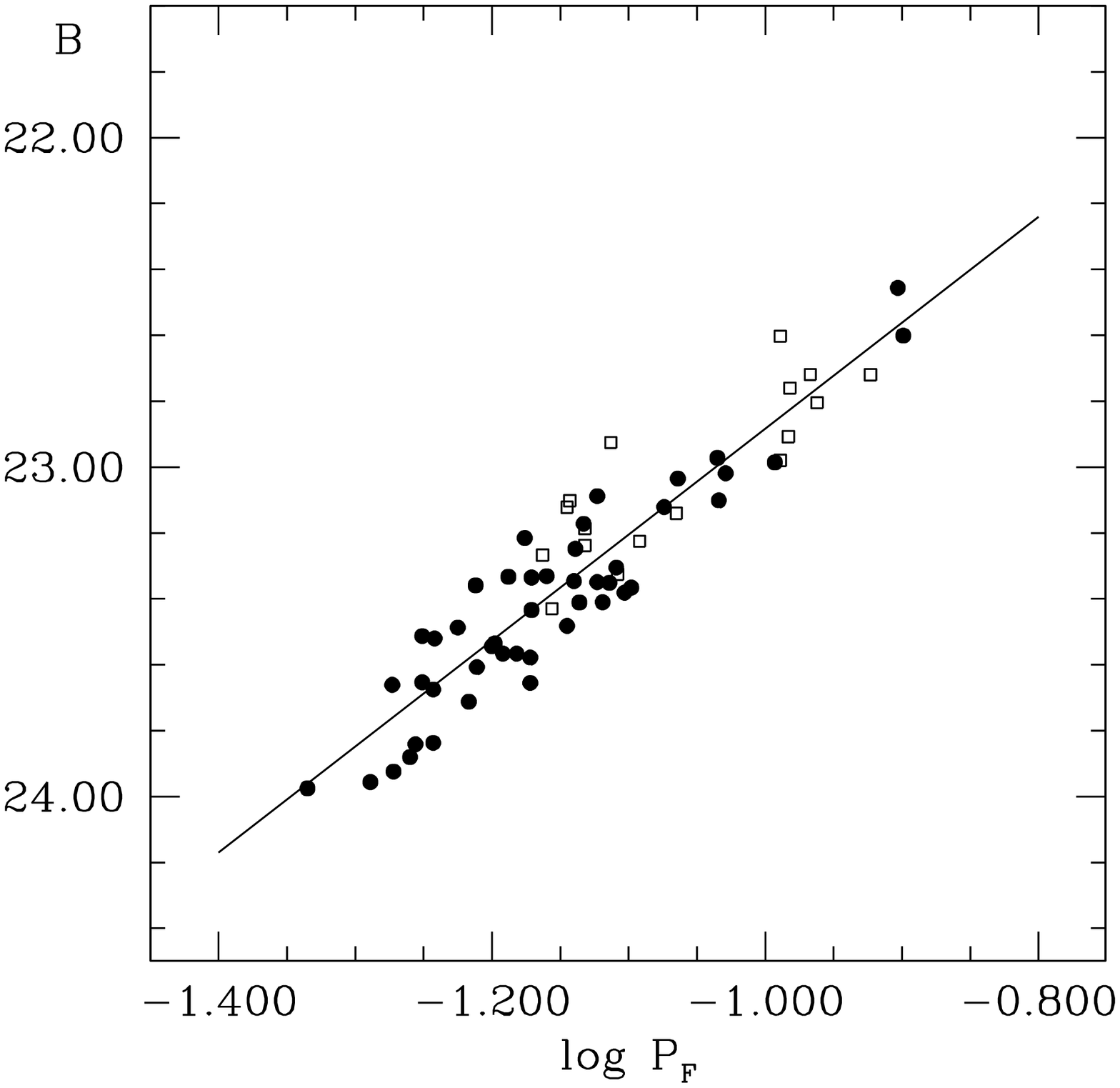}{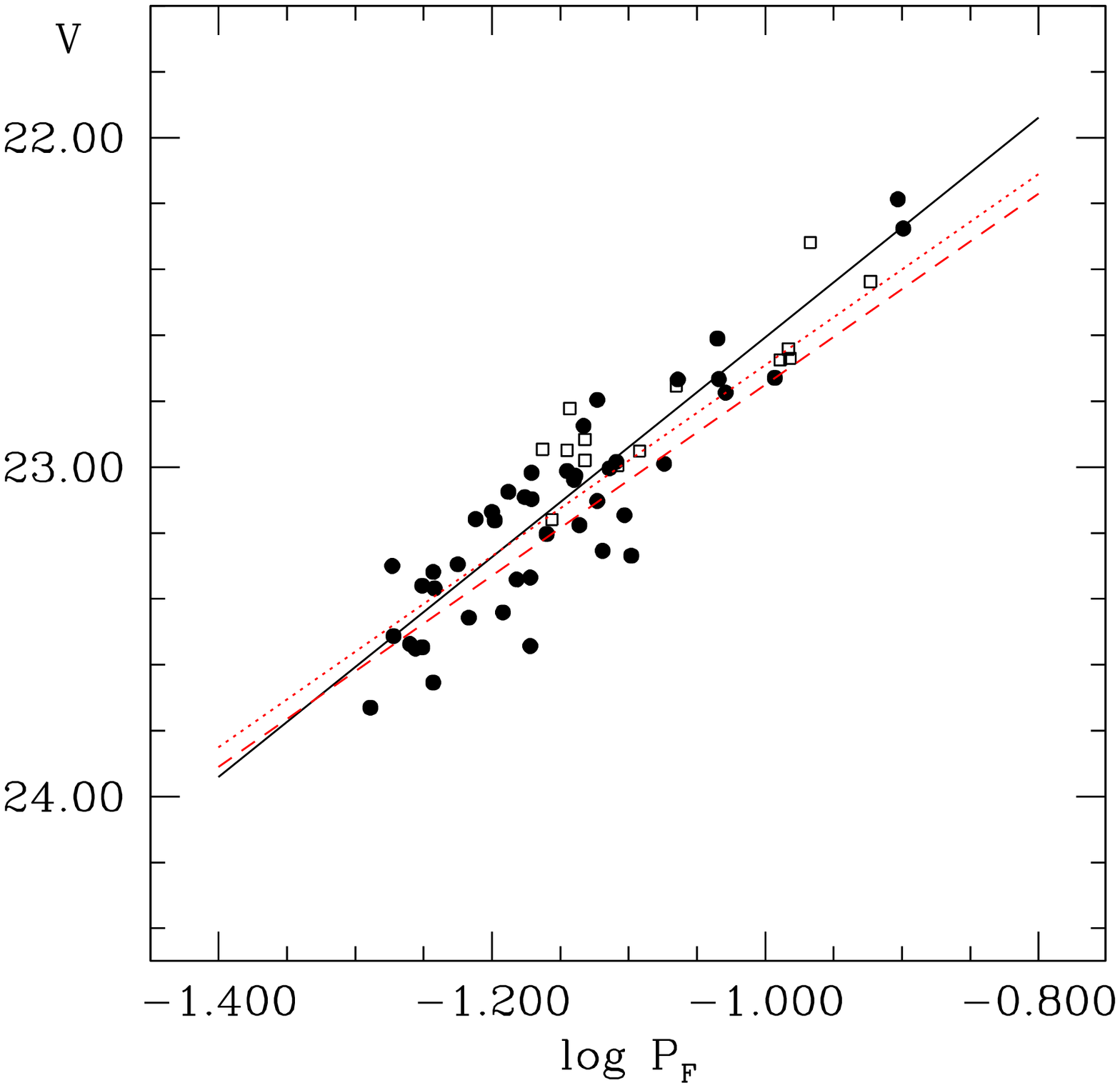}
\caption{
{\it Left panel:} $P-L$ relation of the Fornax SX Phe stars in the $B$-band.
The solid line  is the  $P-L$ relation computed from the Fornax fundamental-mode
(filled circles) and first-overtone (open squares) SX Phe variables. The latter
were ``fundamentalized" assuming $P_{FO}/P_F=0.775$.
{\it Right  panel:} $P-L$ relation of the Fornax SX Phe stars in the $V$-band.
The solid line is the  $P-L$ relation computed from the Fornax SX Phe variables.
Symbols are as in the left panel.
The (red) dashed lines are the $P-L$ relations given by 
\citet{nz} where for Fornax we have assumed $E_{B-V}=+0.02$ mag, $m-M_V$=20.78 mag (reddened value),
and [Fe/H]=$-1.4$ dex
(short dashes) or [Fe/H]=$-$ 2.0 dex (long  dashes).
See the electronic edition of the journal for a color version
of this figure.}
\label{perlum}
\end{figure*}

The large sample of SX Phe stars in Fornax allows us to discuss the $P-L$ relations,
both in the Fornax dSph and in other stellar systems. 
We evaluated the possible relevance of a color term by calculating
the $P-L-C$ relations using the $\langle B \rangle - \langle V \rangle$ values
and different samples 
(all stars, F-mode ones alone, FO-mode ones alone, subluminous ones alone, and F-mode plus fundamentalized
FO-mode ones).  In all cases, 
the least--squares fit of the $P-L-C$ relation yields a non--significant color term.
\citet{pych} obtained a similar null result when trying to improve
the quality of the fit by including a color term in the $P-L$ relation of stars
in M55.
Figure~\ref{bv} shows how there is no trend in the color versus period distribution.
The limited $B-V$ range (0.38~mag) and the
small color excess ($E_{B-V}=+0.02$ mag) of Fornax stars do not introduce significant changes
also when determining the $P-L$ relation using the extinction insensitive
Wesenheit index $W_V$ \citep{wese}.
Probably the color effect is too small to be evidenced in the short period range
covered by the Fornax SX Phe stars. 
In this respect, we note that most of the color dependence in M55  is due to
variable V21. This star has the longest
period and the reddest color in the \citet{pych} sample. However, its light curve shape
and amplitude (only 0.04~mag) suggest that V21 could  be a rotational variable.


\begin{figure}
\plotone{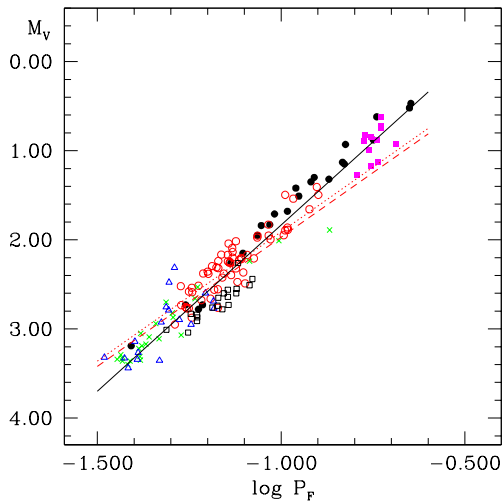}
\caption{$P-L$ distribution of short-period variable stars in different stellar systems:
Milky Way (black filled circles), Large Magellanic Cloud (magenta filled squares),
Fornax dSph (red open circles),
$\omega$ Cen (blue open triangles), M55 (green crosses), Carina dSph (black open squares).
The solid line shows the $P-L$ relation given by \cite{petersen}, i.e., with a slope of --3.73,
the dashed lines are the  $P-L$s obtained by \cite{nz}, i.e., with a slope of --2.90.
See the electronic edition of the journal for a color version
of this figure.
}
\label{pluniv}
\end{figure}

\subsection{The $P-L$ relation in Fornax}
According to our identification of
fundamental (46) and  first overtone (17) pulsators, we have determined
the $P-L$ relations.
First we calculated the fundamental periods of the
first-overtone  pulsators assuming $P_{FO}/P_{F}=0.775$.
Using fundamental and ``fundamentalized" periods (63 in $B$--light and
59 in $V$--light) we then obtained:
\begin{equation}
B=19.66 (\pm0.19)~-3.23(\pm0.17)\times \log P_F
\end{equation}
and
\begin{equation}
V=19.28 (\pm0.21)~-3.33(\pm0.20)\times \log P_F
\end{equation}
                                                                                                                                
The least--squares fits are shown by solid lines in Fig.~\ref{perlum}.
Standard deviations are 0.13~mag in $B$ and 0.15~mag in $V$, respectively.
The slope values are different from those assumed to calculate the $B^\ast$--magnitudes,
but the subdivision
into fundamental, first-overtone and subluminous pulsators does not change significantly:
the main consequence
of changing the slope value is a systematic shift of all the $B^\ast$--magnitudes toward
brighter values.
                                                                                                                                
For comparison purposes,
the $P-L-$[Fe/H] relations given by \cite{nz} are also shown
in the $\log~P-V$ plane of Fig.~\ref{perlum}.
The three lines are nearly coincident in
the $\log~P$ range covered by the Fornax sample. 
It is quite straightforward to use the \cite{nz} relation
to calculate the distance to the Fornax galaxy.
Assuming [Fe/H]=$-$1.4 dex, the 59 SX Phe stars with $V$ light curves yield
a reddened distance modulus  20.76 mag (s.d./$\sqrt{N}$=$\pm$0.02~mag), which
in turn leads to $(m-M)_0$=20.70 mag, a value in excellent agreement with
previous estimates (e.g., \citealt{buonanno,saviane,greco05,for4,rizzi}).
                                                                                                                                
\subsection{Comparison with other $P-L$ relations}\label{plcomp}


\begin{deluxetable*}{l l l l}
\tablewidth{0pt}
\tabletypesize{\scriptsize}
\tablecaption{
SX Phe and HADS stars in different stellar systems. }
\tablehead{
\multicolumn{1}{c}{Stellar system} &
\multicolumn{1}{c}{Period range} &
\multicolumn{1}{c}{SX Phe and HADS stars.} &
\multicolumn{1}{c}{Dereddened distance modulus} \\
\multicolumn{1}{c}{} &
\multicolumn{1}{c}{(days)} &
\multicolumn{1}{c}{Number, list and reference} &
\multicolumn{1}{c}{and reference} \\
}
\startdata
Fornax       &  0.050--0.130   &  59 (44 F and 15 FO), Tab.\ref{basic} in this paper  & 20.72, \cite{rizzi}, \\
       &     &    &  \cite{greco05} \\
\noalign{\smallskip}
Carina       &  0.045--0.075   &  19 (V1-V15,V17-20), in \cite{carina}           & 20.10, \cite{dallora}\\
\noalign{\smallskip}
LMC          &  $\sim$0.17     &  First 12 entries in  \cite{lmc},         & 18.49, \cite{lmc}\\
             &                 &   Tab.~2                                  &   from RR Lyr stars\\
\noalign{\smallskip}
$\omega$ Cen &  0.030--0.060   &  17 in \cite{olech} (V194,V195,  &
$M_V$ values from \cite{olech}\\
             &                 &  V199,V200,V204,V220,V225,V231,V237,&               \\
             &                 &  V238,V250,V252,V253,NV298,NV306,&    \\
             &                 &  NV324,NV326)&    \\
\noalign{\smallskip}
M55          &  0.030--0.100   &  23 (V16-V20,V22-V27,V31-V42) &   13.62, \cite{harris, harris03}\\
             &                 &  in \cite{pych} &  \\
\noalign{\smallskip}
Milky Way    &  0.039--0.225   &  First 23 entries in \cite{vienna},   &
$M_V$ values from \cite{vienna}\\
             &                 &  Tab.~1  &  \\
\label{unique}
\enddata
\end{deluxetable*}

There is observational evidence that short--period pulsators in a
given individual stellar system cover a limited range in period
and that these  period ranges are very different from one system to
another. The narrow period baseline hampers a reliable determination of the
slope of the $P-L$ relation in a single stellar system, where stars are expected to have
similar physical characteristics.
We have therefore
grouped all the SX Phe stars discovered so far in different stellar systems
(Table~\ref{unique}) to span a period range
as large as possible in the $\log~P-M_V$ plane.
To combine data from different systems, (i) we have used $M_V$
  values where available, or calculated $M_V$ values for the short-period
variables using dereddened distance moduli for the host system that are
not based on the variables themselves
(last column in Table~\ref{unique}); (ii) we have ``fundamentalized"
the first--overtone pulsators following the identifications
provided by the
authors\footnote{An unintentional error occurred in Fig.~10 of \cite{pych}.
The shorter period of V35
is used instead of the longer one: if the correct period is used,
the clearness of the first--overtone sequence becomes less evident.}; (iii) we have omitted
the nonradial pulsators; and
(iv) all the uncertain cases were not considered, in the case of both single stars
 (e.g., V16 in \citealt{carina}) and of stellar systems (e.g., NGC~3201, which  is
affected by internal differential reddening; \citealt{anna}).
Finally, we have verified that the adopted distance moduli are
on the same distance scale, i.e., $(m-M)_0$=18.50~mag for the Large Magellanic Cloud.

The combined $P-L$ relation is shown in Fig.~\ref{pluniv}, where
  the SX Phe stars of $\omega$ Cen, M55, Carina, and Fornax 
clearly describe a linear relationship.
When extending the sequence toward longer periods by adding 12 fundamental pulsators
in the Large Magellanic Cloud
(which are classified as $\delta$ Sct variables)
and 23 pulsators in the Milky Way with $P<0.25$~days (mostly $\delta$ Sct stars,
for which $P=0.25$~days is a safe and well established upper limit; \citealt{ogle}),
 the relation continues in  a very regular and natural way.
The outlier located
below the sequence at $\log~P=-0.87$ is variable V21 in M\,55. Its discrepancy
with respect to other stars strengthens our hypothesis that the star is actually
a rotational variable  (see Sect.~\ref{v21}).
The least--squares linear fit calculated over the total sample of stars in Fig.~\ref{pluniv}
and omitting V21 in M55 (153 stars in total) is:
\begin{equation}\label{plsx}
M_V= -1.83 (\pm0.08)~-3.65(\pm0.07)\times \log P_F
\end{equation}
with standard deviation of 0.18~mag.
The slope of $-3.65$ is in
good agreement with first determinations obtained using short--period stars and
no  metallicity term, e.g., $-3.725$ \citep{vienna} and $-3.73$
\citep{petersen}. The three $P-L$ relations are nearly coincident
and for clarity only the line calculated using the \cite{petersen} parameters is shown in Fig.~\ref{pluniv}
as a solid line.
We also plot in Fig.~\ref{pluniv} the  relations derived by \cite{nz}, which
are practically coincident with those reported by \cite{fernie} and \cite{laney},
and use a slope close to $-$2.90.
The Fornax and the Milky Way stars with $\log P>-1.0$ and
the M55 stars at the shortest periods seem to fit the relation with
slope of $-$3.73 better than those with a $-$2.90 slope.
                                                                                                                                 
We stress that when $P-L$ relations with a given slope (no matter whether $-$2.90
or $-$3.73) are used to fit periods and magnitudes of variable stars
in individual clusters or galaxies, results are generally satisfactory, since
their periods span too narrow ranges to put a strong constraint on the slope.

\section{Summary and Discussion}
The detection of  a large and homogenous sample of short--period
variables in the Fornax dSph galaxy allowed us both to
discuss the properties of SX Phe stars 
and to revisit their $P-L$ relation.

\subsection{Origin of the subluminous variables} 
The main issue of our analysis is the detection of a 
large scatter around the $P-L$ relation. We propose to explain such a 
scatter with 
the occurrence of different pulsation modes and the presence of
a group of subluminous stars. 
A mix of pulsational modes among SX Phe stars 
\citep{matcarina, carina, pych, olech, lmc}
and in other pulsating stars (Red giants: \citealt{kiss1, kiss2};
Cepheids: \citealt{udalski}) is quite common. The 
subluminous stars, shown here for the first time to be a significant
population, appear to have been detectable because they exhibit  
high--amplitude radial pulsations and are possibly related to the complex
stellar populations of Fornax.
We suggest that the scatter in luminosity of short--period variable stars
in Fornax  has a physical rather than instrumental explanation, since
observational or geometrical effects are producing a much smaller dispersion.
We will attempt a preliminary discussion of possible
explanations for the observed scatter, including a metallicity effect and  
different star--formation processes.

The metallicity effect has already been suggested by \cite{nz} to
reproduce the $P-L$ relations in different environments. 
In Fornax  the subluminous
stars,  which are about 25\% of the total SX Phe population,  
should be very metal--deficient stars.
In this context, the very low metal content could
explain the lack of long periods, since acoustic waves propagate faster into lighter media.
If we adopt the  [Fe/H] dependence of
$-0.190$ \citep{nz}, a 1.8~dex difference would be needed to produce the observed 
0.35~mag luminosity difference between subluminous variables and 
the bulk of SX Phe stars. \citet{saviane} and
\citet{battaglia06} report on two distinct populations, a metal--rich
([Fe/H]$>-1.3$) component and a metal--poor ([Fe/H]$<-1.3$) one. 
\citet{coleman} suggest three peaks in the metallicity distribution
([Fe/H]$\simeq$--1.0, --1.5, and $\leq$--2.0). Therefore, the SX Phe variables
should cover the whole range of metallicities spanned by the Fornax stars
and the subluminous stars should belong to the extremely metal--poor population. 
However, we note that the results obtained in Sect.~\ref{plcomp} 
suggest that the metallicity dependence is not as strong as previously inferred.

A related hypothesis is that the SX Phe stars' population in
Fornax arises from different star formation processes.  SX Phe stars are
located in a CMD region where intermediate-age stars (the dominant
population in Fornax) and old blue-stragglers may co-exist
\citep[e.g.,][]{momany07}.
The isochrones shown in Fig.~\ref{cmd} suggest (adopting a metallicity
$Z=0.001$) an age of $\sim$3 to 6~Gyr (corresponding to 
a range in mass from 1.0 to 1.2 M$_\sun$ at the turnoff point) for
the SX Phe stars arising from a population of single intermediate--age
stars. 
A sizable ancient population ($>$10~Gyr) is also present  
\citep{saviane,battaglia06,coleman},
as demonstrated by the large number of RR Lyr stars \citep{greco05} and
by the fit of the position of RR Lyr stars in the CMD 
with the mean ridge line of the Galactic globular 
cluster M3, certainly older than 10~Gyr (Fig.~\ref{cmd}). 
Pulsating stars resulting from the normal evolution of a
single star in a finite range of ages and from the merging 
of a close binary system could
co--exist below the horizontal branch.  We can speculate that 
the subluminous variables are the latter ones: after the coalescence
stage, the stellar structure can be different from that produced 
by the normal evolution of a single intermediate-age star.
Since differences in the $B-V$ indices (i.e., temperature
effects) are not observed, we must infer that the blue stragglers would
have a smaller radius.
To explain a 0.35~mag difference, the radius of a subluminous star
should be 86\% that of a normal star having same temperature. In turn, a
smaller radius explains the lack of long periods.
In a similar way, the blue stragglers could slightly deviate 
from the classical mass--luminosity relation, also taking into account that stars of different
age are populating the instability strip below the horizontal branch (Fig.~\ref{cmd}).
These explanations, still tentative,  seem attractive, since they would provide 
observational constraints on discriminating blue straggler stars in dwarf
spheroidal galaxies and on theoretical models of blue stragglers' formation.

Both hypotheses need future investigations. 
The presence
of a group of subluminous variables may offer the possibility to use SX Phe
stars as star formation tracers, since either in the case of very metal poor stars
or in the case of the final stage of the evolution of close binary systems they should belong
to a very old population. 
Spectroscopic observations 
of the SX Phe stars will be very useful to verify if the spread in the observed magnitudes 
is due to a metallicity effect as well as to study the kinematics of the different groups. 
The subluminous and the normal stars show the same spatial distribution,
but we also reconsider the possibility that the subluminous
stars are related to a depth effect along the line of sight,
i.e., to streams associated with the Fornax dSph galaxy.
In such a case, the 0.35~mag difference 
implies as much as 17\% in distance, which seems unlikely \citep{coleman}. 
Moreover, we did not 
detect any subluminous RR Lyr variable able to support a depth effect.

Have subluminous short-period stars been observed in other
  environments ?
 When modeling the SX Phe pulsating stars in $\omega$ Cen, \cite{olech}
found at least four cases of  SX Phe stars (with amplitude from 0.03 to 0.16 mag)
whose periods can be explained by  masses significantly lower than expected.
The SX Phe variables in Carina seem to be systematically below the $P-L$
relations in Fig.~\ref{pluniv}. This shift is small, but it could become
more relevant if Carina has a distance modulus smaller than the adopted one
(i.e., 20.10~mag); we note that the literature values 
range from 19.87 mag \citep{mighell} to 20.19 mag \citep{dolphin}.

\subsection{The slope of the $P-L$ relation}                                                                                                               
We obtained a slope
value of $-$3.65 by combining stars in six different stellar systems. 
This slope differs significantly from the $-$2.90 value currently used,
but agrees very well with previous estimates ($-3.725$,
\citealt{vienna}; 
$-3.73$, \citealt{petersen}).
The disagreement between the two slope values
can be ascribed to different reasons. The steeper value has been (re--)obtained here as a
straightforward result (Fig.~\ref{perlum}), without considering
a metallicity term.
The importance of the metallicity effect is still a debated matter even in the case of
Cepheids \citep{storm}.  Moreover, since
the metal content of individual stars in
galaxies such as Carina, Fornax, and the Magellanic Clouds is still rather
difficult to measure, is common procedure to use average metallicity values, which
may not be very reliable. Thus neglecting the metallicity term could still be reasonable
as long as accurate metal abundance of individual stars are not available.
The shallower value of $-$2.90 is also supported by the
assumption that short--period stars should be linked to Cepheids \citep{fernie}.
The application of this constraint to the Milky Way stars \citep{lmc} is based on the
identification of the fundamental period in low--amplitude, multiperiodic pulsators
(which are the only $\delta$ Sct stars having  reliable {\sc hipparcos} parallaxes).
However,  this identification
is not obvious and can be made only by combining photometric 
and spectroscopic observations \citep{mante}.  A general improvement in the
$P-L$ relation of SX Phe and $\delta$ Sct stars (as the reduction of the scatter
and the introduction of  a second term) will be possible when accurate 
multicolor photometry and metallicity values for individual stars in different 
stellar systems are available. 

\section{Conclusions}

The intensive observation of a large part of the Fornax dSph galaxy 
down to 3.0~mag below the horizontal branch yielded us  a suitable
tool to investigate the properties of the Fornax stellar populations.
The peculiar characteristics of the sample of SX Phe variables,
as the group of subluminous variables, provide
useful clues on the stellar formation processes.
The numerous sample of
Fornax variables significantly increased the number of SX Phe variables known
in different stellar systems,
allowing us the possibility to obtain a comprehensive $P-L$ relation.

We plan to determine the physical parameters of a subset of  stars
fitting their well defined light curves  by means of  theoretical pulsational models.
Moreover, the  Fornax project  will be completed by 
future works on the other variables (namely, RR Lyr
stars, anomalous and Pop.~II Cepheids, eclipsing binaries).

\section*{Acknowledgments}
Part of this work has been the subject of the Laurea thesis of LDA. His supervisor
and our colleague, Prof.~Laura E.~Pasinetti, suddendly passed away on September 13, 2006.
Several astronomers have been trained under her tutelage  and we gratefully honor
her memory. 
The authors also wish to thank H.~McNamara for useful comments on a first draft
of the manuscript, as well as M.~Catelan and H.A.~Smith for the constant support
to the Fornax project. We thank the anonymous referee for comments that have
helped to improve the paper.
The research was funded by PRIN INAF 39/2005 (P.I. M.~Tosi)
and by PRIN INAF 2006 (P.I. G.~Clementini).

\newpage

\section{ON LINE MATERIAL}

Tables 2 and 3 can be requested from the first author as an ASCII 
and a tar file, respectively.

\setcounter{figure}{2}
\setcounter{table}{1}
\begin{longtable}[10pt]{l r r rcrrc l } 
\caption{Identification and properties of the Fornax dSph SX Phe stars}\\
\hline
\hline
\noalign{\smallskip}
\multicolumn{1}{c}{Name\tablenotemark{a}}  & \multicolumn{1}{c}{$\alpha$ (2000.0)} & \multicolumn{1}{c}{$\delta$ (2000.0)} &  \multicolumn{1}{c}{Period} &
\multicolumn{1}{c}{Epoch (T$_{\rm max}$)} &
\multicolumn{1}{c}{$\langle B \rangle$} & A$_B$ &  $\langle B \rangle$ - $\langle V \rangle$ & \multicolumn{1}{c}{Type\tablenotemark{b}}\\
&\multicolumn{1}{c}{} & \multicolumn{1}{c}{}   & \multicolumn{1}{c}{(days)} &  \multicolumn{1}{c}{(HJD-2452222)} & & & \\
\noalign{\smallskip}
\hline
\endfirsthead
\noalign{\smallskip}
\caption{continued}\\
\hline
\hline
\noalign{\smallskip}
\multicolumn{1}{c}{Name\tablenotemark{a}}  & \multicolumn{1}{c}{$\alpha$ (2000.0)} & \multicolumn{1}{c}{$\delta$ (2000.0)} &  \multicolumn{1}{c}{Period} &
\multicolumn{1}{c}{Epoch (T$_{\rm max}$)} &
\multicolumn{1}{c}{$\langle B \rangle$} & A$_B$ &  $\langle B \rangle$ - $\langle V \rangle$ & \multicolumn{1}{c}{Type\tablenotemark{b}}\\
&\multicolumn{1}{c}{} & \multicolumn{1}{c}{}   & \multicolumn{1}{c}{(days)} &  \multicolumn{1}{c}{(HJD-2452222)} & & & \\
\noalign{\smallskip}
\hline
\noalign{\smallskip}
\endhead
\endfoot
\noalign{\smallskip}
8\_V9899       &  2 41 14.2  & --34 23 02.7  &   0.04619  &  0.266  &  23.98  &   0.53  &   -      &  F    \\
7\_V8618       &  2 40 20.2  & --34 23 58.6  &   0.04917  &  0.170  &  24.14  &   0.99  &   0.34  &  SL   \\
2\_V3796       &  2 40 12.1  & --34 09 07.3  &   0.05137  &  0.253  &  23.96  &   0.77  &   0.23  &  F    \\
5\_V6170       &  2 38 51.9  & --34 24 46.7  &   0.05326  &  0.289  &  23.27  &   0.23  &   0.32  &  FO   \\
5\_V15474      &  2 39 02.6  & --34 20 26.4  &   0.05338  &  0.238  &  23.66  &   0.51  &   0.36  &  F    \\
5\_V16343      &  2 38 42.1  & --34 20 00.3  &   0.05339  &  0.256  &  23.92  &   0.88  &   0.41  &  F    \\
7\_V5215       &  2 40 12.2  & --34 25 23.7  &   0.05413  &  0.262  &  23.43  &   0.49  &   0.27  &  FO   \\
2\_V16907      &  2 40 18.5  & --34 01 19.6  &   0.05490  &  0.309  &  23.88  &   0.44  &   0.34  &  F    \\
6\_V15390  &  2 39 51.8  & --34 23 29.6  &   0.05493  &  0.229  &  24.13  &   0.69  &   0.16  &  SL   \\
2\_V4387   &  2 40 19.8  & --34 08 49.4  &   0.05540  &  0.276  &  23.84  &   0.58  &   0.29  &  F    \\
7\_V4659   &  2 40 22.2  & --34 25 37.7  &   0.05545  &  0.314  &  23.12  &   0.46  &   0.17  &  FO   \\
7\_V6056   &  2 40 28.8  & --34 25 02.3  &   0.05569  &  0.230  &  23.10  &   0.24  &   0.28  &  FO   \\
7\_V7347   &  2 40 04.0  & --34 24 30.9  &   0.05596  &  0.245  &  23.97  &   0.98  &   0.17  &  SL   \\
7\_V19493  &  2 39 59.2  & --34 19 24.8  &   0.05613  &  0.234  &  23.51  &   0.40  &   0.15  &  F    \\
6\_V8812   &  2 39 31.0  & --34 25 14.8  &   0.05616  &  0.322  &  23.65  &   0.76  &   0.11  &  F    \\
3\_V5563   &  2 39 19.6  & --34 07 57.8  &   0.05704  &  0.219  &  24.05  &   0.57  &   0.19  &  SL   \\
7\_V13345  &  2 40 32.7  & --34 21 59.7  &   0.05713  &  0.228  &  23.24  &   0.37  &   0.26  &  FO   \\
7\_V19309  &  2 40 14.9  & --34 19 29.3  &   0.05716  &  0.323  &  23.19  &   0.50  &   0.27  &  FO   \\
6\_V44543  &  2 39 26.1  & --34 14 37.1  &   0.05717  &  0.330  &  23.67  &   0.66  &   0.36  &  F    \\
2\_V11324  &  2 40 04.1  & --34 04 59.9  &   0.05717  &  0.290  &  23.84  &   0.67  &   0.18  &  F    \\
7\_V33186  &  2 40 16.5  & --34 13 29.4  &   0.05726  &  0.321  &  23.52  &   0.27  &   0.15  &  F    \\
2\_V1606   &  2 40 20.9  & --34 10 15.5  &   0.05957  &  0.292  &  23.49  &   0.58  &   0.19  &  F    \\
8\_V1625   &  2 40 44.1  & --34 26 45.7  &   0.05971  &  0.278  &  22.92  &   0.35  &   -     &  FO   \\
1\_V6694   &  2 41 06.3  & --34 07 22.7  &   0.06041  &  0.325  &  23.33  &   0.59  &   0.33  &  FO   \\
6\_V40765  &  2 39 47.5  & --34 15 53.3  &   0.06073  &  0.323  &  23.71  &   0.78  &   0.25  &  F    \\
6\_V49021  &  2 39 43.0  & --34 12 52.0  &   0.06134  &  0.290  &  23.36  &   0.54  &   0.20  &  F    \\
8\_V28121  &  2 41 06.8  & --34 14 48.1  &   0.06147  &  0.316  &  23.61  &   0.53  &   -  &  F    \\
3\_V6718  &  2 39 27.8  & --34 07 16.3  &   0.06230  &  0.316  &  23.81  &   0.40  &   -  &  SL   \\
6\_V18889  &  2 39 42.3  & --34 22 33.2  &   0.06265  &  0.266  &  23.23  &   0.46  &   0.27  &  FO   \\
5\_V26561  &  2 38 41.8  & --34 14 36.8  &   0.06303  &  0.329  &  23.54  &   0.81  &   0.41  &  F    \\
3\_V10893  &  2 39 40.3  & --34 04 37.1  &   0.06340  &  0.336  &  23.53  &   0.43  &   0.37  &  F    \\
2\_V7944   &  2 40 15.4  & --34 06 54.1  &   0.06371  &  0.302  &  23.90  &   0.61  &   0.26  &  SL   \\
6\_V35797  &  2 39 26.7  & --34 17 33.4  &   0.06420  &  0.338  &  23.57  &   0.85  &   0.13  &  F    \\
6\_V37637  &  2 39 53.2  & --34 16 56.0  &   0.06481  &  0.276  &  23.33  &   0.56  &   0.26  &  F    \\
6\_V39151  &  2 39 31.6  & --34 16 26.3  &   0.06529  &  0.371  &  23.85  &   0.69  &   -  &  SL   \\
2\_V4169   &  2 40 31.3  & --34 08 55.3  &   0.06581  &  0.308  &  23.57  &   0.80  &   0.22  &  F    \\
5\_V510    &  2 39 07.2  & --34 27 19.8  &   0.06596  &  0.330  &  23.90  &   0.76  &   0.15  &  SL   \\
7\_V2872   &  2 40 19.9  & --34 26 22.6  &   0.06661  &  0.285  &  23.92  &   0.97  &   0.34  &  SL   \\
7\_V34972  &  2 40 25.8  & --34 12 40.7  &   0.06662  &  0.311  &  23.22  &   0.52  &   0.12  &  F    \\
8\_V4532   &  2 40 41.7  & --34 25 29.2  &   0.06668  &  0.344  &  23.14  &   0.25  &   0.39  &  FO   \\
7\_V22094  &  2 40 26.2  & --34 18 17.5  &   0.06717  &  0.369  &  22.19  &   0.22  &   0.42  &       \\
6\_V22746  &  2 39 50.7  & --34 21 26.7  &   0.06722  &  0.336  &  23.58  &   0.71  &   0.24  &  F    \\
7\_V15830  &  2 40 05.8  & --34 20 58.3  &   0.06729  &  0.278  &  23.72  &   0.82  &   0.22  &  SL   \\
2\_V8650   &  2 40 11.9  & --34 06 30.3  &   0.06734  &  0.321  &  23.66  &   0.79  &   0.11  &  F    \\
7\_V18358  &  2 40 07.0  & --34 19 52.6  &   0.06739  &  0.312  &  23.33  &   0.40  &   0.32  &  F    \\
8\_V9427   &  2 40 51.0  & --34 23 16.5  &   0.06741  &  0.238  &  23.43  &   0.38  &   0.34  &  F    \\
2\_V18380  &  2 40 11.8  & --34 00 13.5  &   0.06817  &  0.435  &  23.75  &   0.67  &   0.25  &  SL   \\
6\_V44201  &  2 39 23.1  & --34 14 44.9  &   0.06849  &  0.252  &  23.70  &   0.49  &   0.23  &  SL   \\
7\_V6259   &  2 40 09.5  & --34 24 57.5  &   0.06921  &  0.410  &  23.33  &   0.38  &   0.13  &  F    \\
7\_V16899  &  2 40 35.6  & --34 20 30.4  &   0.06945  &  0.394  &  23.96  &   0.91  &   0.31  &  SL   \\
3\_V17080  &  2 39 35.3  & --33 59 20.1  &   0.07096  &  0.334  &  23.64  &   0.56  &   0.30  &  SL   \\
6\_V5831   &  2 39 46.1  & --34 26 02.4  &   0.07168  &  0.309  &  23.48  &   0.31  &   0.47  &  F    \\
2\_V9849   &  2 40 23.6  & --34 05 50.3  &   0.07240  &  0.386  &  23.35  &   0.70  &   0.31  &  F    \\
6\_V3640   &  2 39 28.4  & --34 26 36.7  &   0.07256  &  0.294  &  23.25  &   0.29  &   0.22  &  F    \\
6\_V38875  &  2 39 29.1  & --34 16 31.6  &   0.07316  &  0.409  &  23.41  &   0.84  &   0.23  &  F    \\
6\_V37380  &  2 39 20.6  & --34 17 01.1  &   0.07362  &  0.316  &  23.17  &   0.39  &   0.30  &  F    \\
8\_V10941  &  2 40 42.0  & --34 22 36.7  &   0.07373  &  0.352  &  23.74  &   0.59  &   0.32  &  SL   \\
8\_V18853  &  2 41 11.4  & --34 19 02.4  &   0.07531  &  0.346  &  23.09  &   0.25  &   0.29  &  F    \\
6\_V38403  &  2 39 46.8  & --34 16 40.7  &   0.07532  &  0.392  &  23.35  &   0.74  &   0.25  &  F    \\
2\_V881   &  2 40 14.6  & --34 10 37.5  &   0.07597  &  0.346  &  23.72  &   0.40  &   0.34  &  SL   \\
7\_V35654  &  2 40 02.8  & --34 12 22.7  &   0.07598  &  0.435  &  23.41  &   0.36  &   0.16  &  F    \\
7\_V9132   &  2 40 01.4  & --34 23 46.0  &   0.07623  &  0.365  &  23.89  &   0.65  &   0.27  &  SL   \\
8\_V22313  &  2 41 10.9  & --34 17 28.1  &   0.07669  &  0.385  &  23.69  &   0.37  &   0.31  &  SL   \\
8\_V9363   &  2 41 16.5  & --34 23 17.0  &   0.07694  &  0.448  &  23.35  &   0.34  &   0.35  &  F    \\
8\_V7020   &  2 40 44.3  & --34 24 21.8  &   0.07776  &  0.346  &  23.31  &   0.58  &   0.32  &  F    \\
4\_V813    &  2 39 08.3  & --34 10 29.3  &   0.07887  &  0.341  &  23.38  &   0.72  &   0.24  &  F    \\
6\_V816    &  2 39 54.0  & --34 27 21.4  &   0.07942  &  0.441  &  22.60  &   0.26  &   0.32  &  FO   \\
5\_V2834   &  2 38 59.2  & --34 26 18.8  &   0.07952  &  0.353  &  22.98  &   0.70  &   0.31  &  FO   \\
6\_V33971  &  2 39 34.2  & --34 18 08.2  &   0.07977  &  0.462  &  23.37  &   0.85  &   0.10  &  F    \\
5\_V2985   &  2 38 58.8  & --34 26 14.7  &   0.08045  &  0.323  &  22.91  &   0.29  &   0.27  &  FO   \\
7\_V25610  &  2 40 27.7  & --34 16 47.3  &   0.08064  &  0.327  &  22.76  &   0.34  &   0.09  &  FO   \\
2\_V5582   &  2 40 00.0  & --34 08 10.6  &   0.08129  &  0.377  &  23.63  &   0.73  &   0.19  &  SL   \\
6\_V19268  &  2 39 52.7  & --34 22 26.6  &   0.08365  &  0.305  &  22.72  &   0.33  &   0.40  &  FO   \\
7\_V1671   &  2 40 08.4  & --34 26 52.5  &   0.08427  &  0.340  &  23.12  &   0.80  &   0.13  &  F    \\
6\_V14592  &  2 39 25.6  & --34 23 42.2  &   0.08458  &  0.395  &  22.81  &   0.41  &   -   &  FO   \\
8\_V20859  &  2 41 15.2  & --34 18 08.1  &   0.08515  &  0.361  &  23.42  &   0.62  &   -  &  SL   \\
6\_V10166  &  2 39 41.2  & --34 24 53.7  &   0.08639  &  0.513  &  23.03  &   0.61  &   0.30  &  F    \\
8\_V10823  &  2 40 44.5  & --34 22 39.9  &   0.09221  &  0.492  &  22.97  &   0.20  &   0.36  &  F    \\
8\_V30859  &  2 41 16.4  & --34 13 30.6  &   0.09245  &  0.472  &  23.10  &   0.29  &   0.37  &  F    \\
7\_V19699  &  2 40 21.0  & --34 19 19.2  &   0.09254  &  0.507  &  22.72  &   0.29  &   0.28  &  FO   \\
7\_V21373  &  2 40 05.0  & --34 18 36.0  &   0.09345  &  0.495  &  23.02  &   0.43  &   0.25  &  F    \\
7\_V32007  &  2 40 08.5  & --34 14 01.1  &   0.10151  &  0.494  &  22.99  &   0.43  &   0.26  &  F    \\
2\_V7941   &  2 40 30.8  & --34 06 53.9  &   0.11348  &  0.541  &  23.13  &   0.41  &   0.31  &  SL   \\
8\_V33331  &  2 40 47.8  & --34 12 20.4  &   0.12508  &  0.631  &  22.46  &   0.51  &   0.27  &  F    \\
6\_V19089  &  2 39 42.2  & --34 22 29.8  &   0.12619  &  0.648  &  22.60  &   0.61  &   0.32  &  F    \\
%
\noalign{\smallskip}
\hline
\noalign{\smallskip}\\
\noalign{
Units of right ascension are hours, minutes, and seconds,
and units of declination are degrees, arcminutes, and arcseconds.
}
\noalign{$^a$~~~
Variable stars are ordered by increasing period, the
first digit of the ID gives the number of the CCD in the WFI mosaic, the
second part is the DAOPHOT identifier.}
\noalign{\smallskip}
\noalign{$^b$~~~F: fundamental radial-mode pulsator; FO: first overtone
radial-mode pulsator; SL: subluminous variable.}
\end{longtable}
\normalsize
                                                                                                                                
\begin{figure*}
\caption{Atlas of the $B$ and $V$ light curves of the SX Phe discovered in the Fornax survey}
\includegraphics[width=0.329\columnwidth,height=0.27\columnwidth]{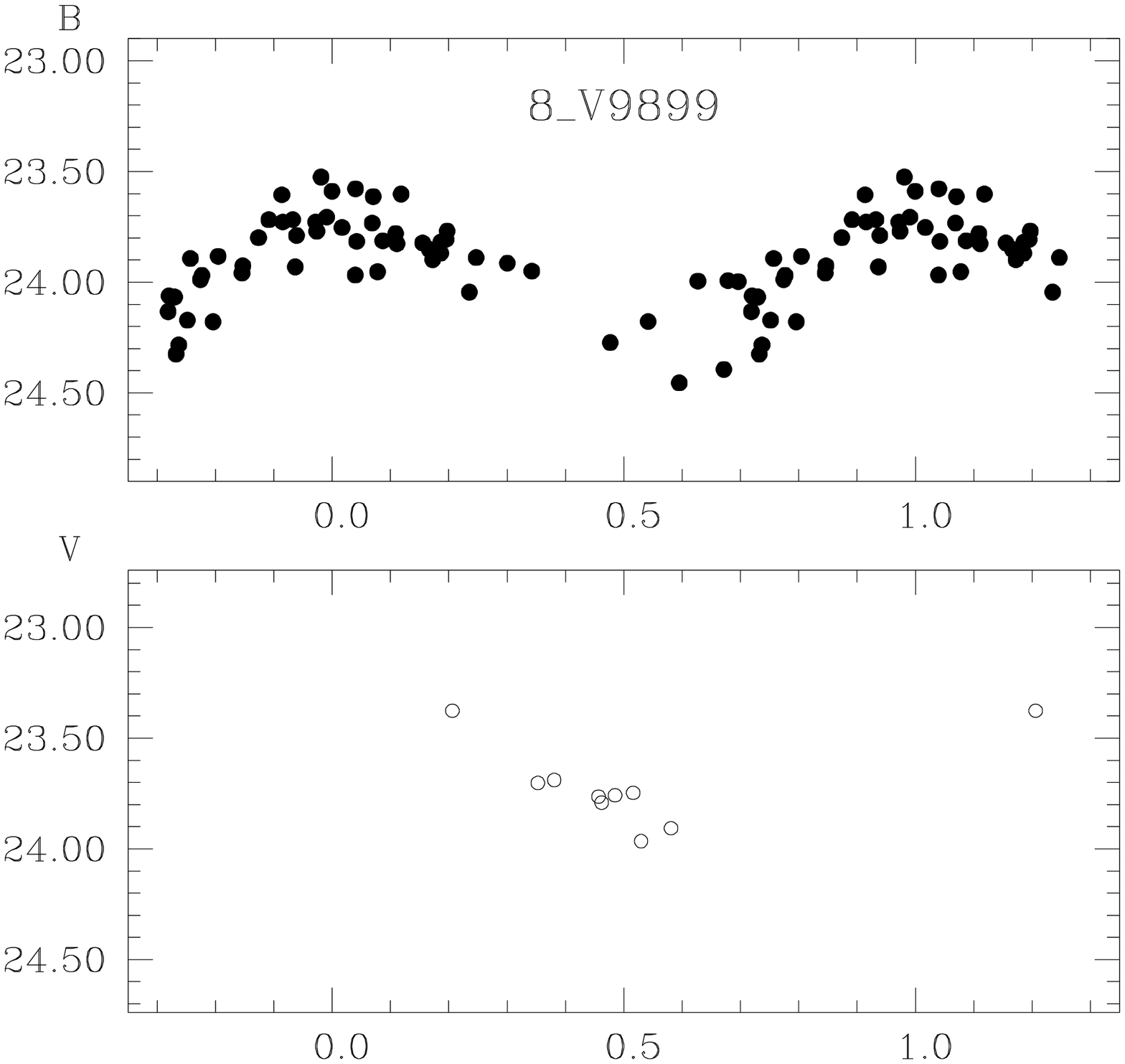}
\includegraphics[width=0.329\columnwidth,height=0.27\columnwidth]{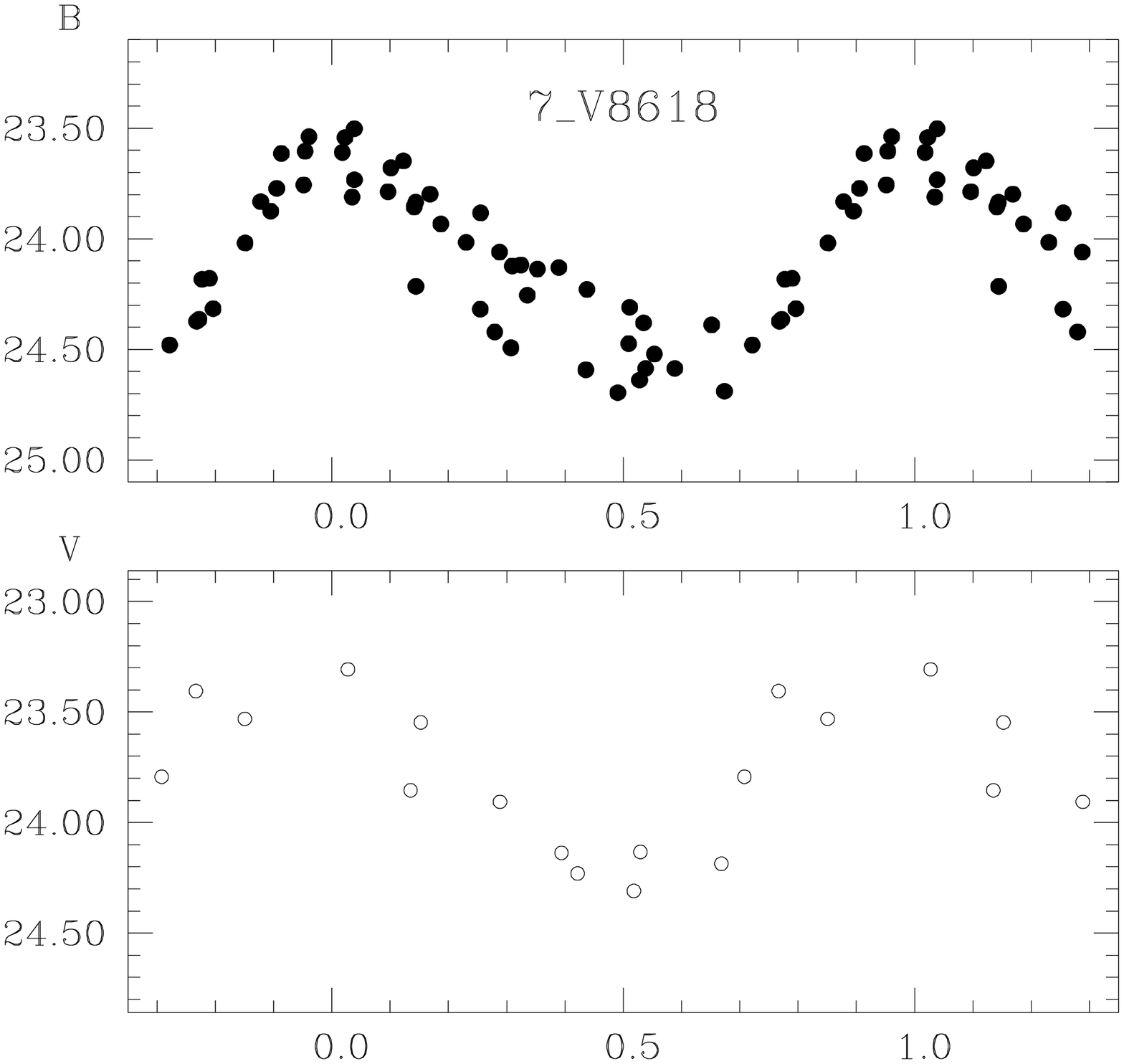}
\includegraphics[width=0.329\columnwidth,height=0.27\columnwidth]{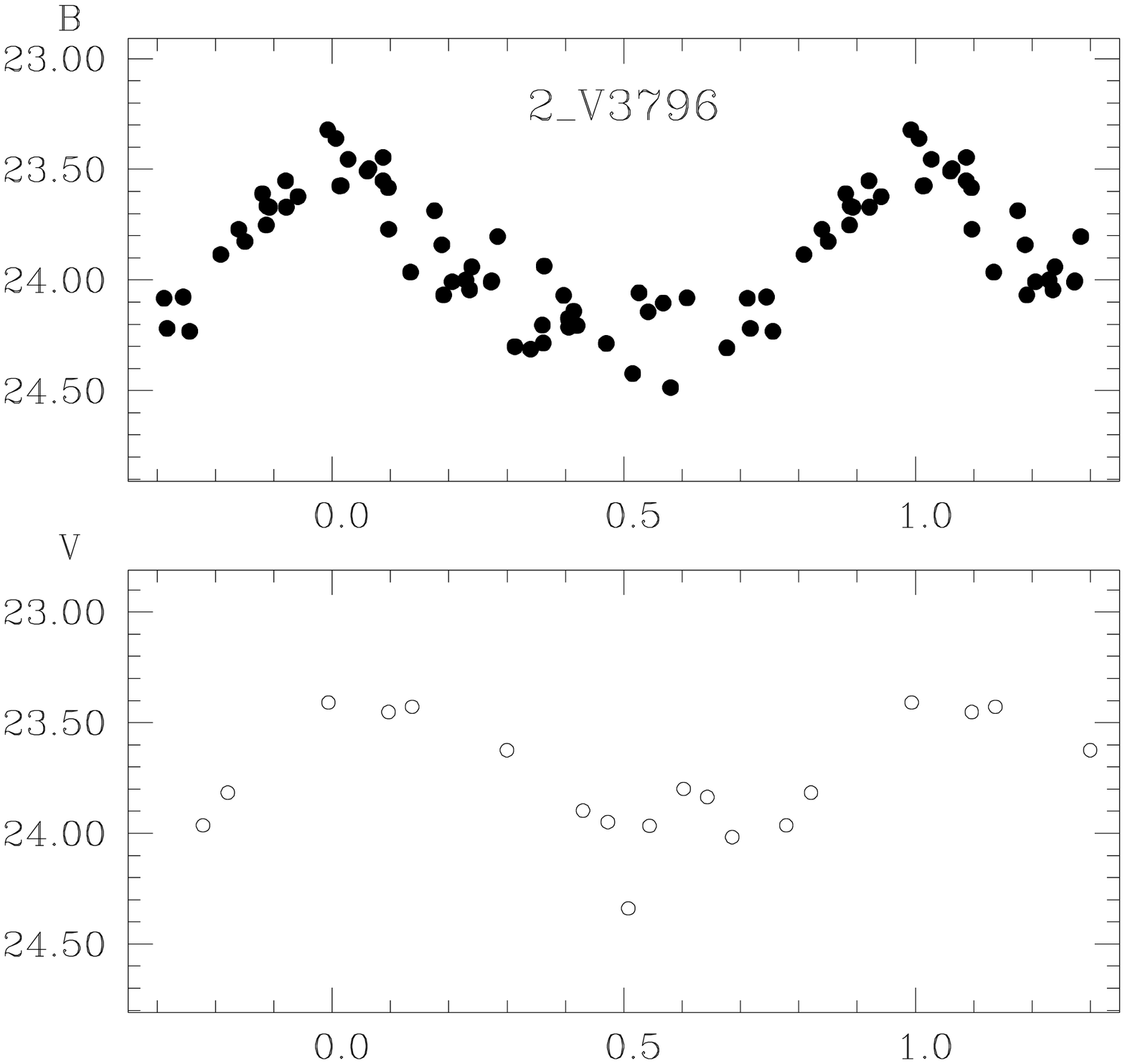}
\includegraphics[width=0.329\columnwidth,height=0.27\columnwidth]{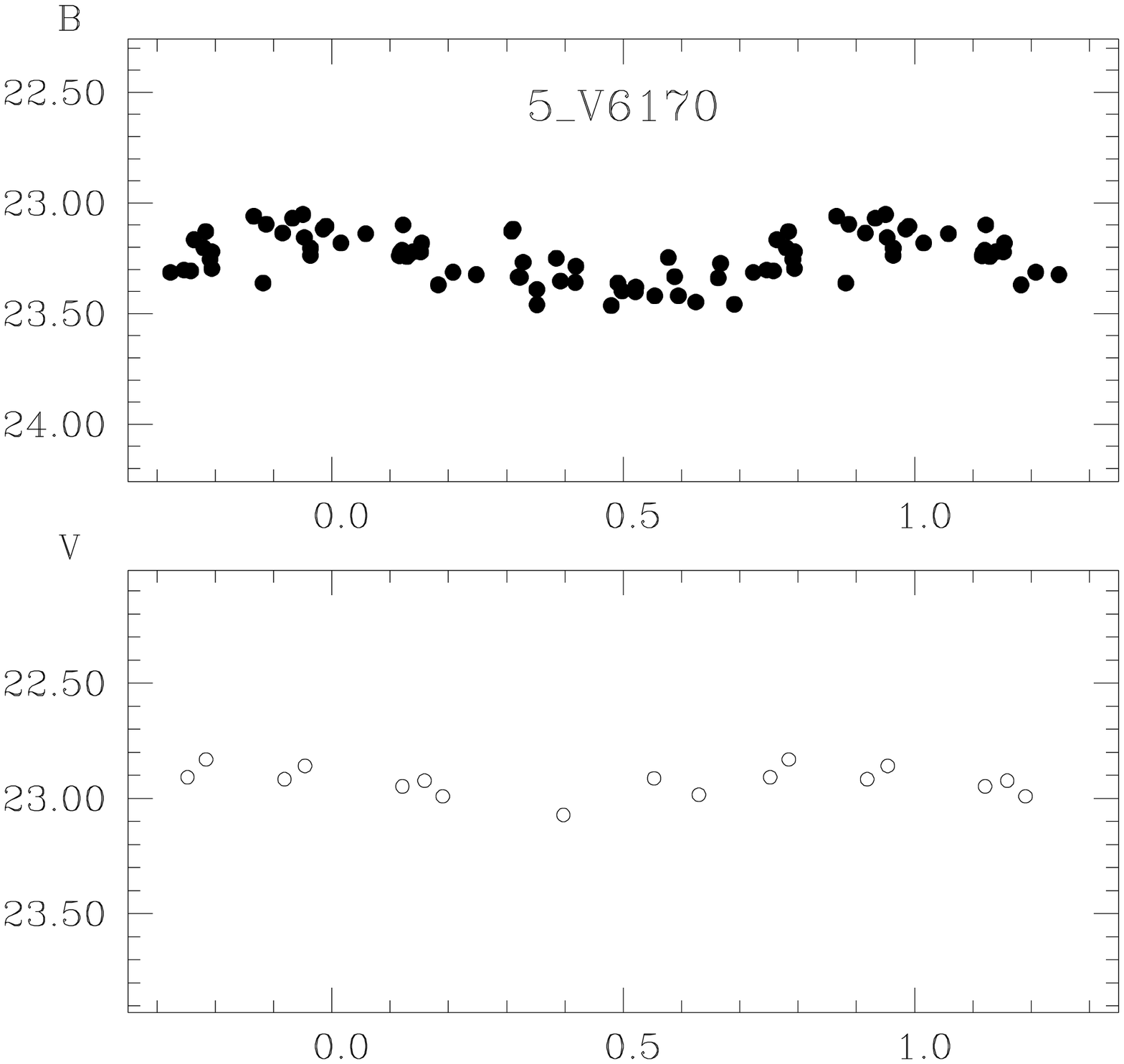}
\includegraphics[width=0.329\columnwidth,height=0.27\columnwidth]{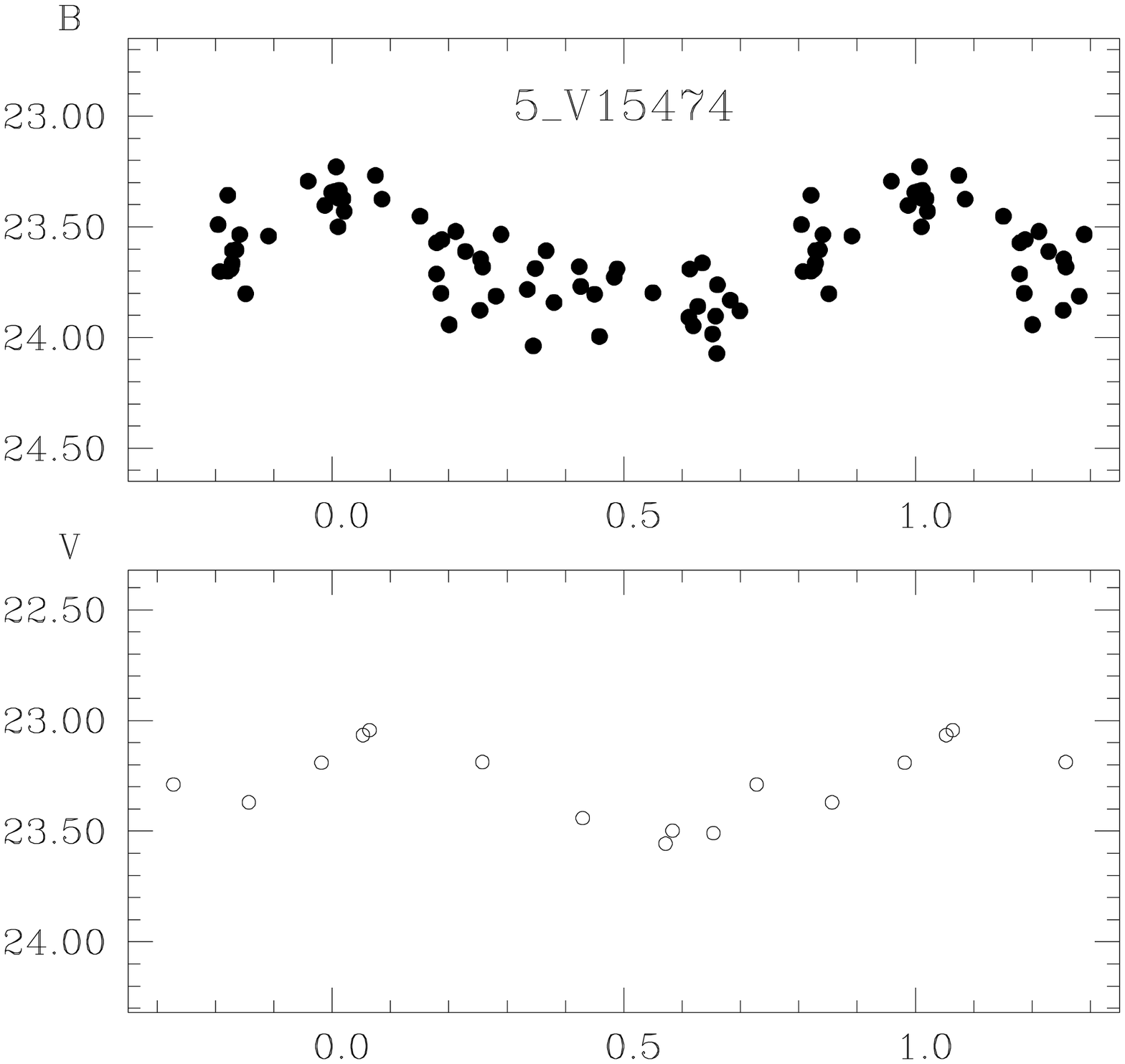}
\includegraphics[width=0.329\columnwidth,height=0.27\columnwidth]{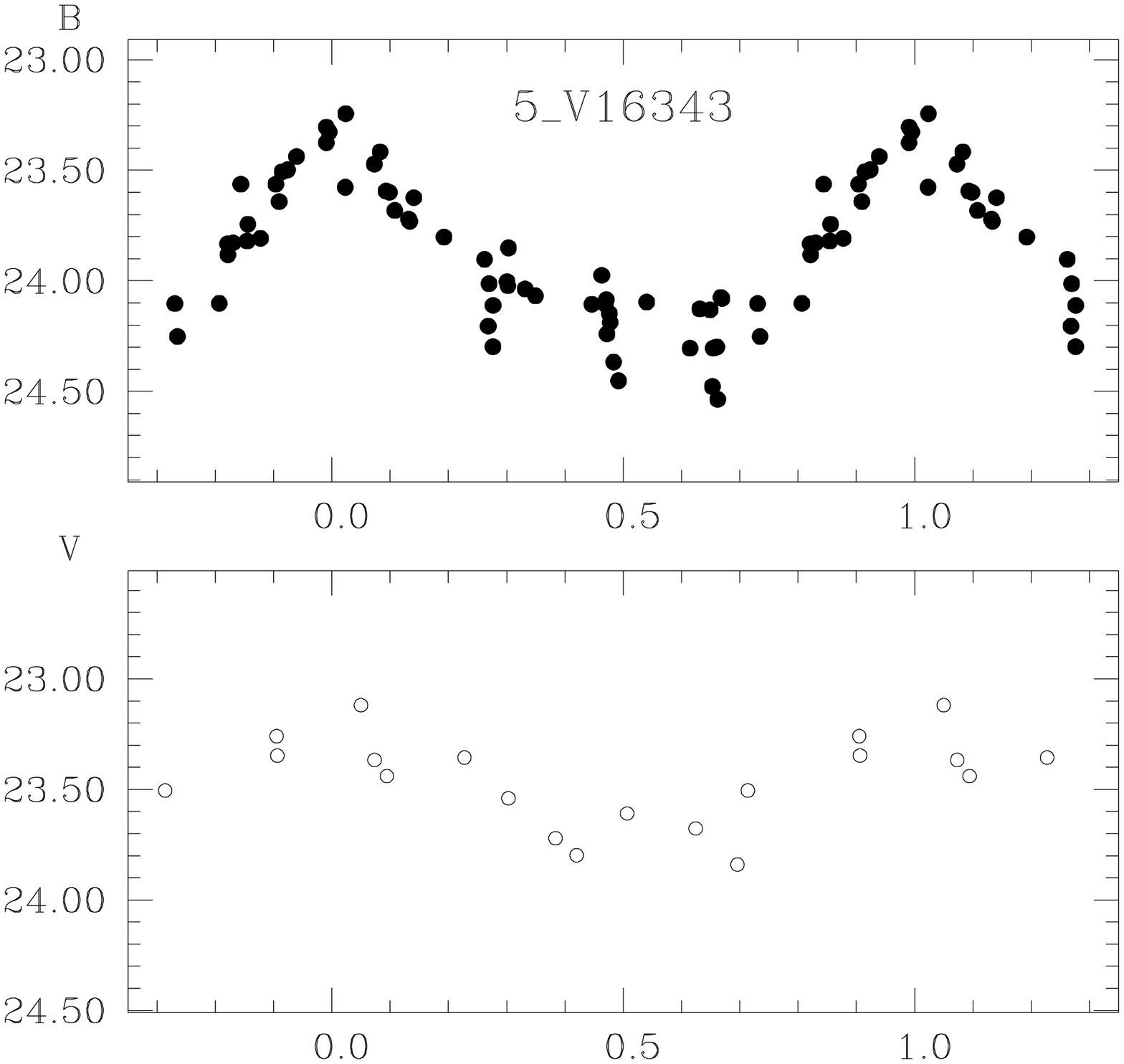}
\includegraphics[width=0.329\columnwidth,height=0.27\columnwidth]{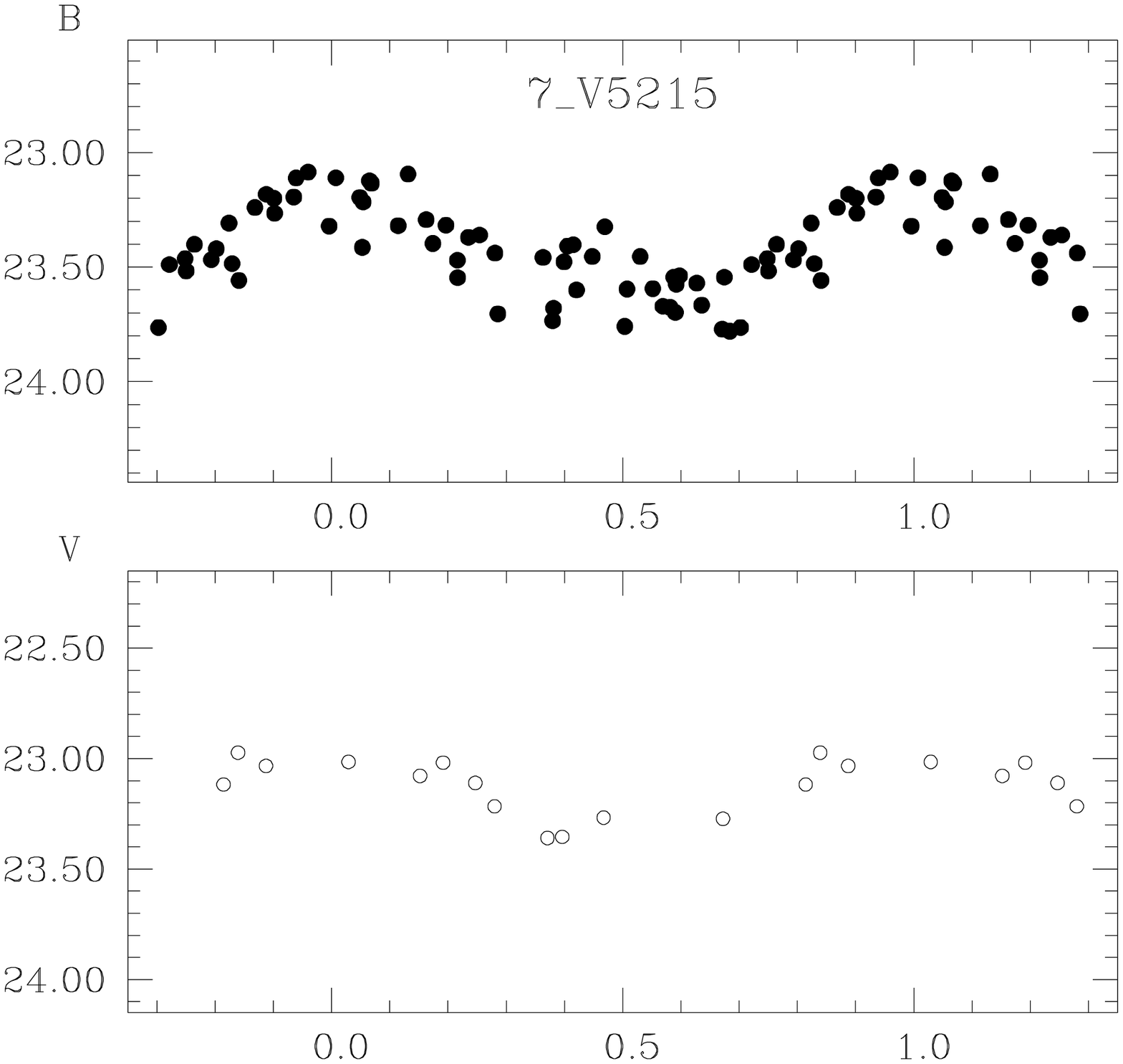}
\includegraphics[width=0.329\columnwidth,height=0.27\columnwidth]{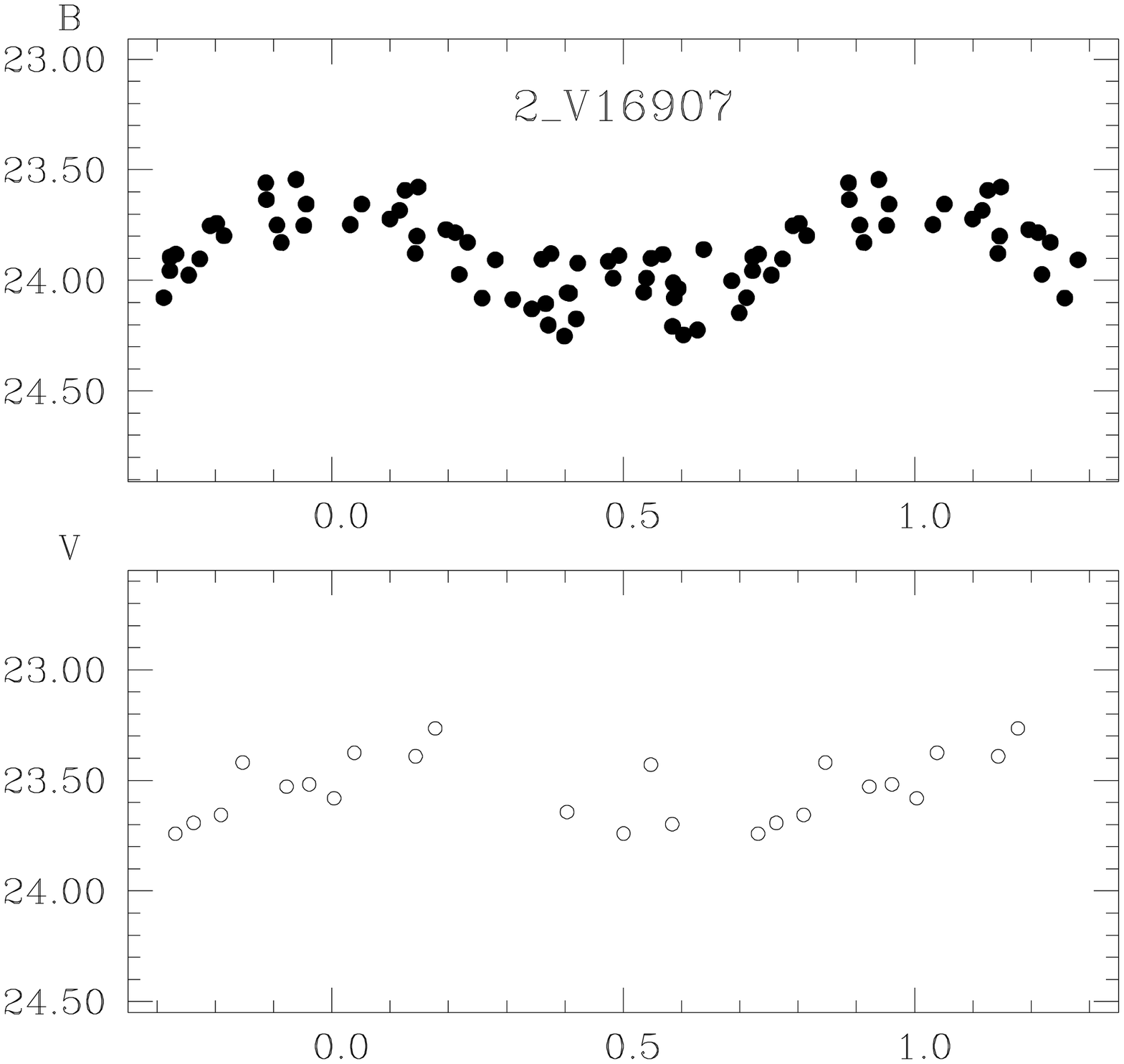}
\includegraphics[width=0.329\columnwidth,height=0.27\columnwidth]{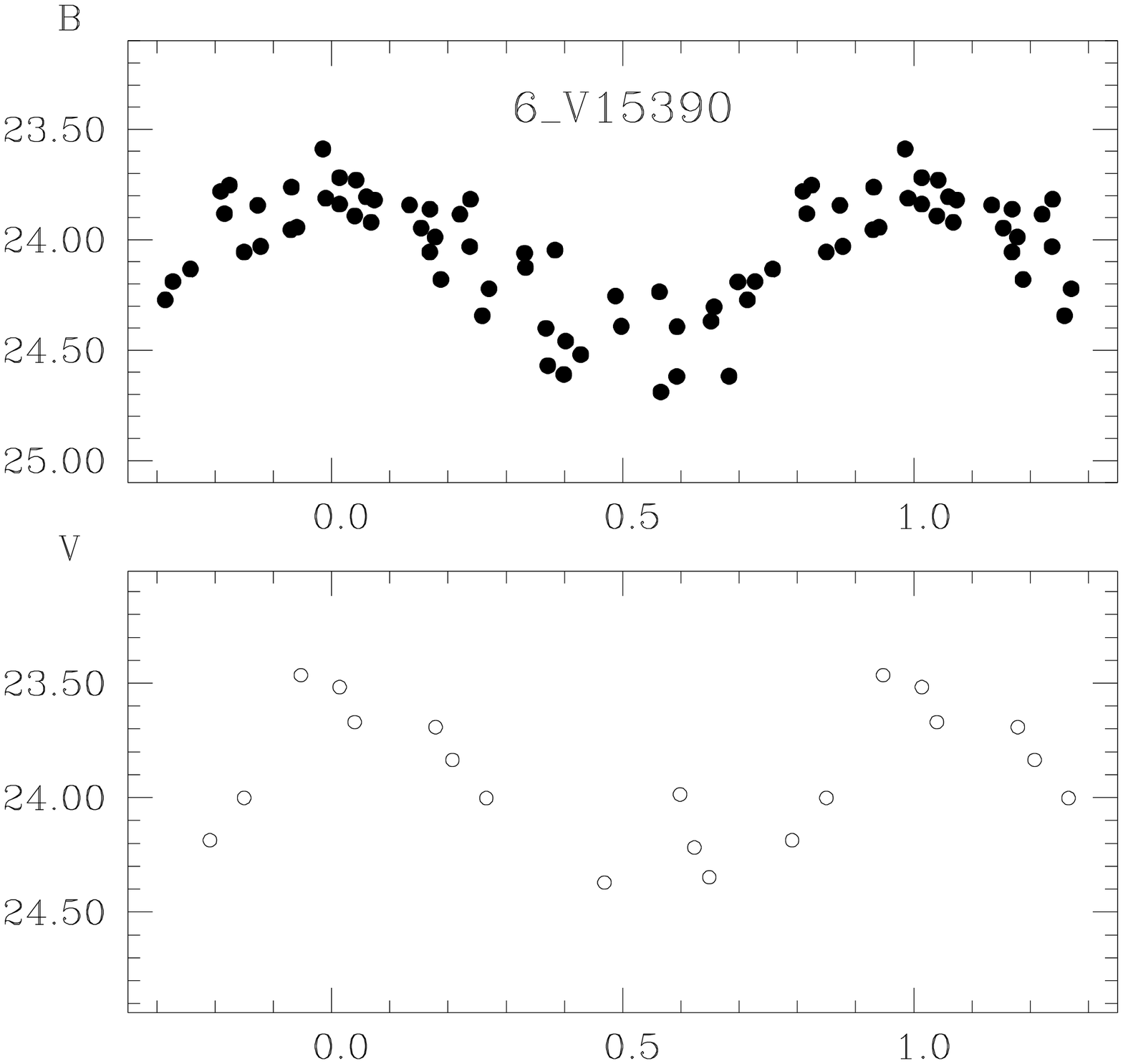}
\includegraphics[width=0.329\columnwidth,height=0.27\columnwidth]{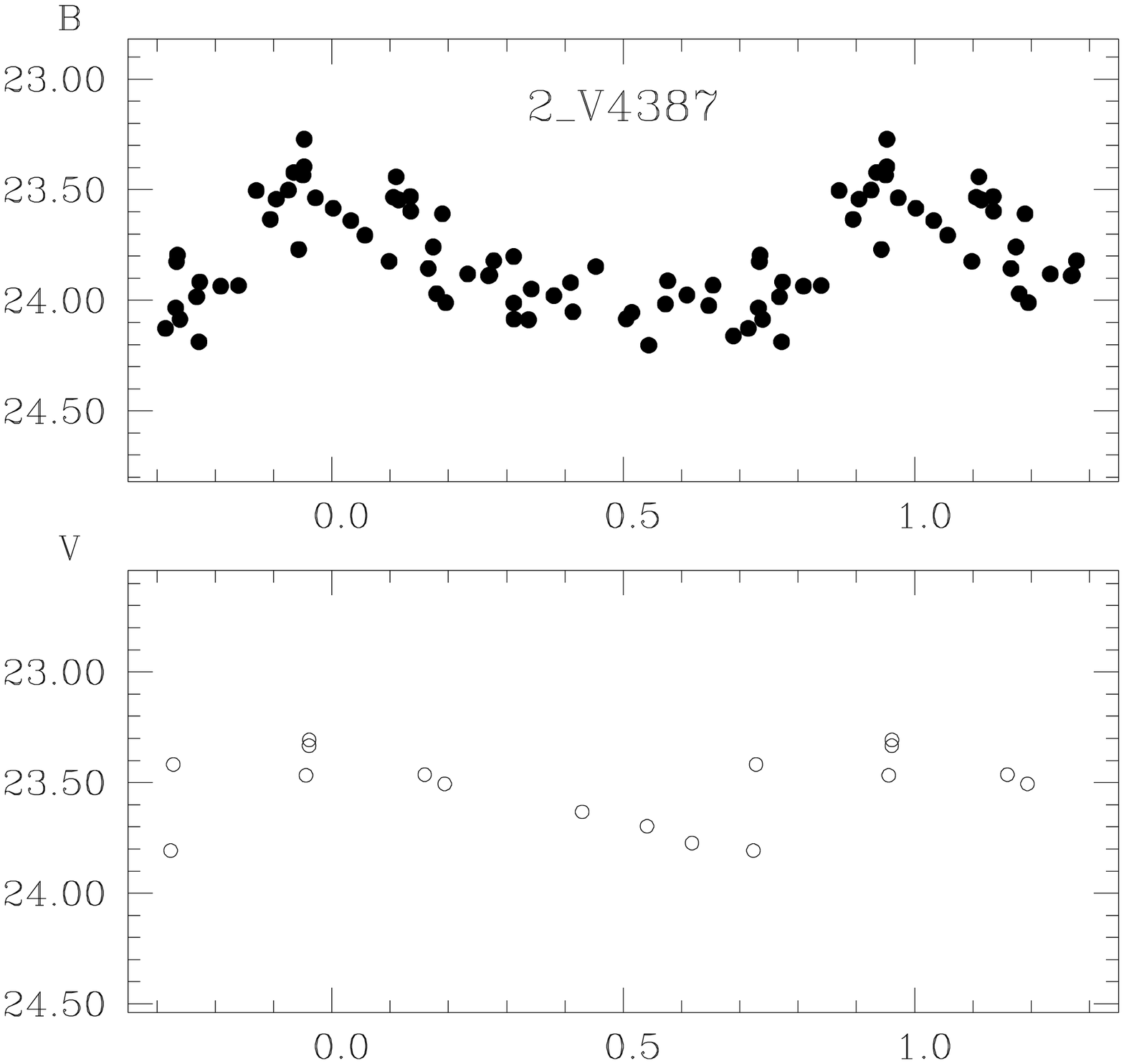}
\includegraphics[width=0.329\columnwidth,height=0.27\columnwidth]{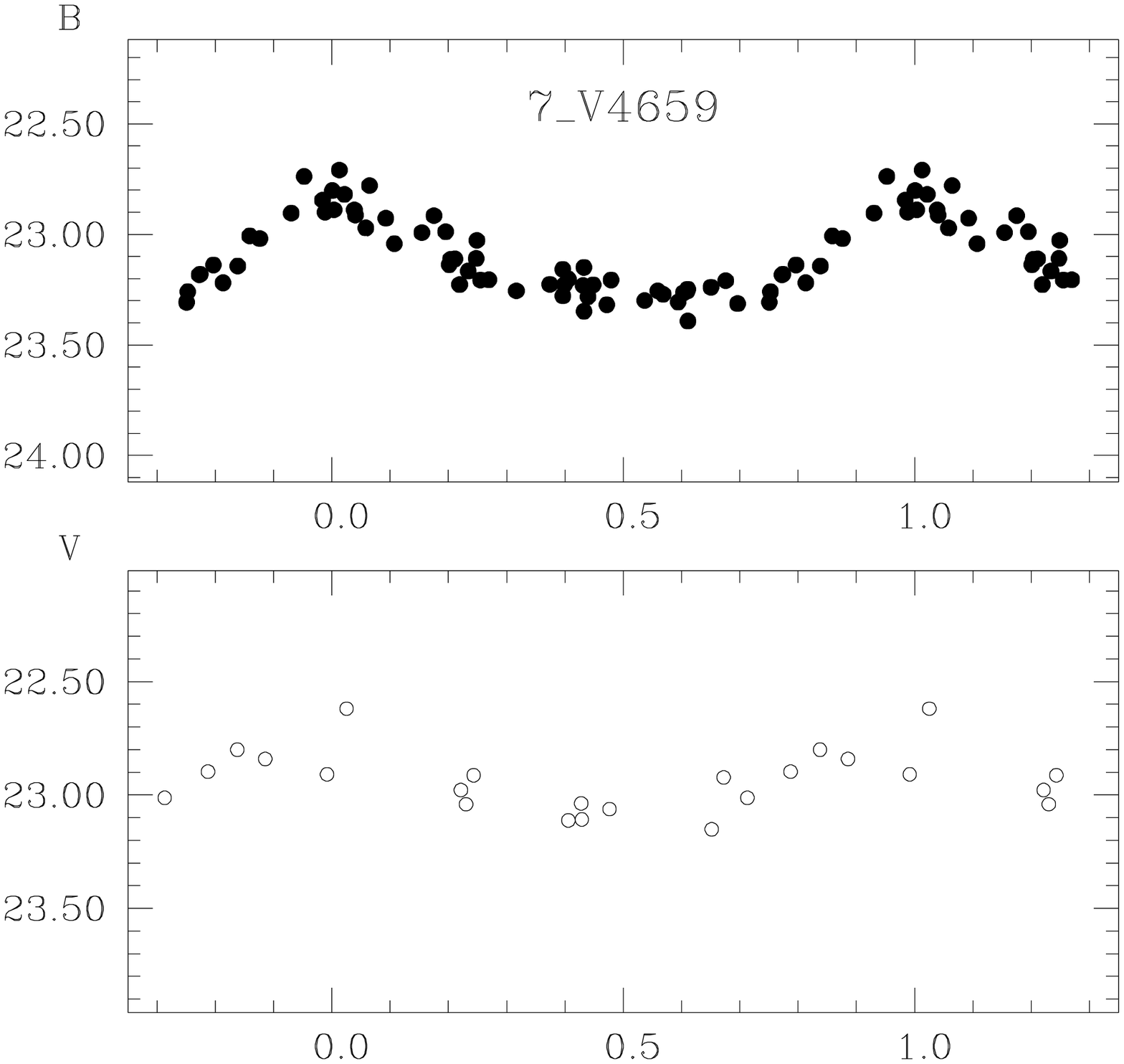}
\includegraphics[width=0.329\columnwidth,height=0.27\columnwidth]{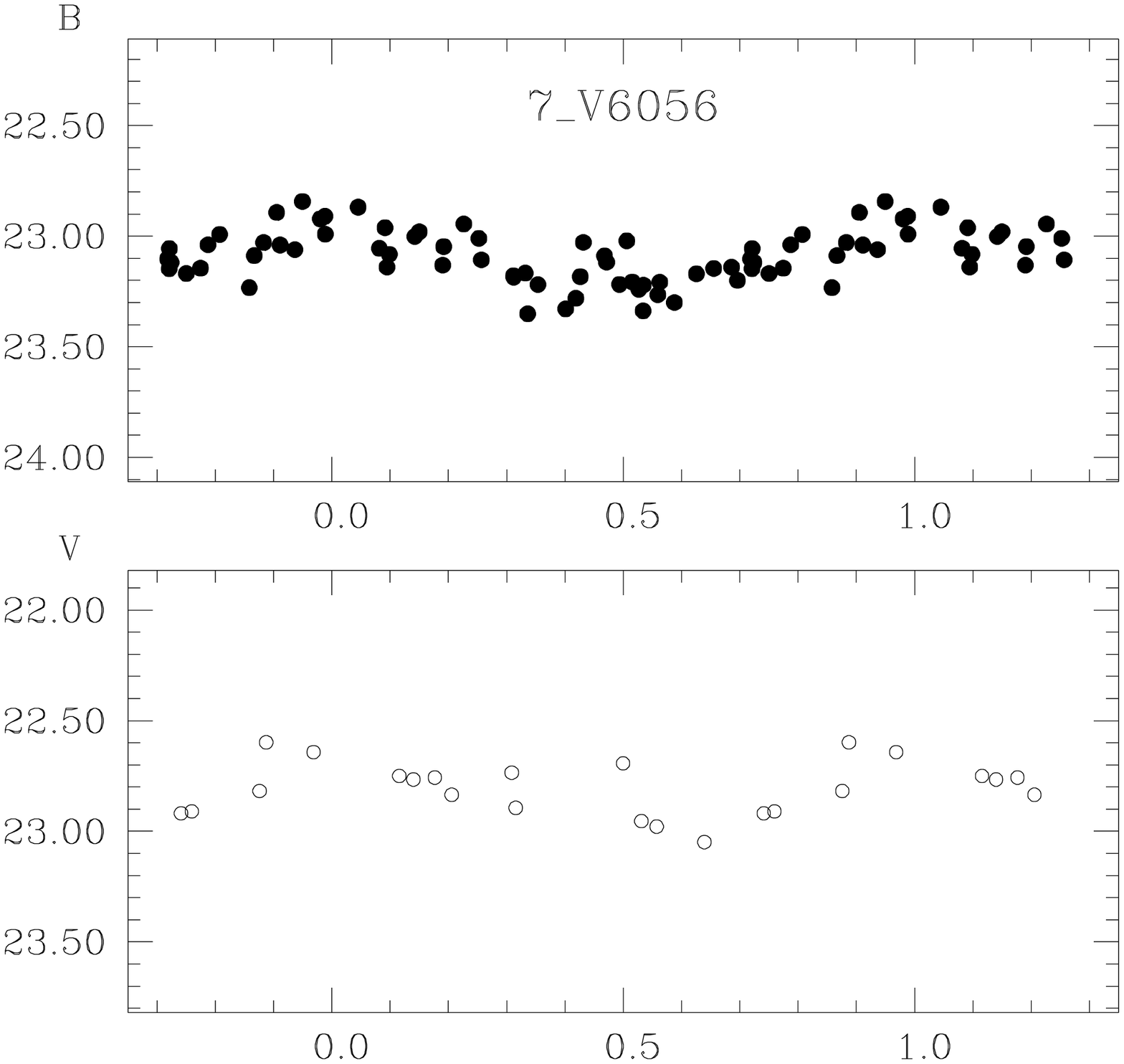}
\includegraphics[width=0.329\columnwidth,height=0.27\columnwidth]{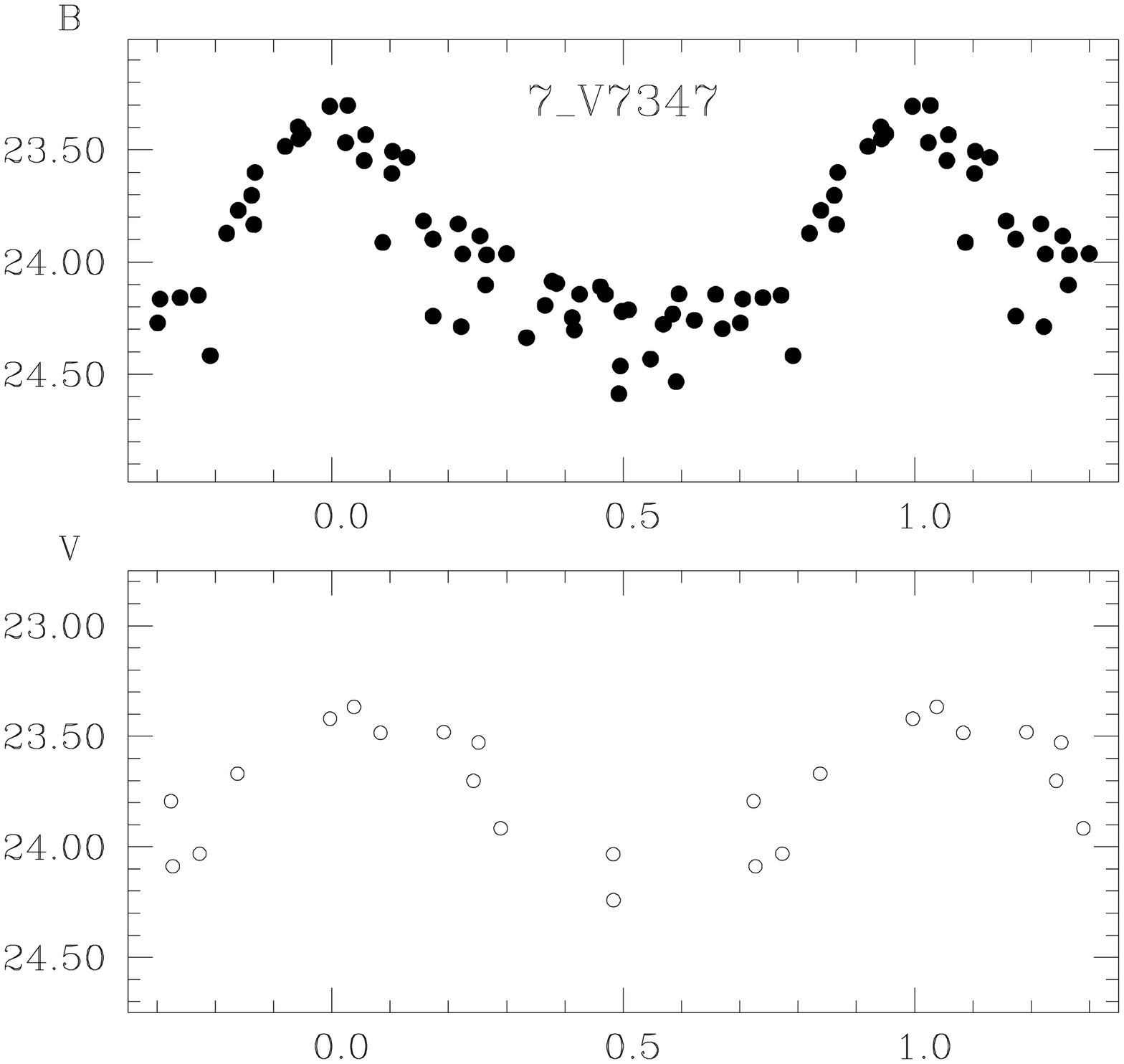}
\includegraphics[width=0.329\columnwidth,height=0.27\columnwidth]{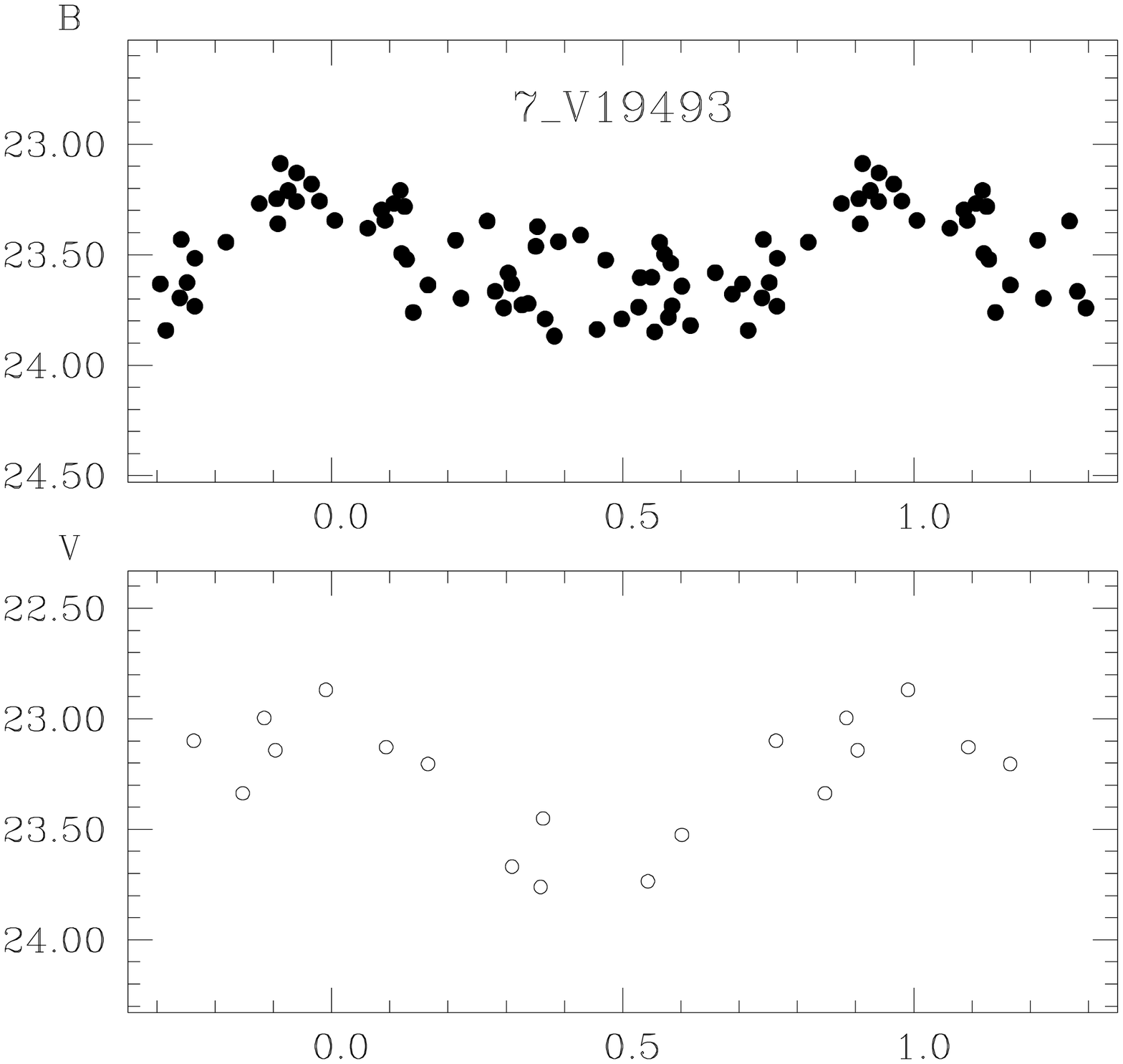}
\includegraphics[width=0.329\columnwidth,height=0.27\columnwidth]{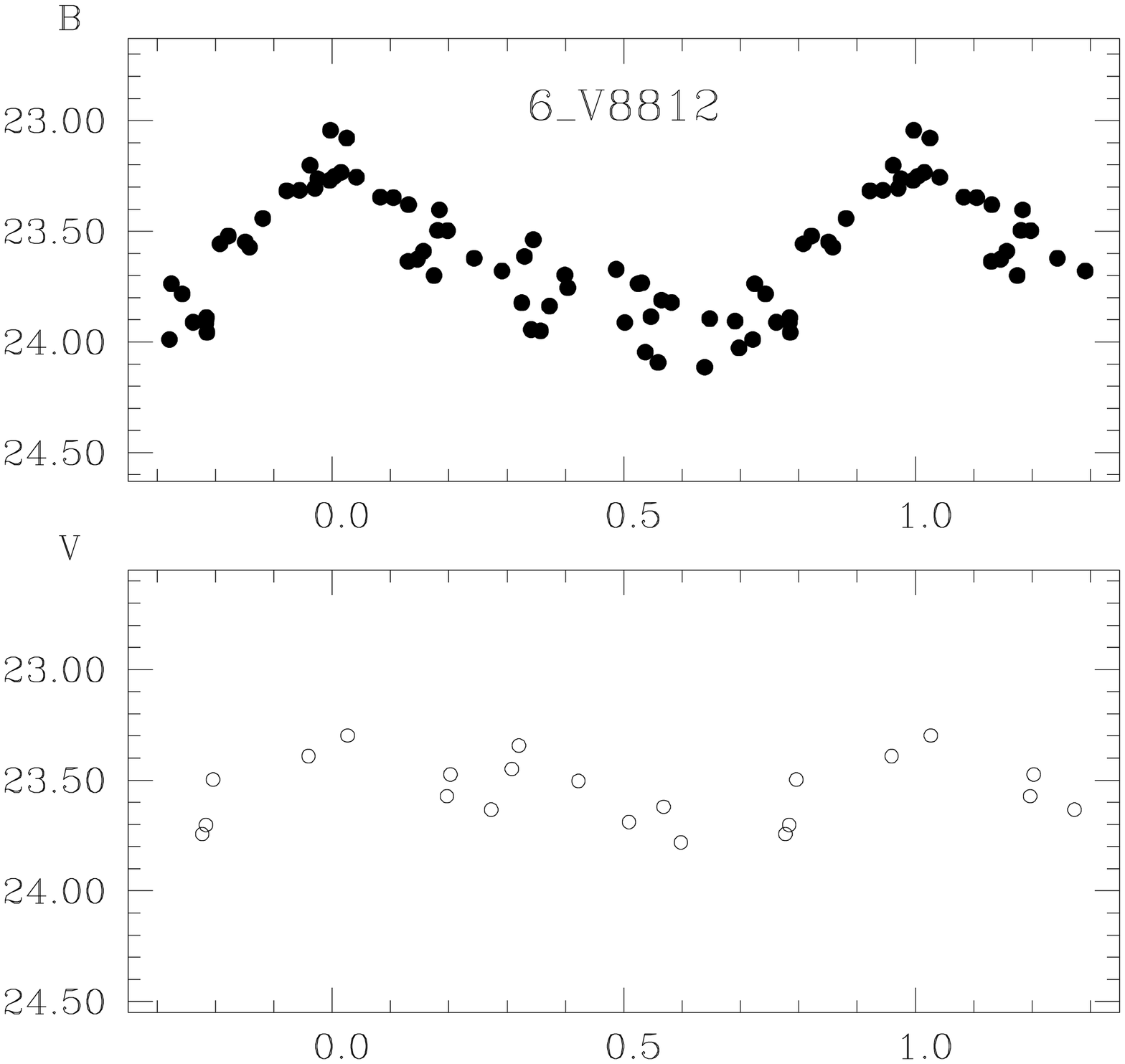}
\end{figure*}
\begin{figure*}
\includegraphics[width=0.329\columnwidth,height=0.27\columnwidth]{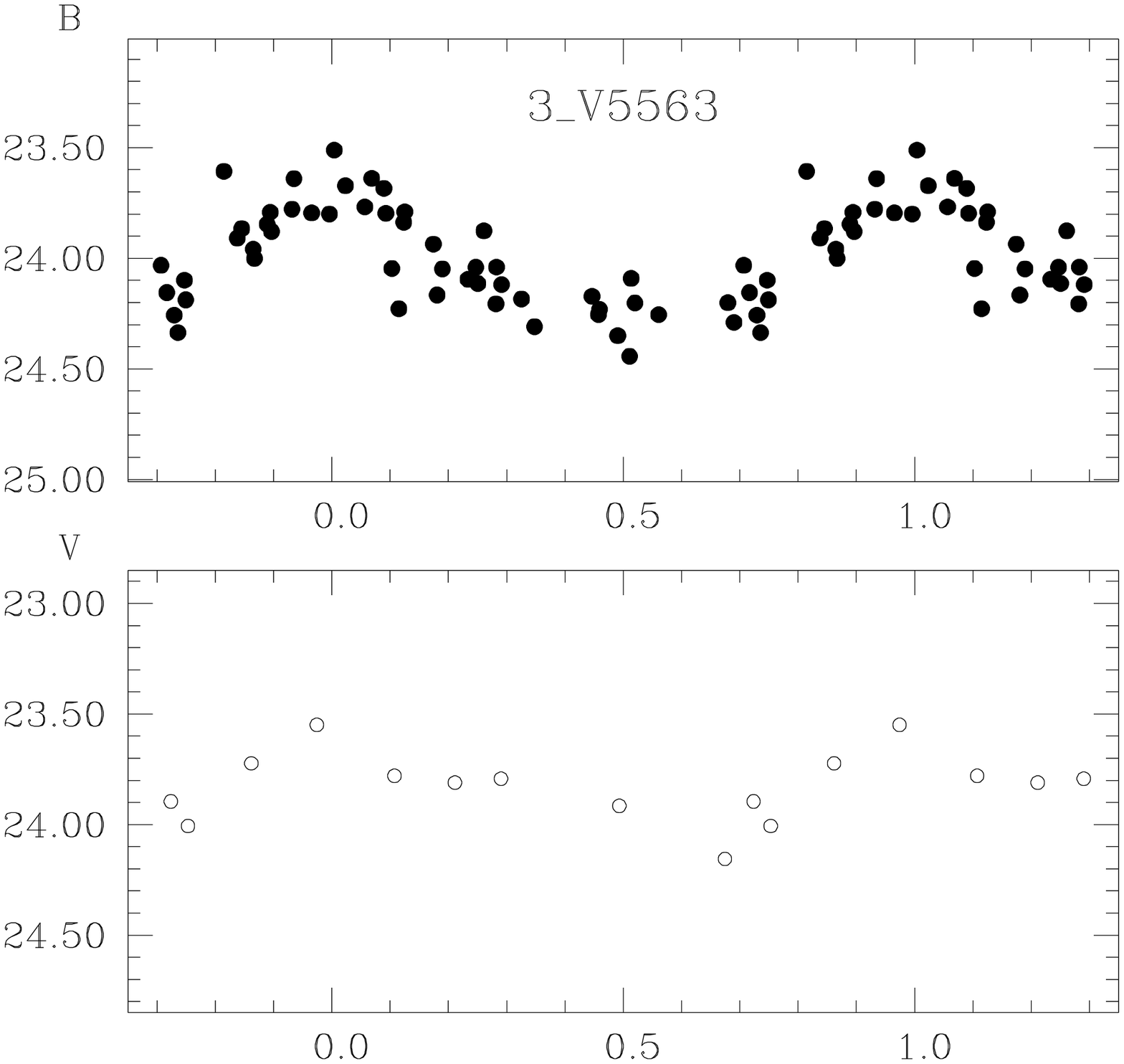}
\includegraphics[width=0.329\columnwidth,height=0.27\columnwidth]{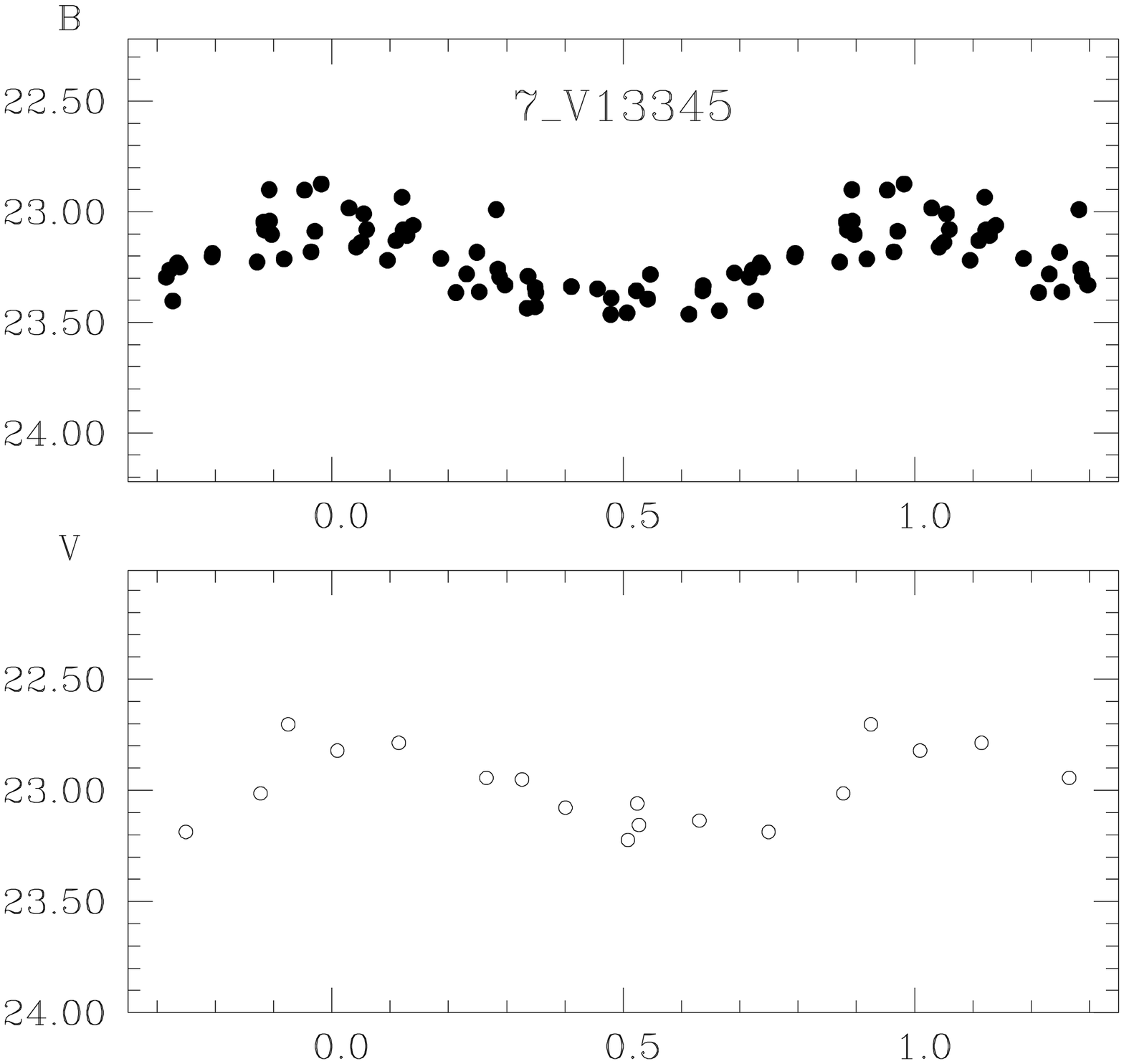}
\includegraphics[width=0.329\columnwidth,height=0.27\columnwidth]{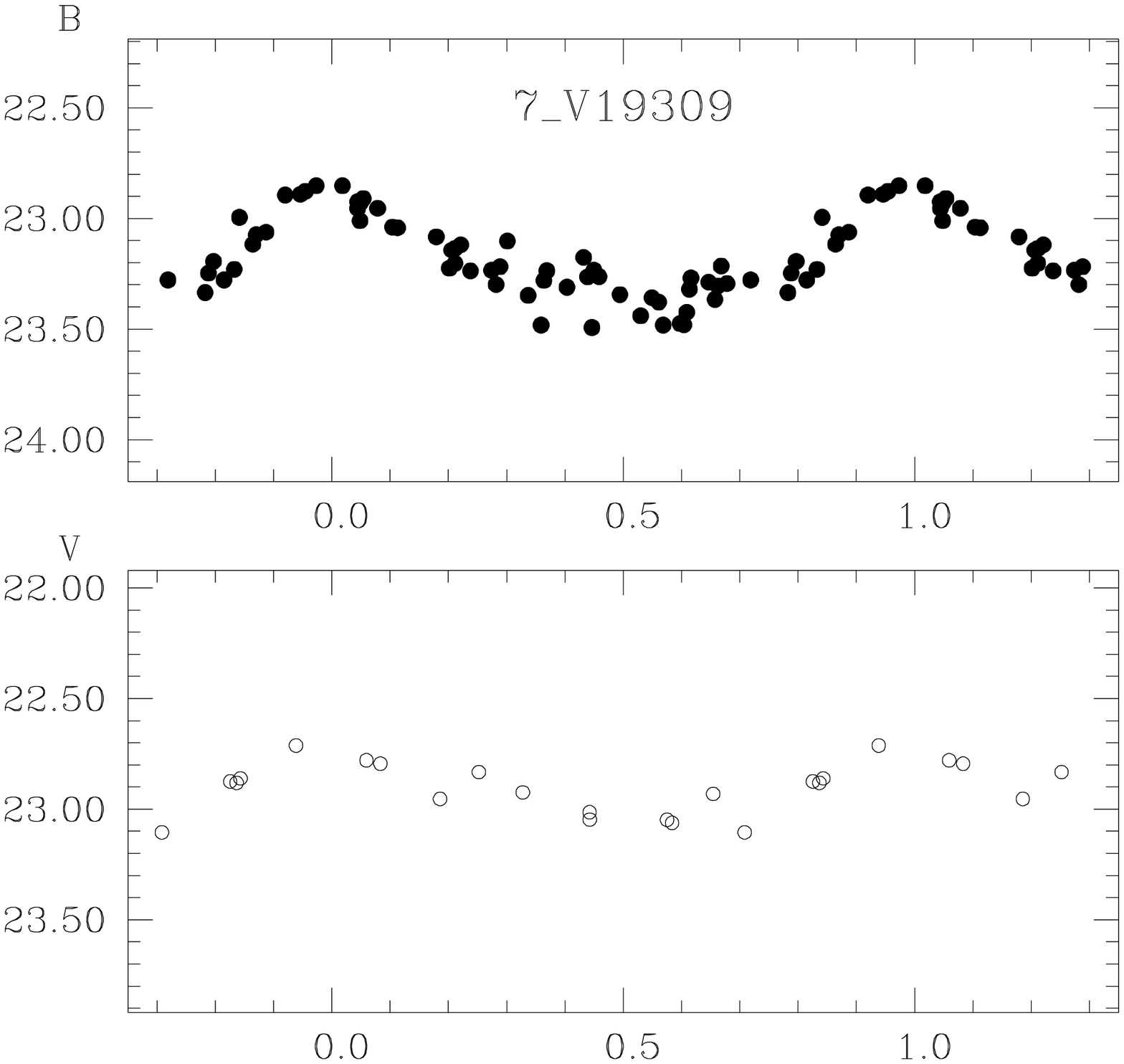}
\includegraphics[width=0.329\columnwidth,height=0.27\columnwidth]{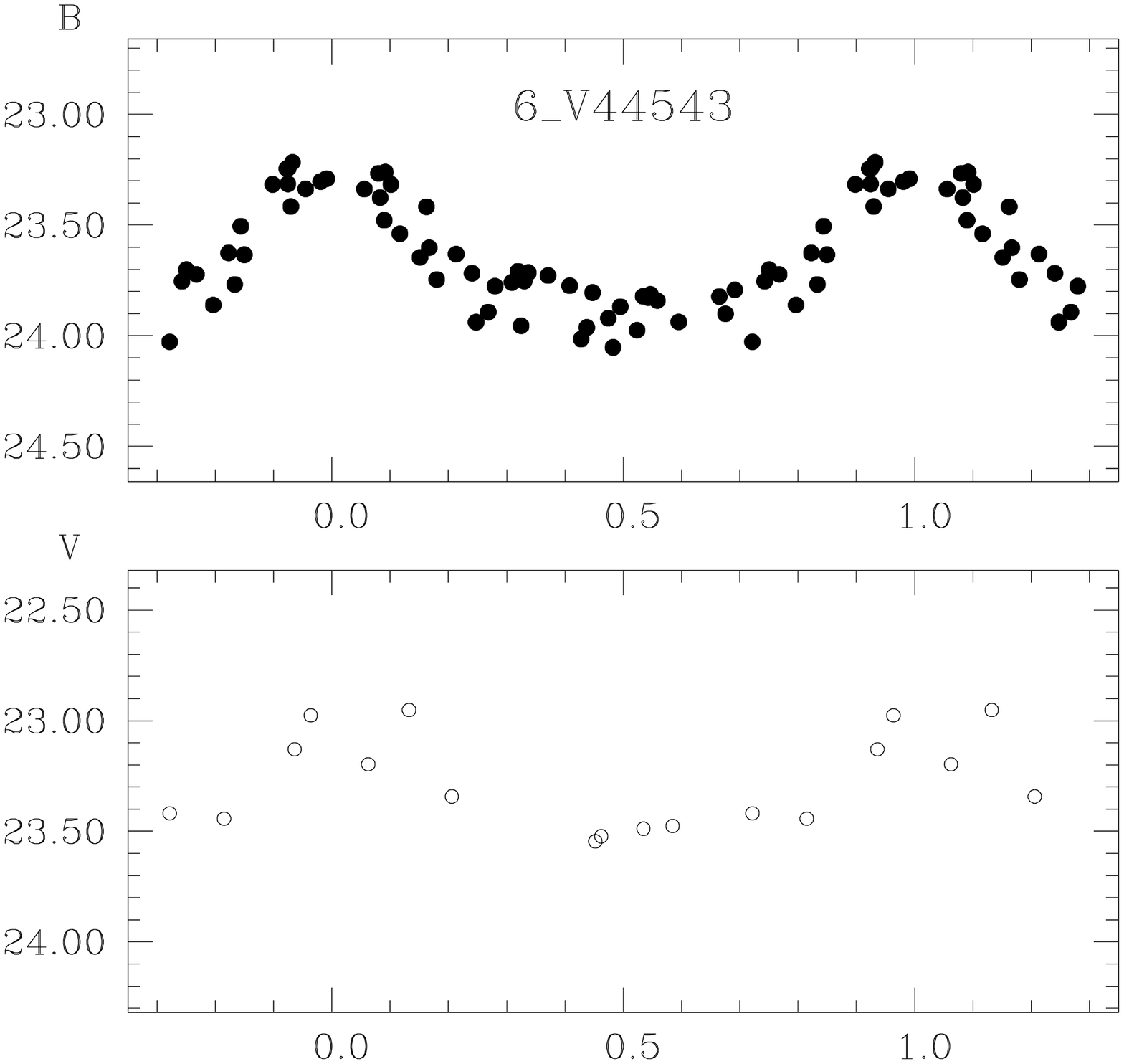}
\includegraphics[width=0.329\columnwidth,height=0.27\columnwidth]{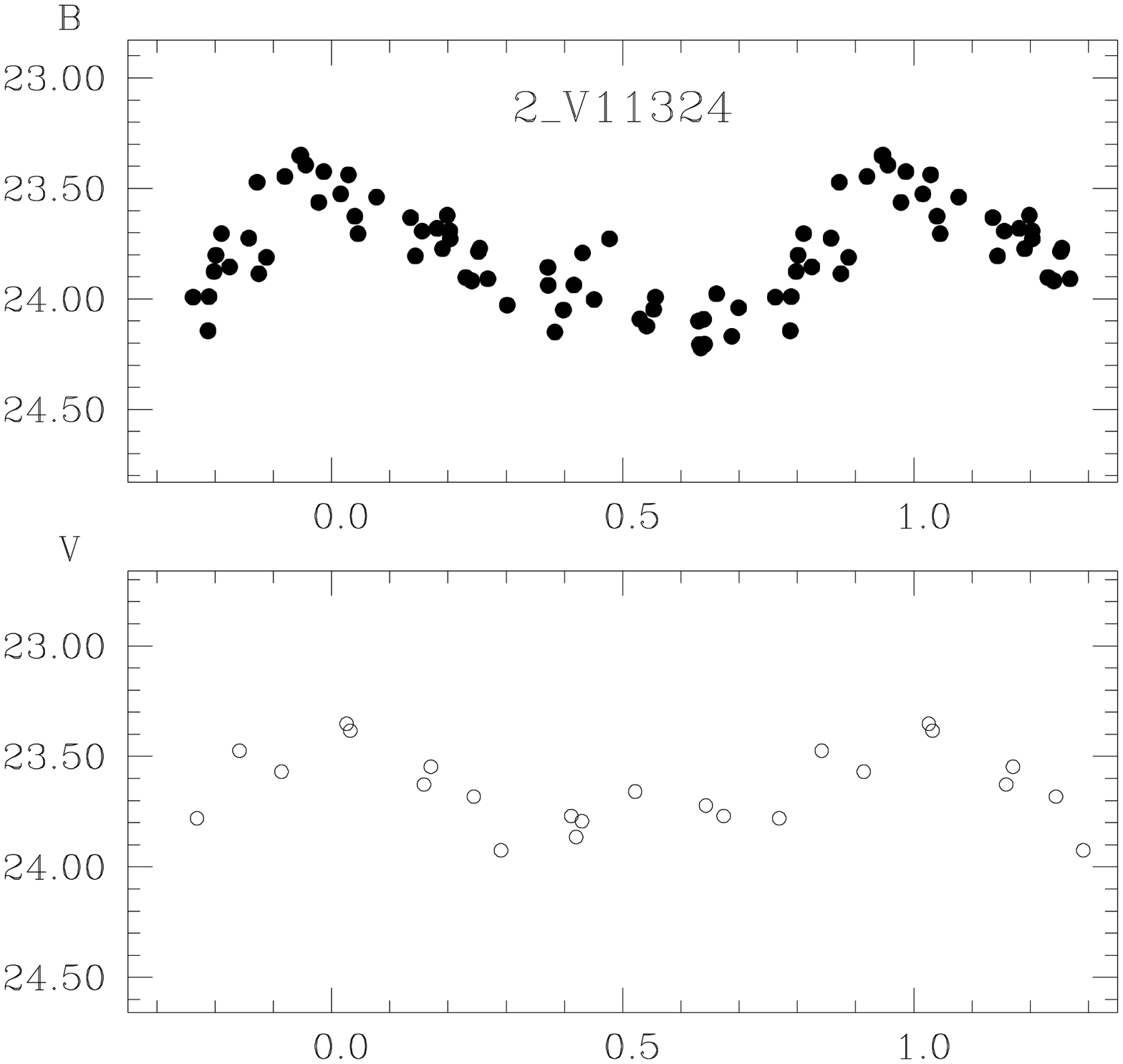}
\includegraphics[width=0.329\columnwidth,height=0.27\columnwidth]{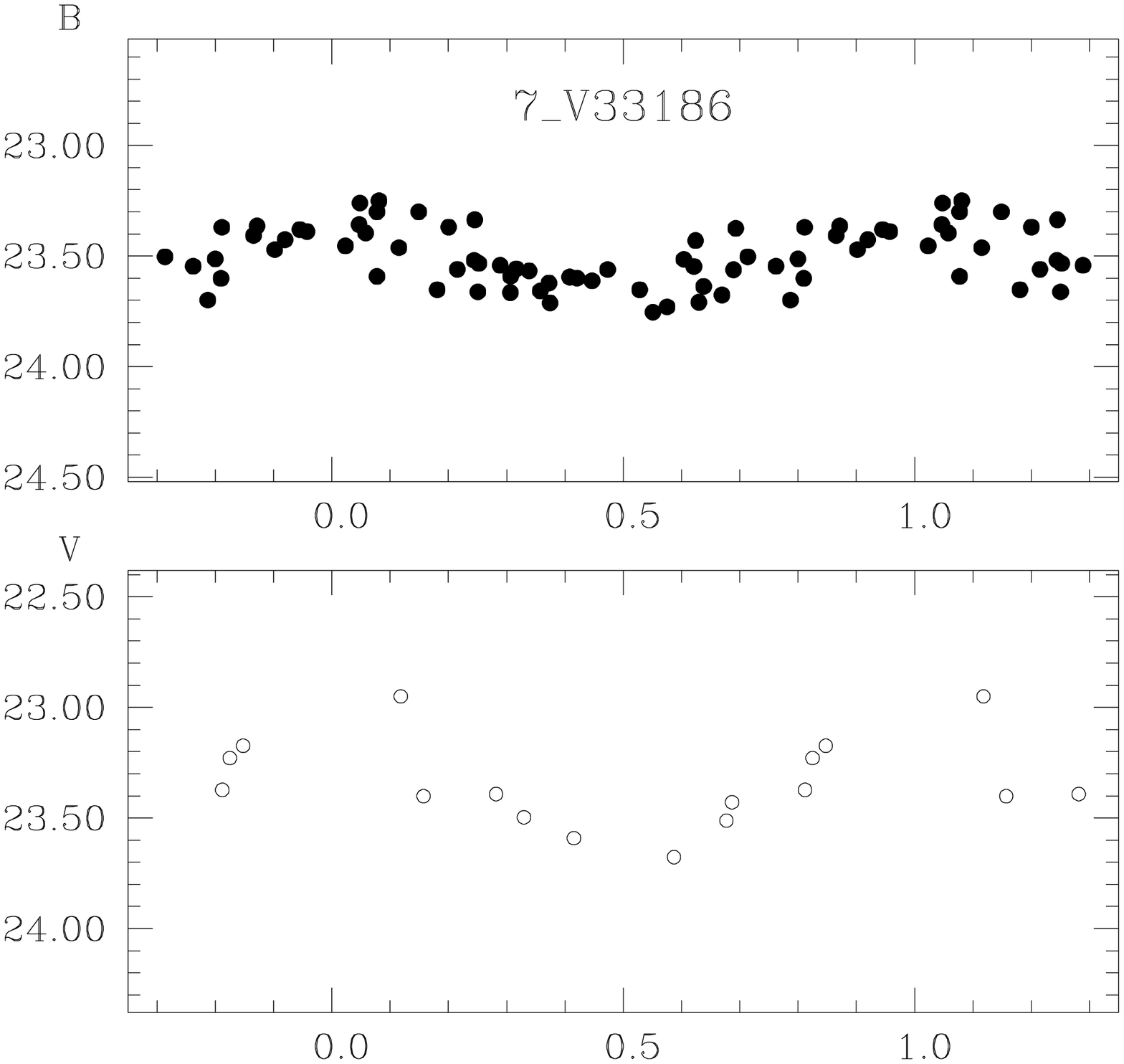}
\includegraphics[width=0.329\columnwidth,height=0.27\columnwidth]{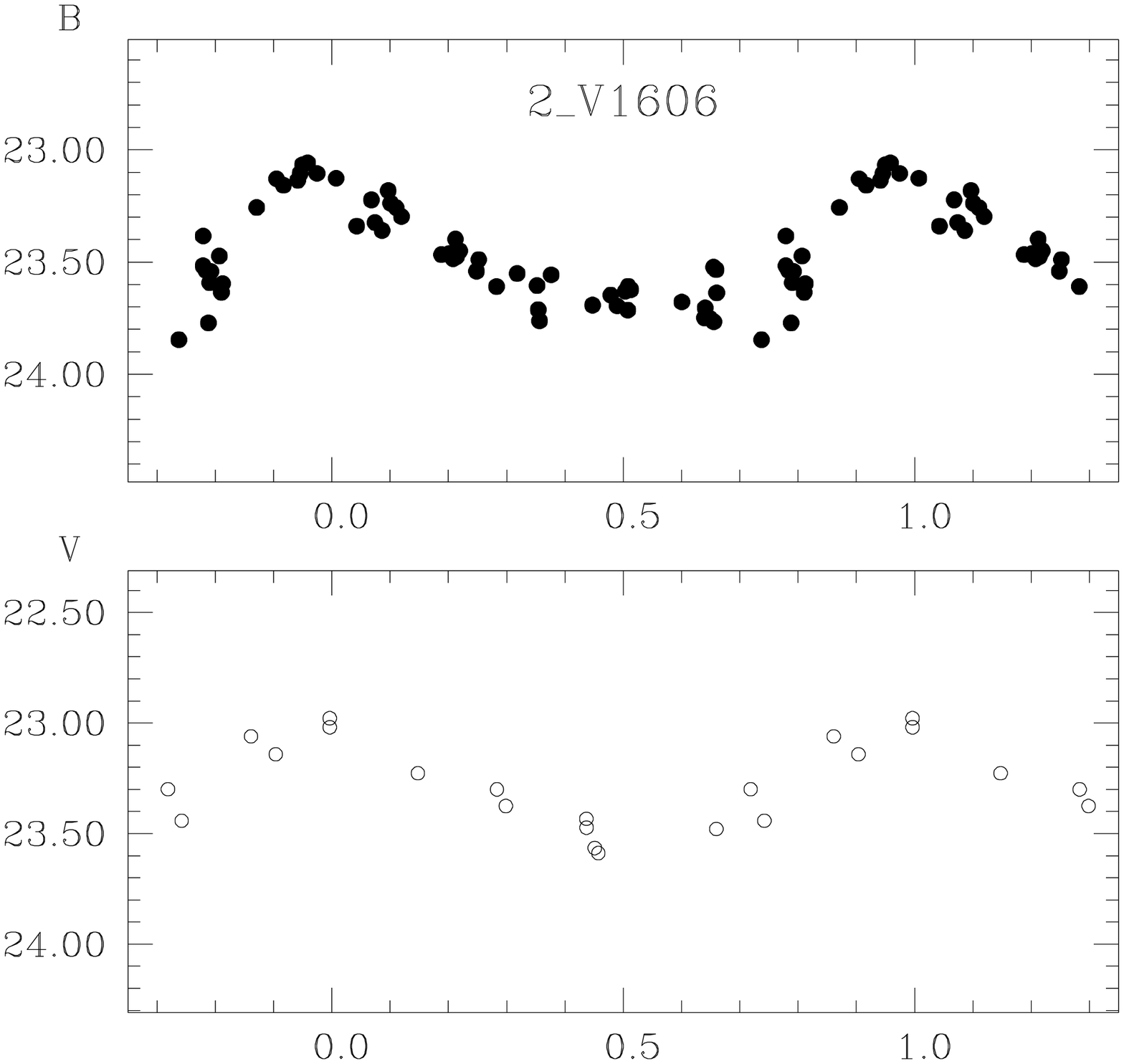}
\includegraphics[width=0.329\columnwidth,height=0.27\columnwidth]{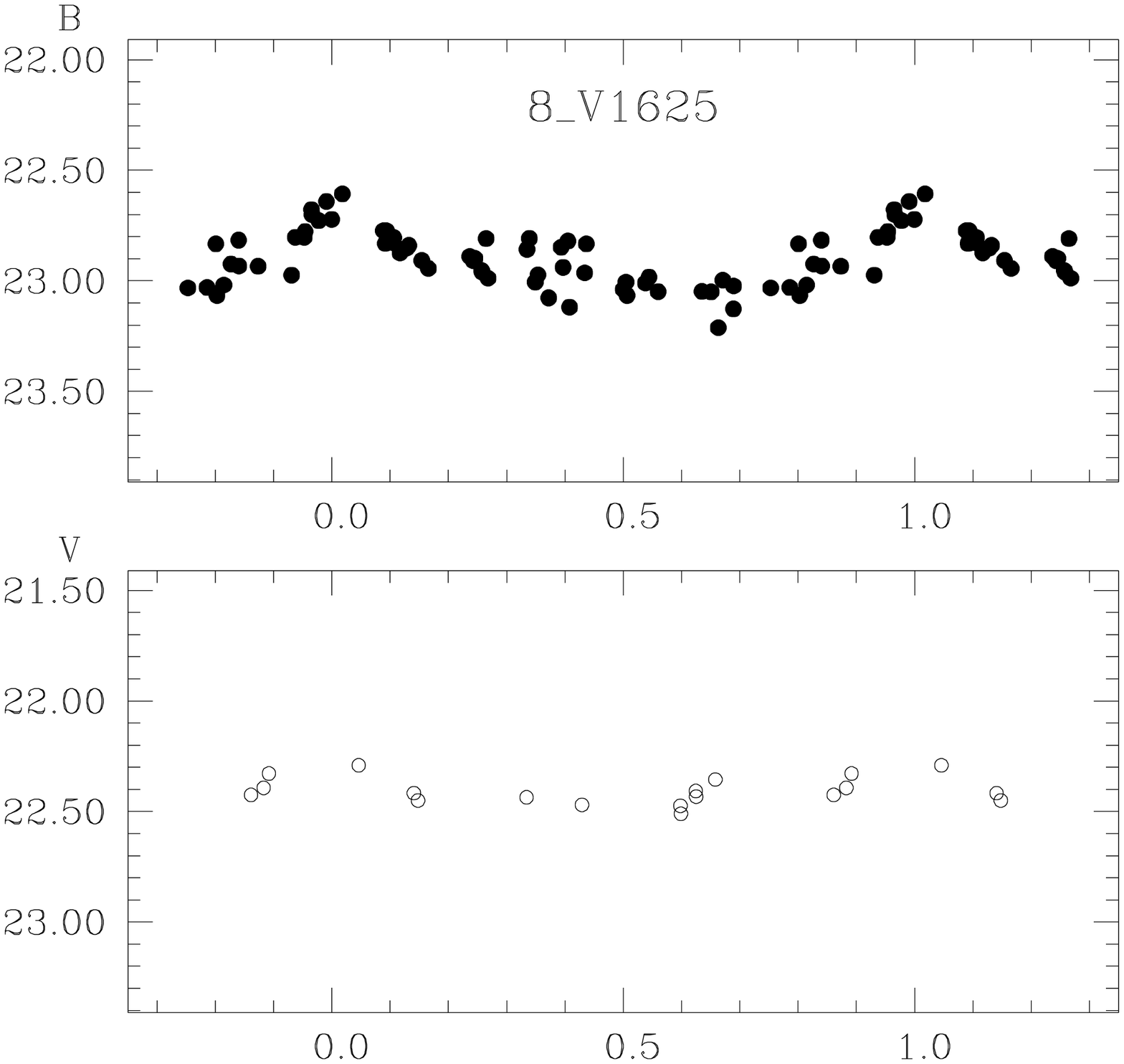}
\includegraphics[width=0.329\columnwidth,height=0.27\columnwidth]{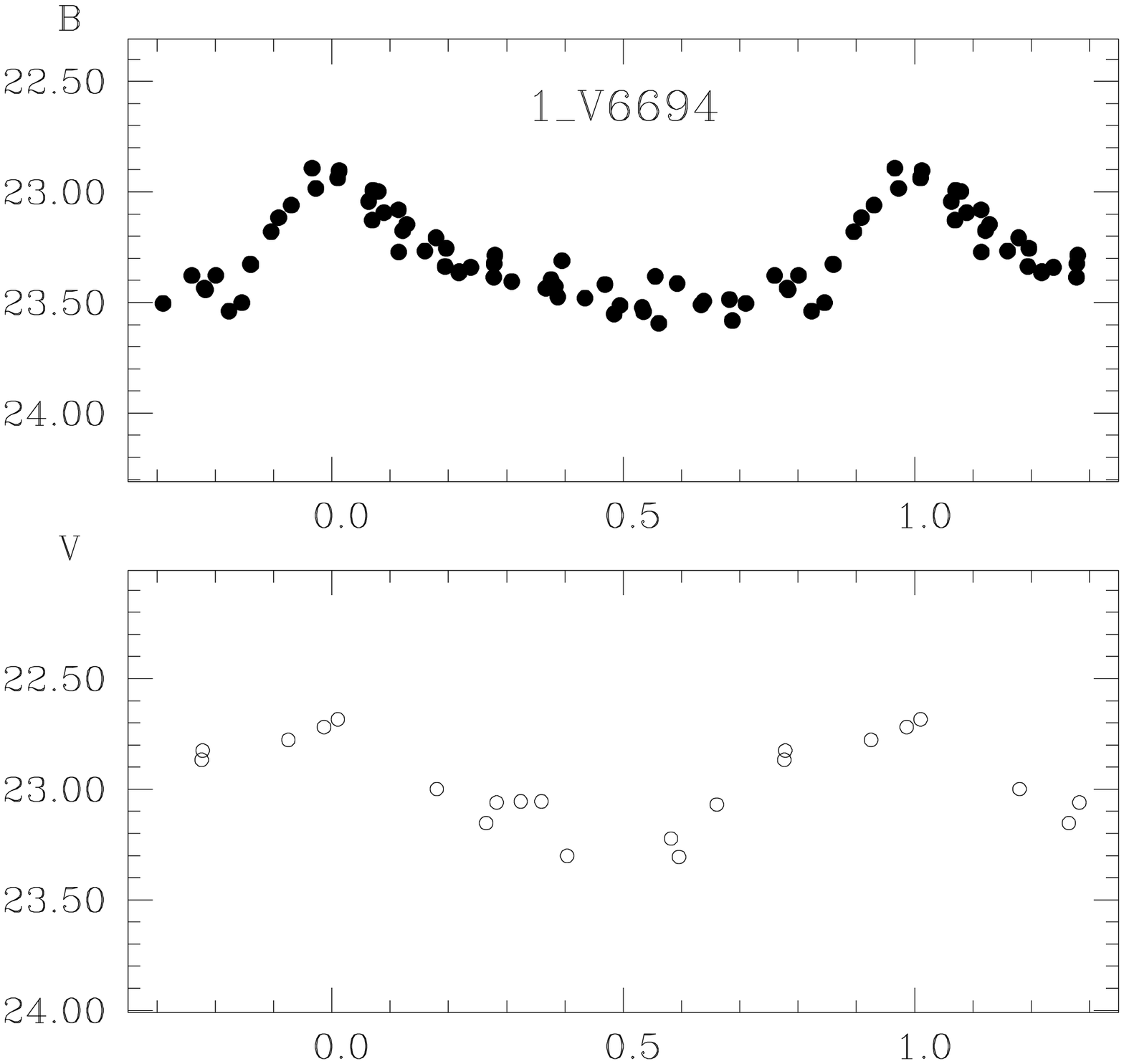}
\includegraphics[width=0.329\columnwidth,height=0.27\columnwidth]{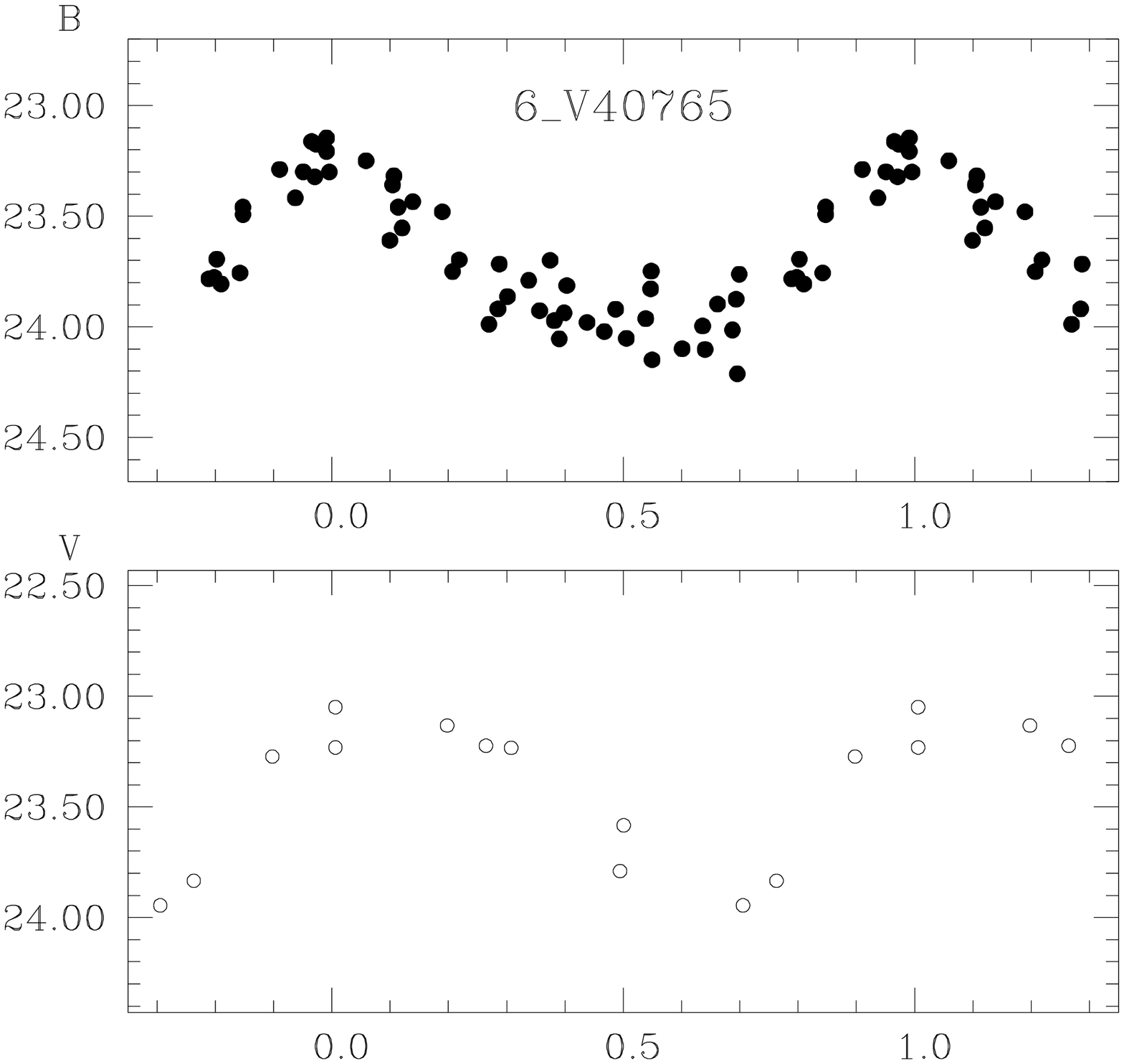}
\includegraphics[width=0.329\columnwidth,height=0.27\columnwidth]{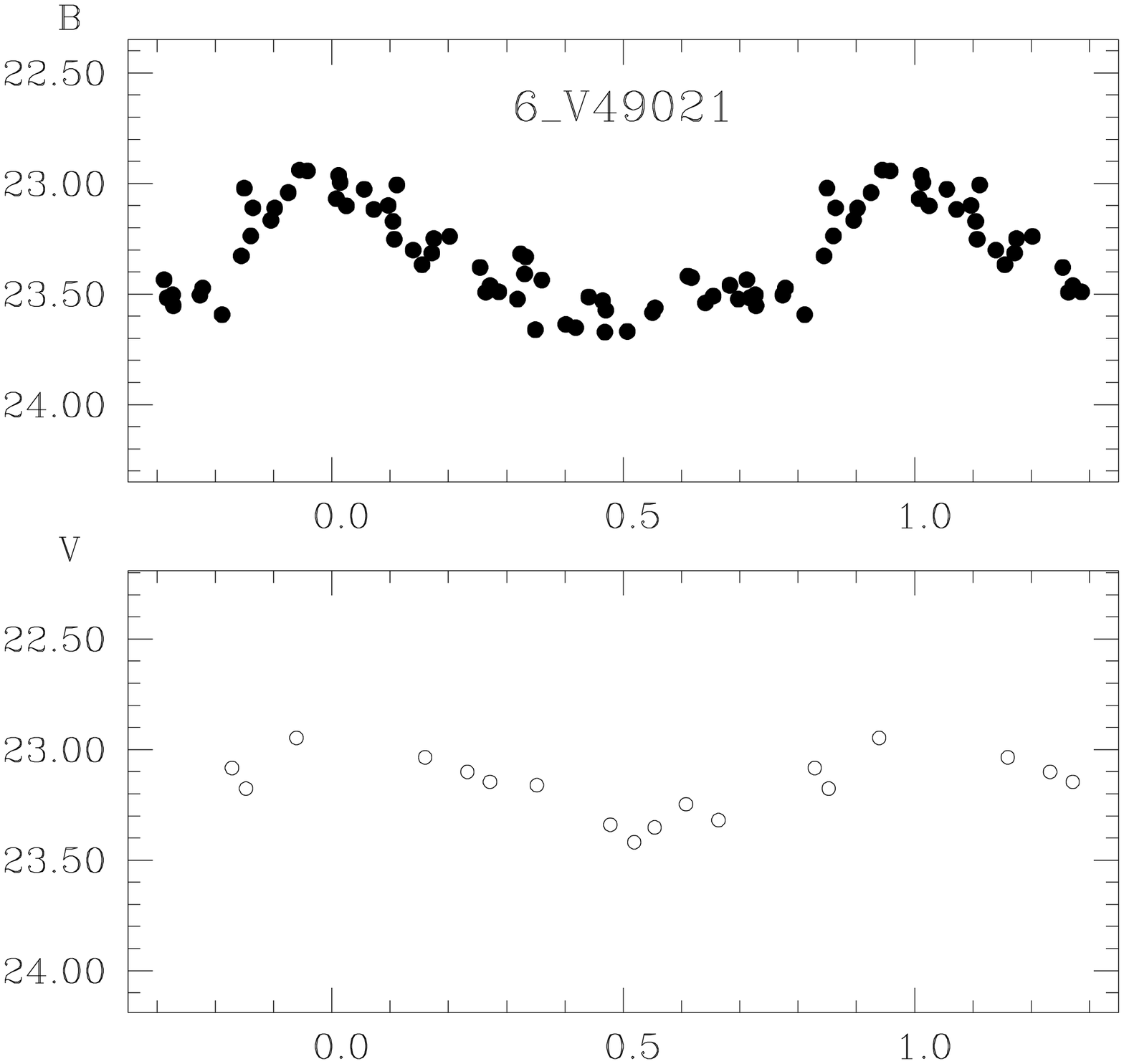}
\includegraphics[width=0.329\columnwidth,height=0.27\columnwidth]{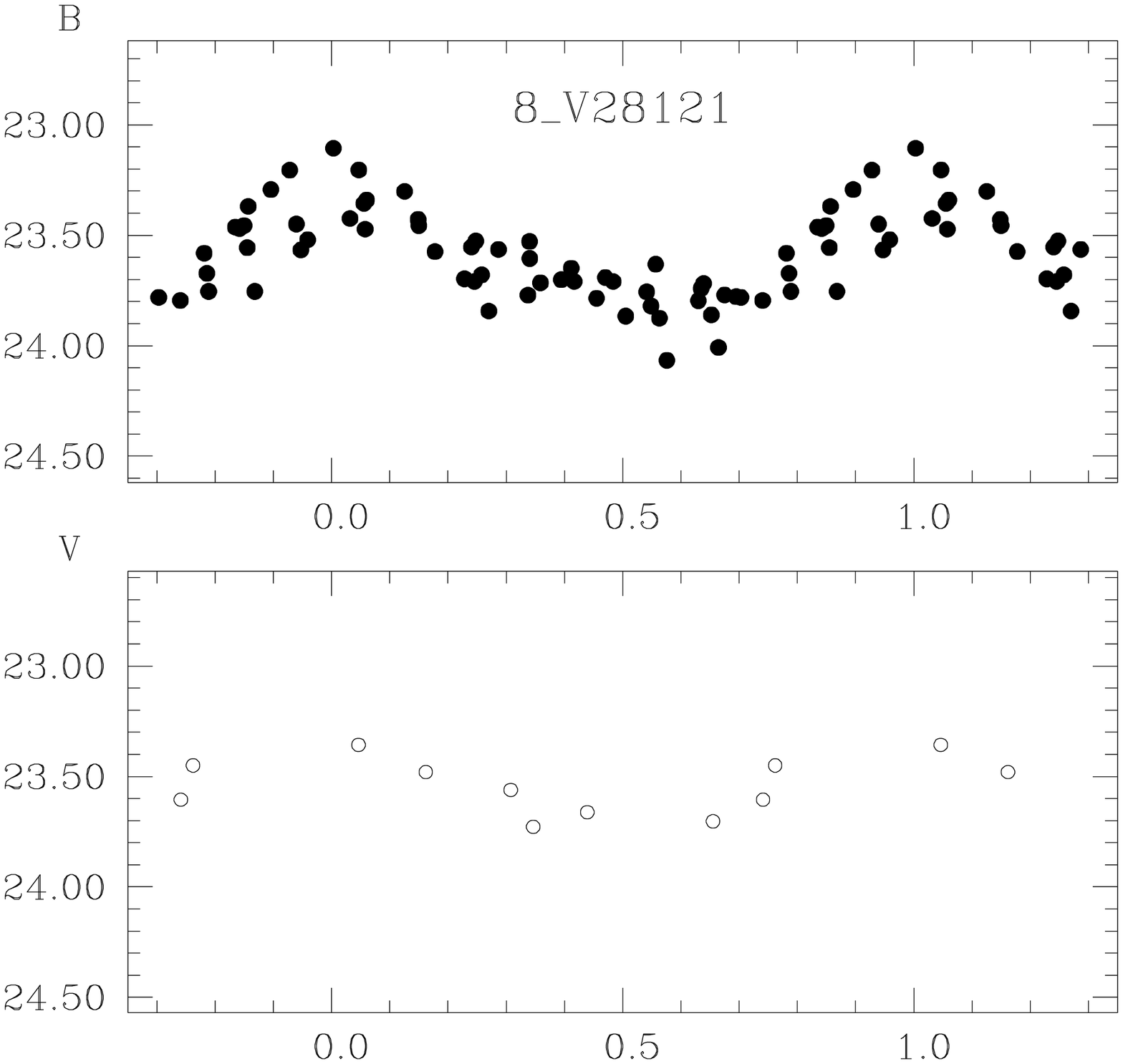}
\includegraphics[width=0.329\columnwidth,height=0.27\columnwidth]{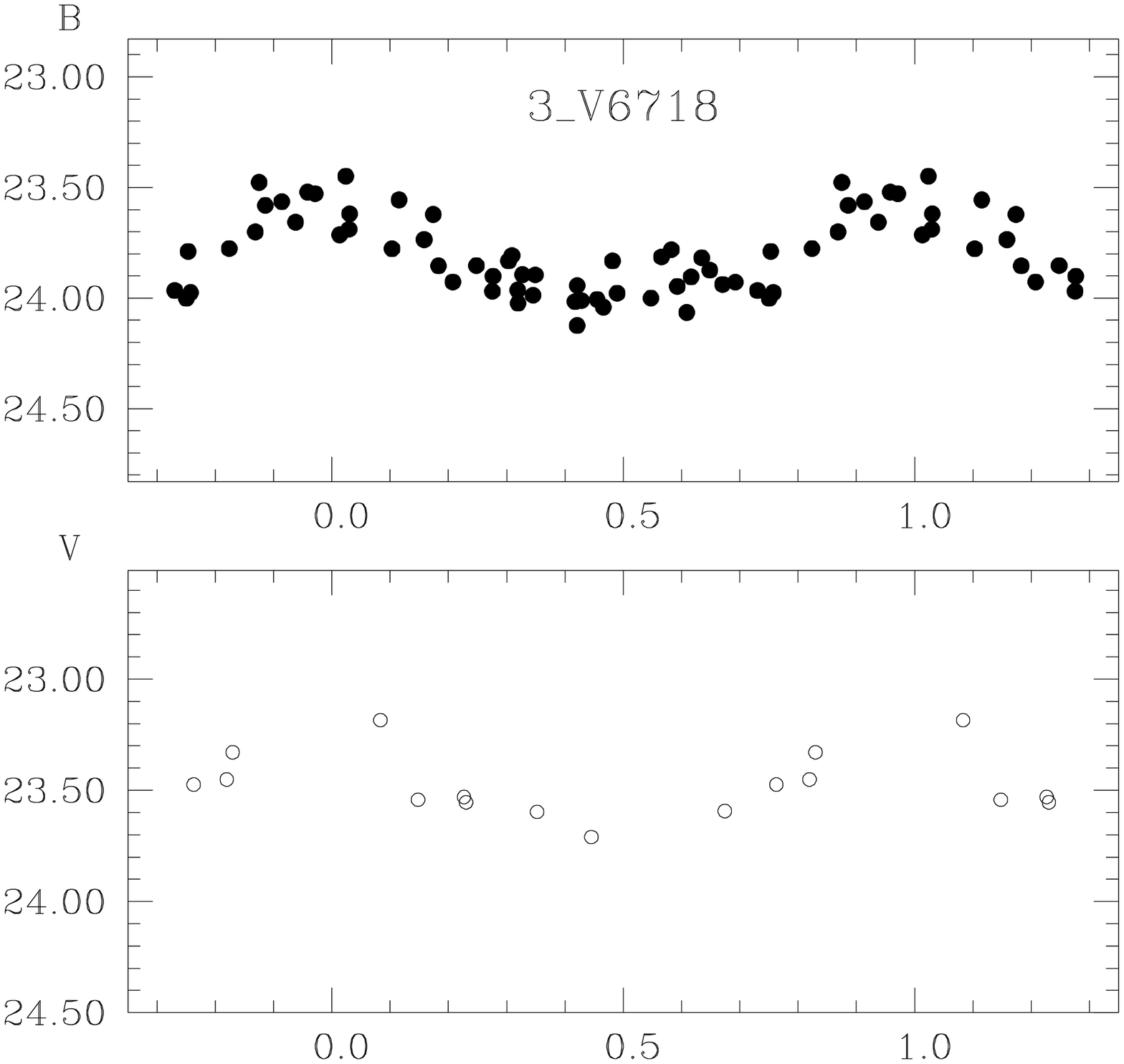}
\includegraphics[width=0.329\columnwidth,height=0.27\columnwidth]{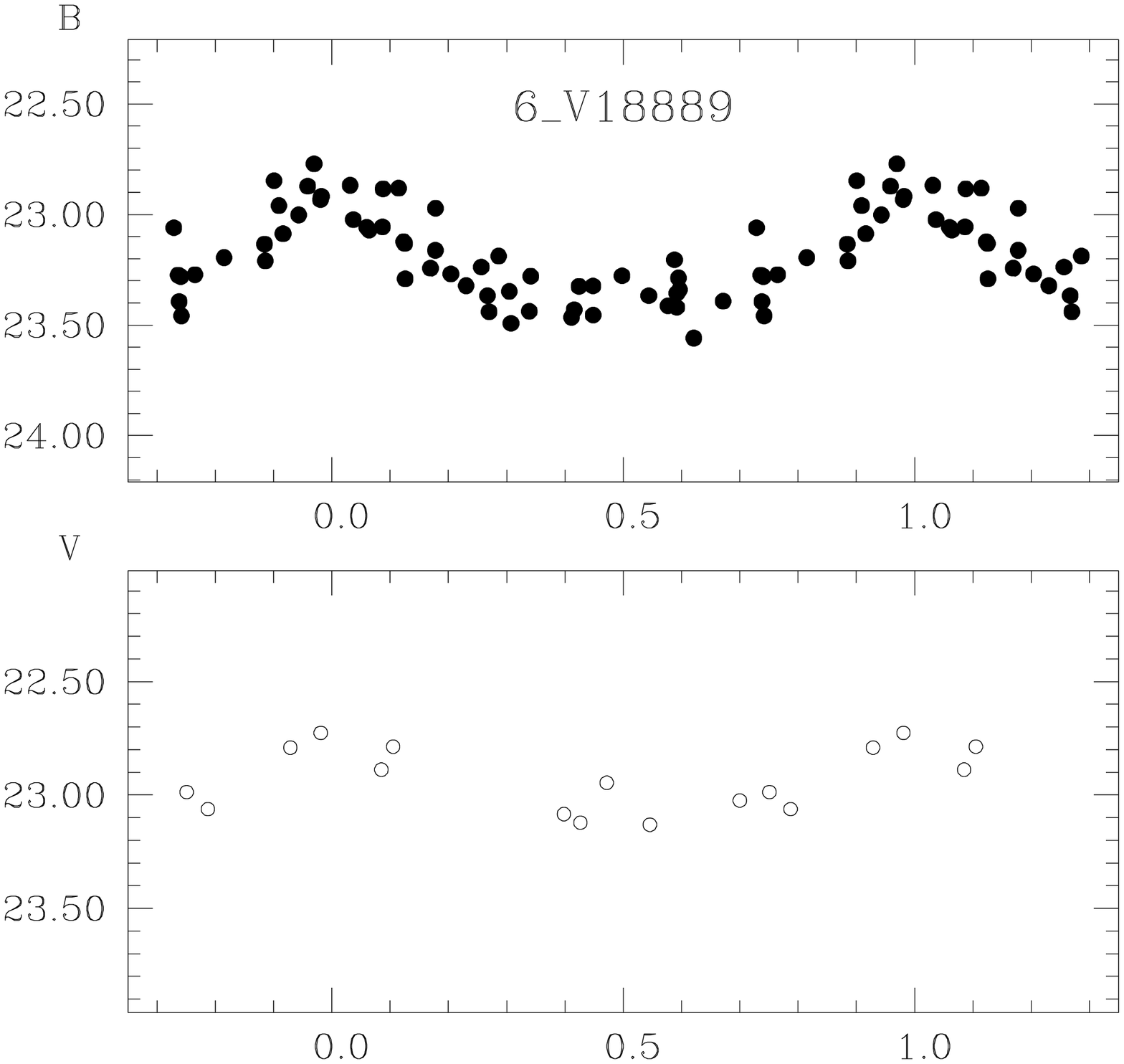}
\includegraphics[width=0.329\columnwidth,height=0.27\columnwidth]{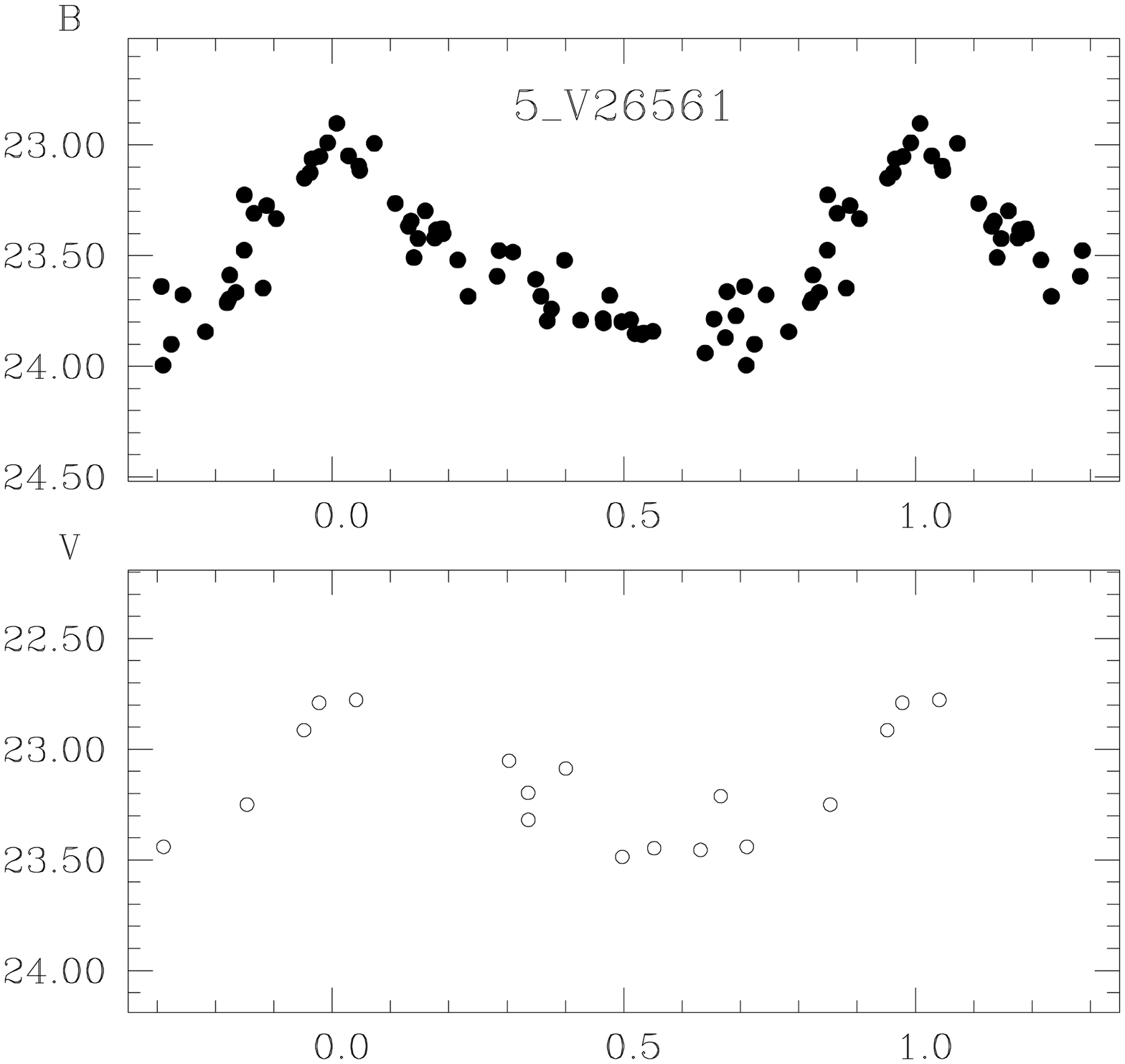}
\end{figure*}
\begin{figure*}
\includegraphics[width=0.329\columnwidth,height=0.27\columnwidth]{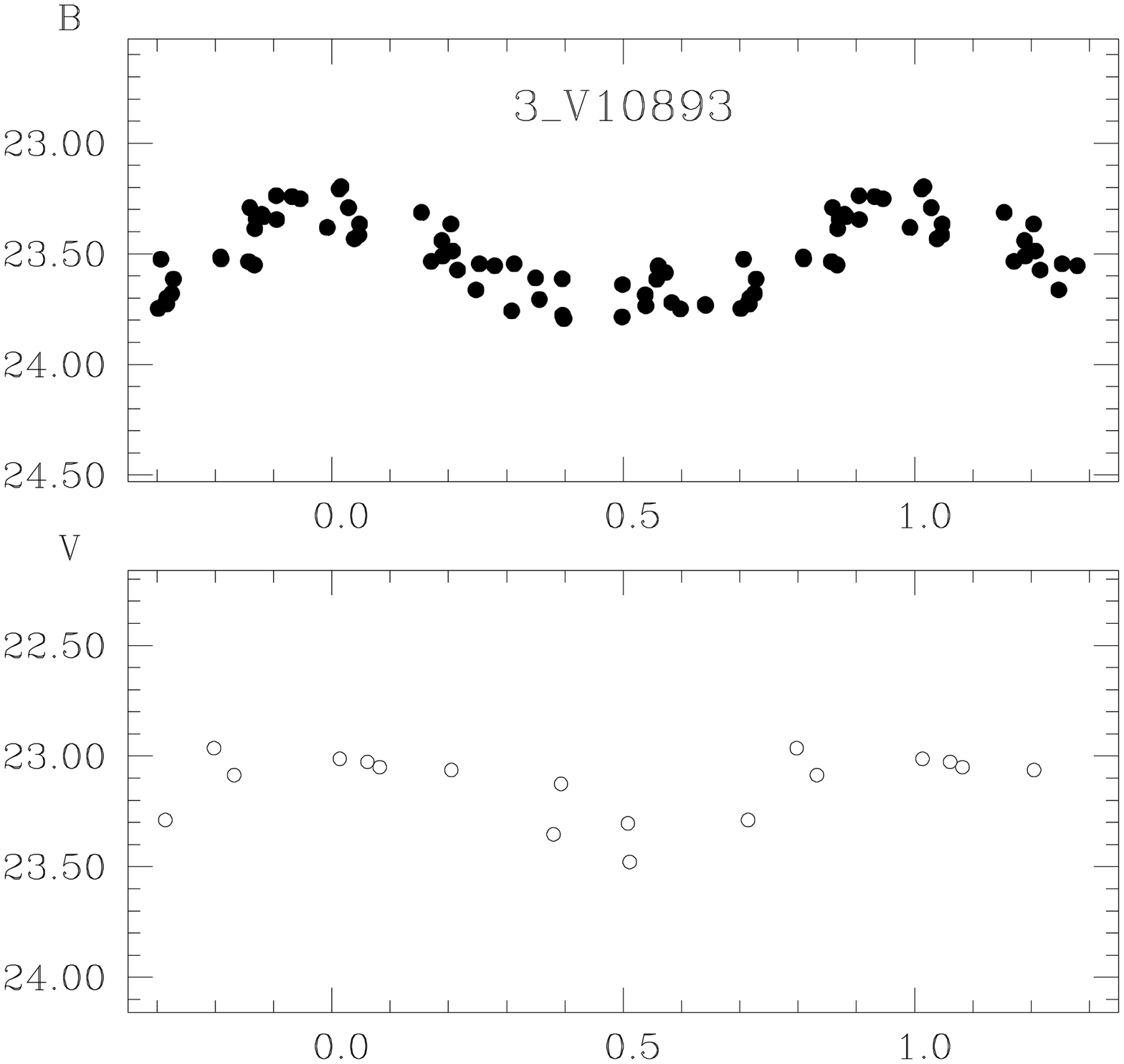}
\includegraphics[width=0.329\columnwidth,height=0.27\columnwidth]{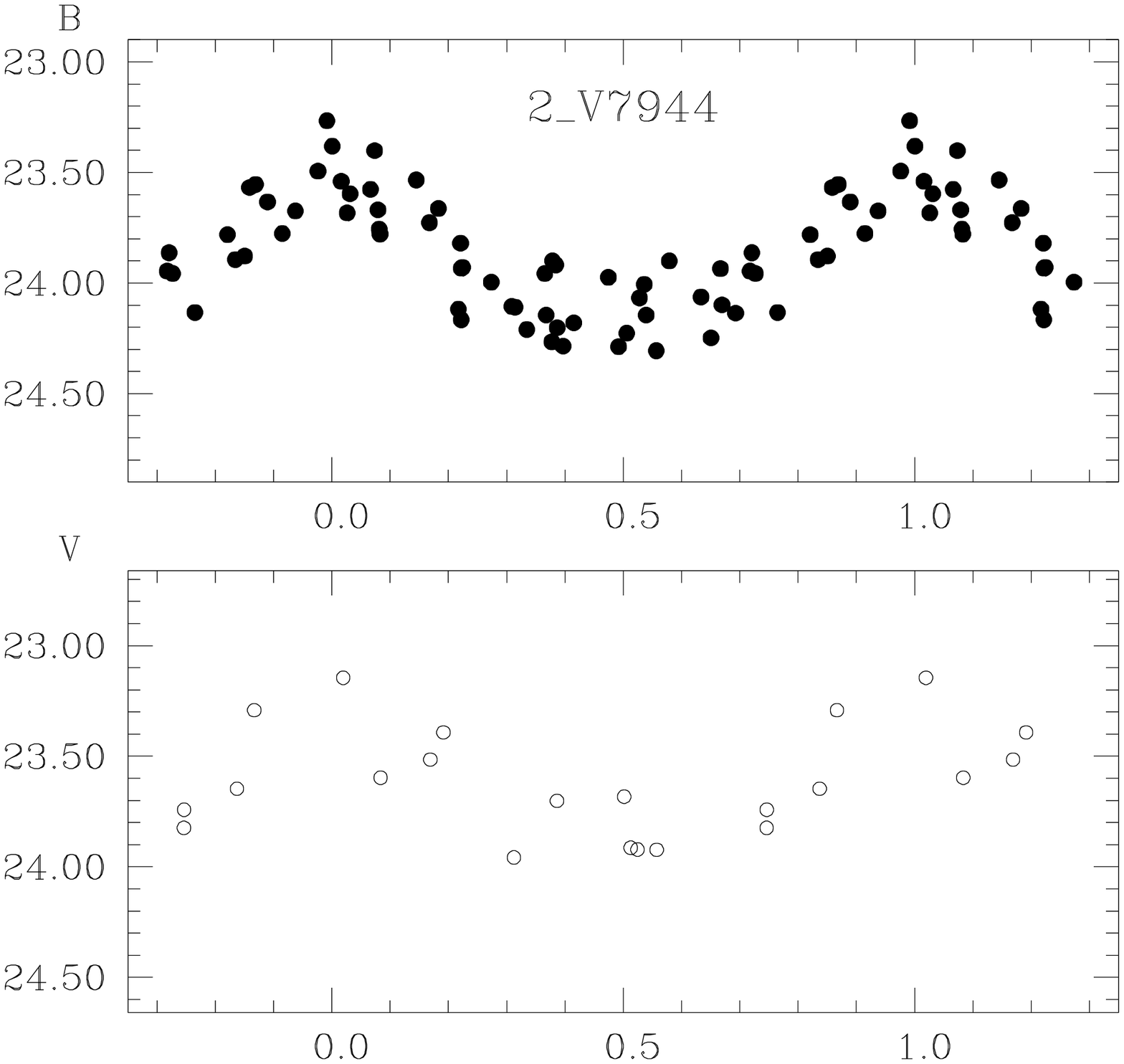}
\includegraphics[width=0.329\columnwidth,height=0.27\columnwidth]{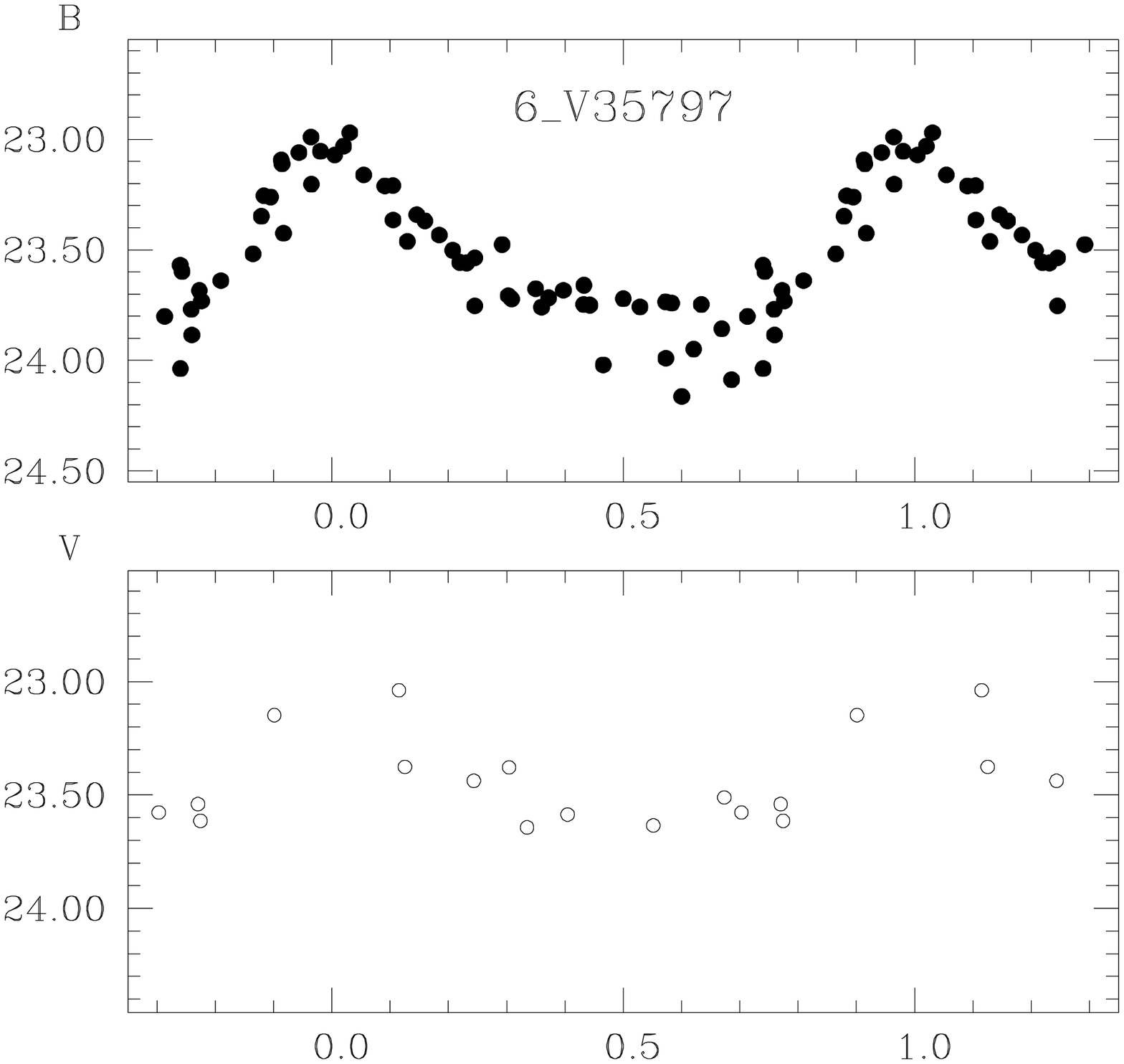}
\includegraphics[width=0.329\columnwidth,height=0.27\columnwidth]{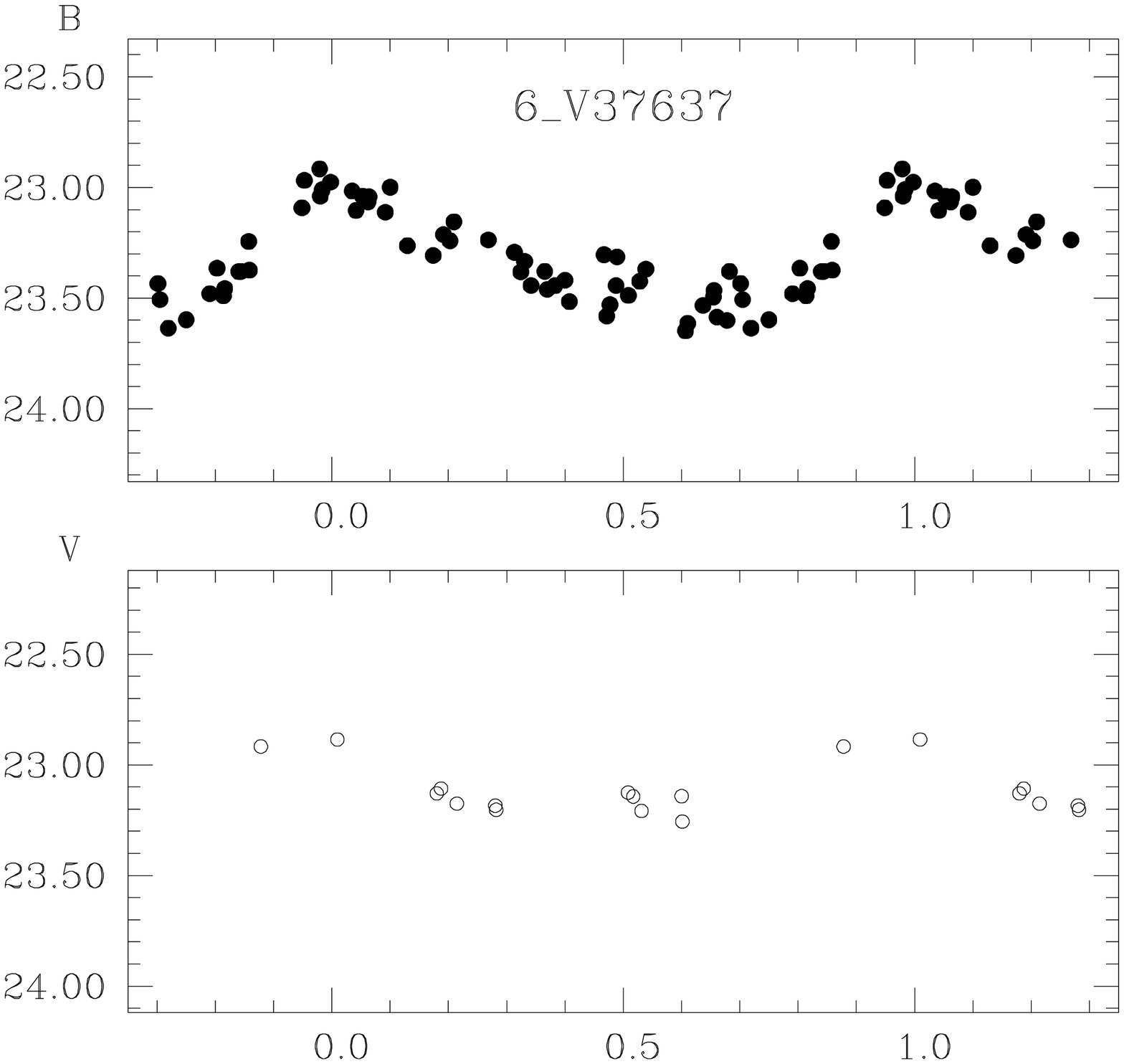}
\includegraphics[width=0.329\columnwidth,height=0.27\columnwidth]{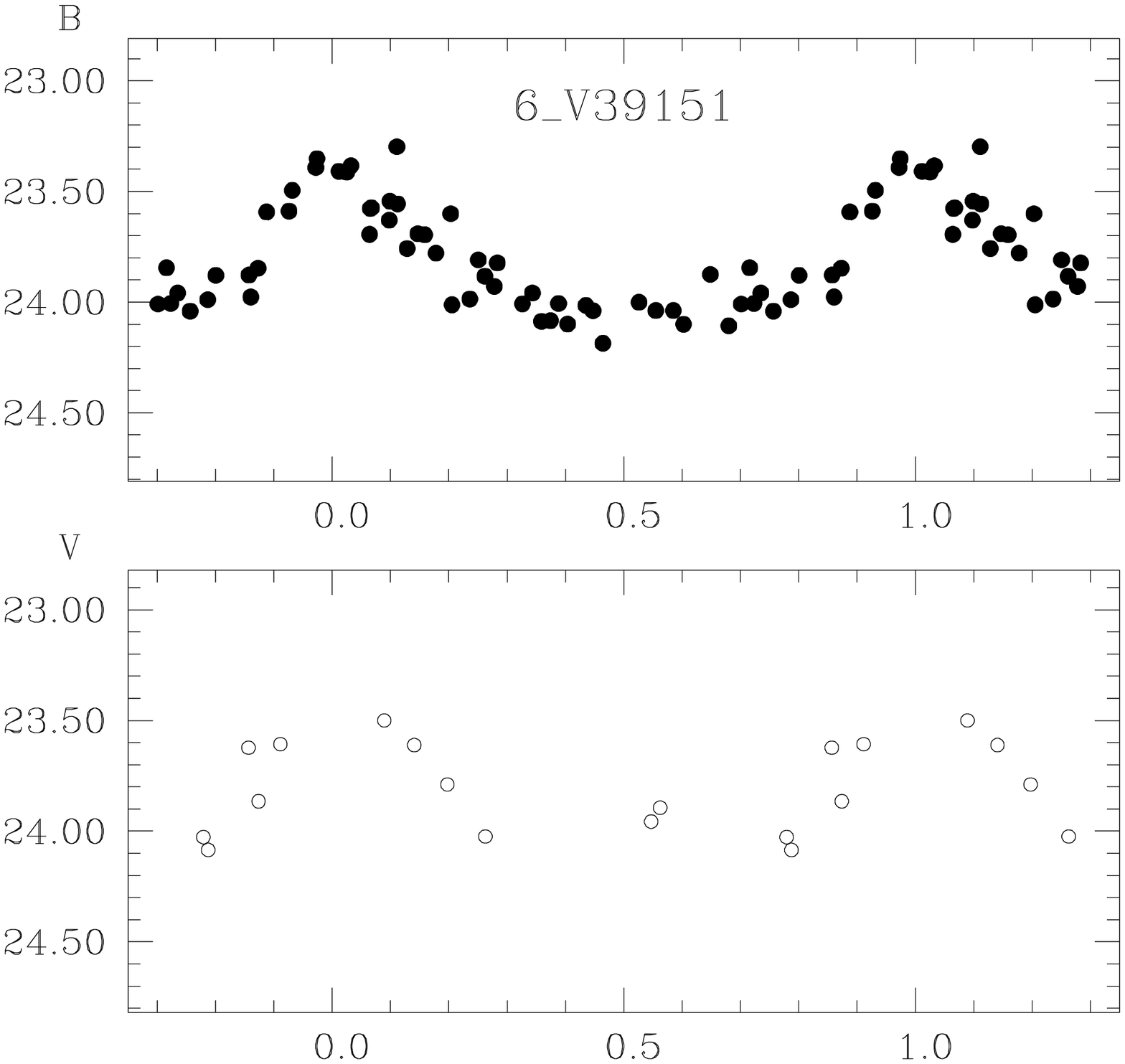}
\includegraphics[width=0.329\columnwidth,height=0.27\columnwidth]{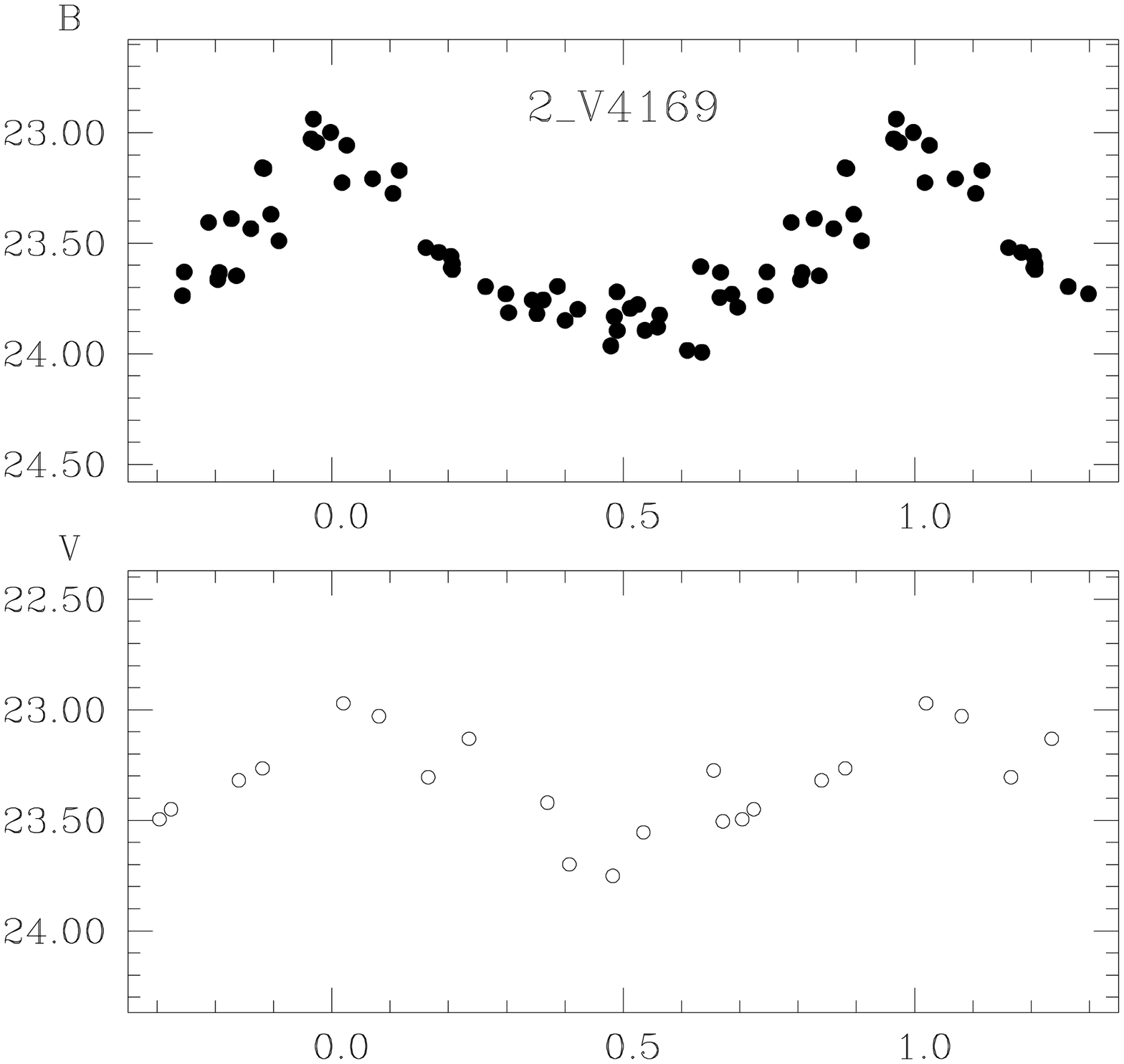}
\includegraphics[width=0.329\columnwidth,height=0.27\columnwidth]{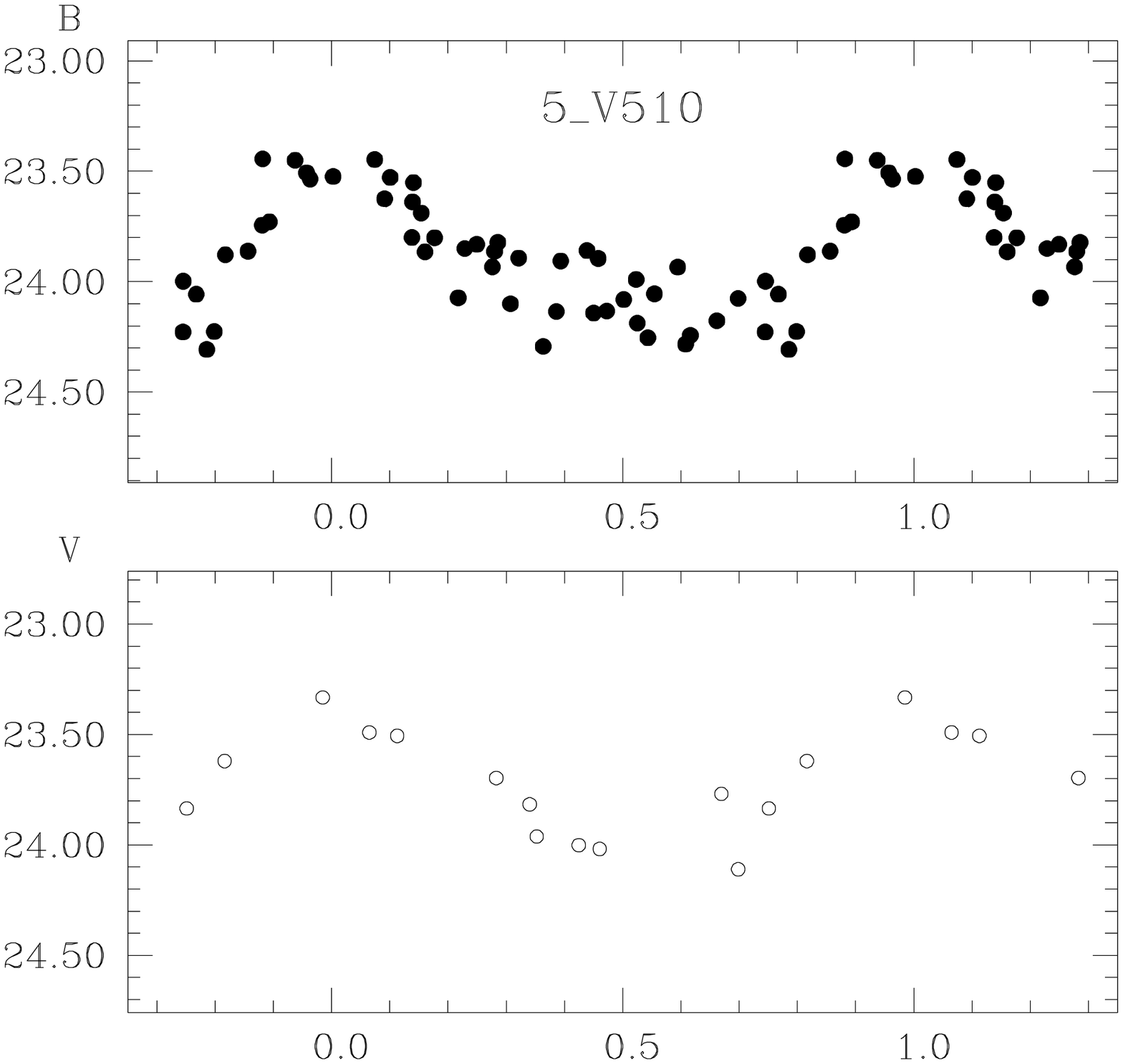}
\includegraphics[width=0.329\columnwidth,height=0.27\columnwidth]{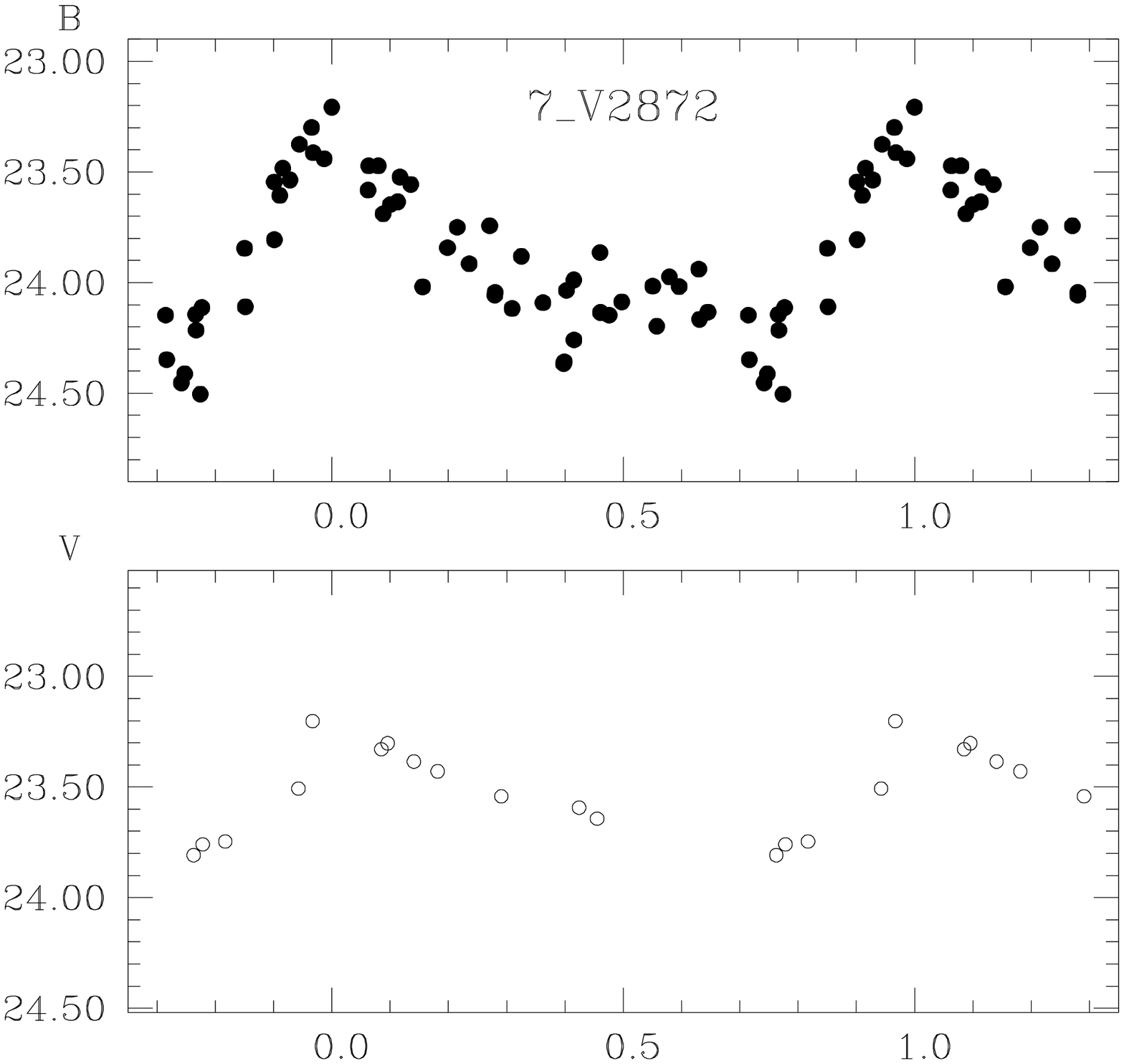}
\includegraphics[width=0.329\columnwidth,height=0.27\columnwidth]{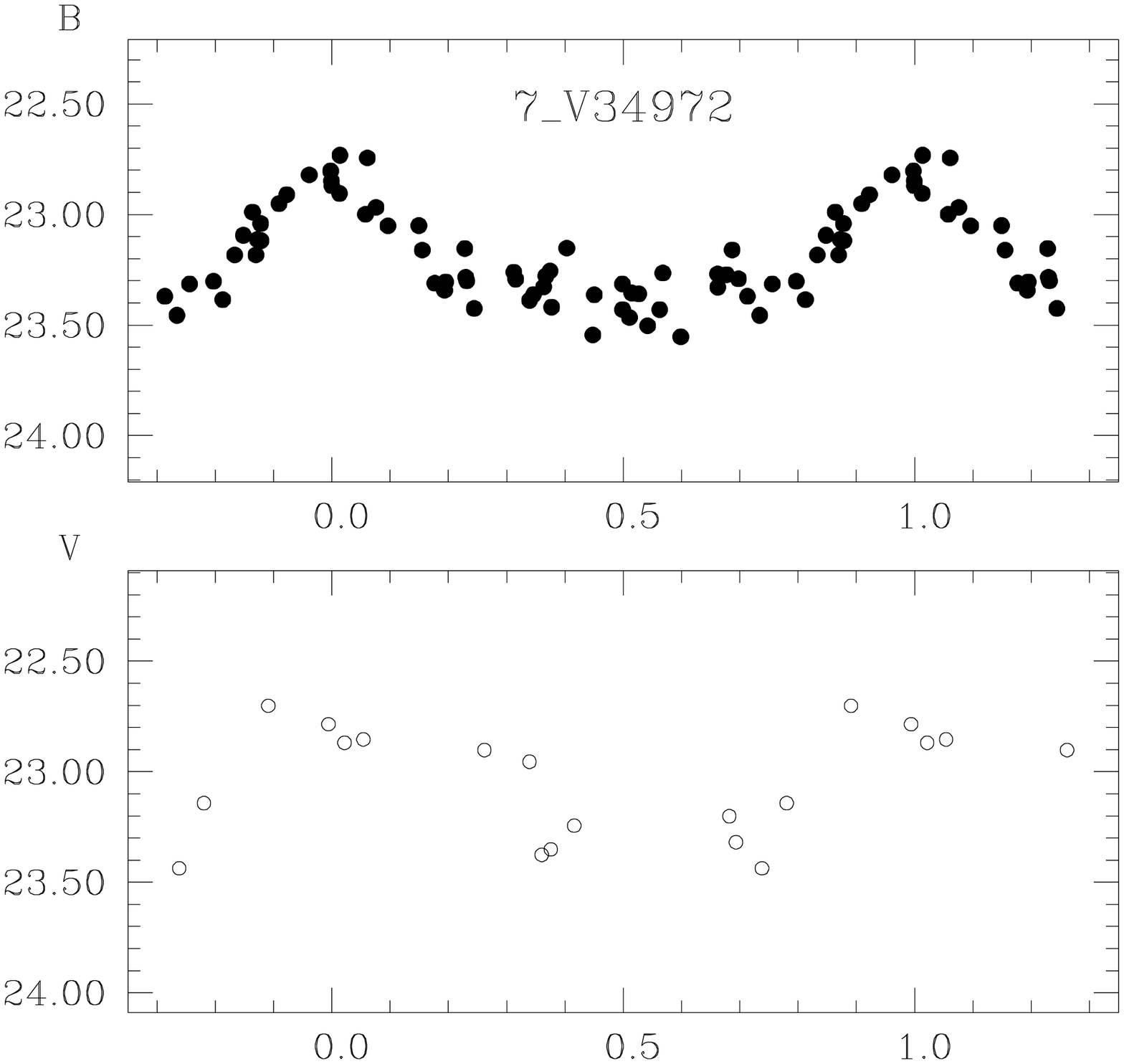}
\includegraphics[width=0.329\columnwidth,height=0.27\columnwidth]{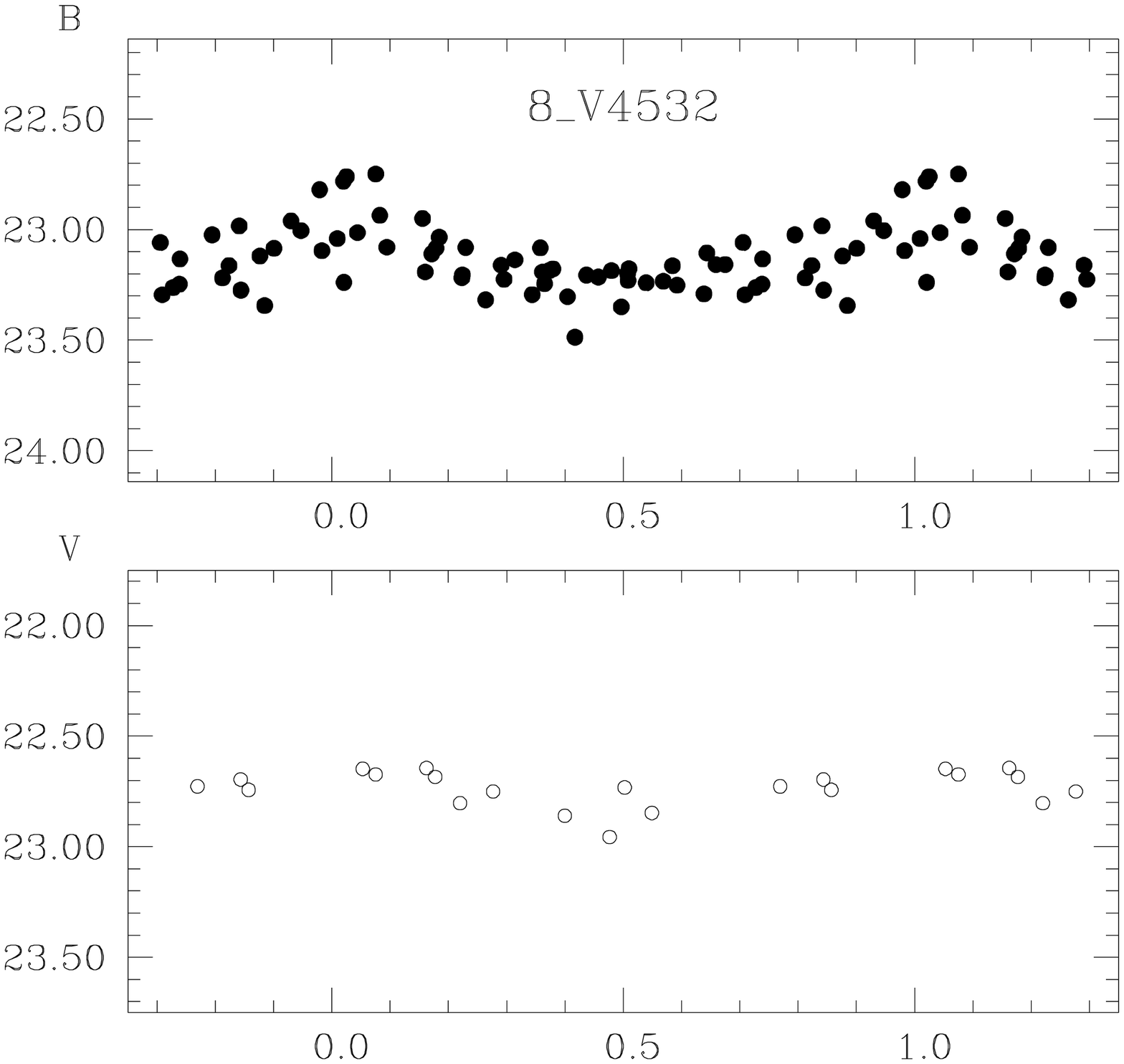}
\includegraphics[width=0.329\columnwidth,height=0.27\columnwidth]{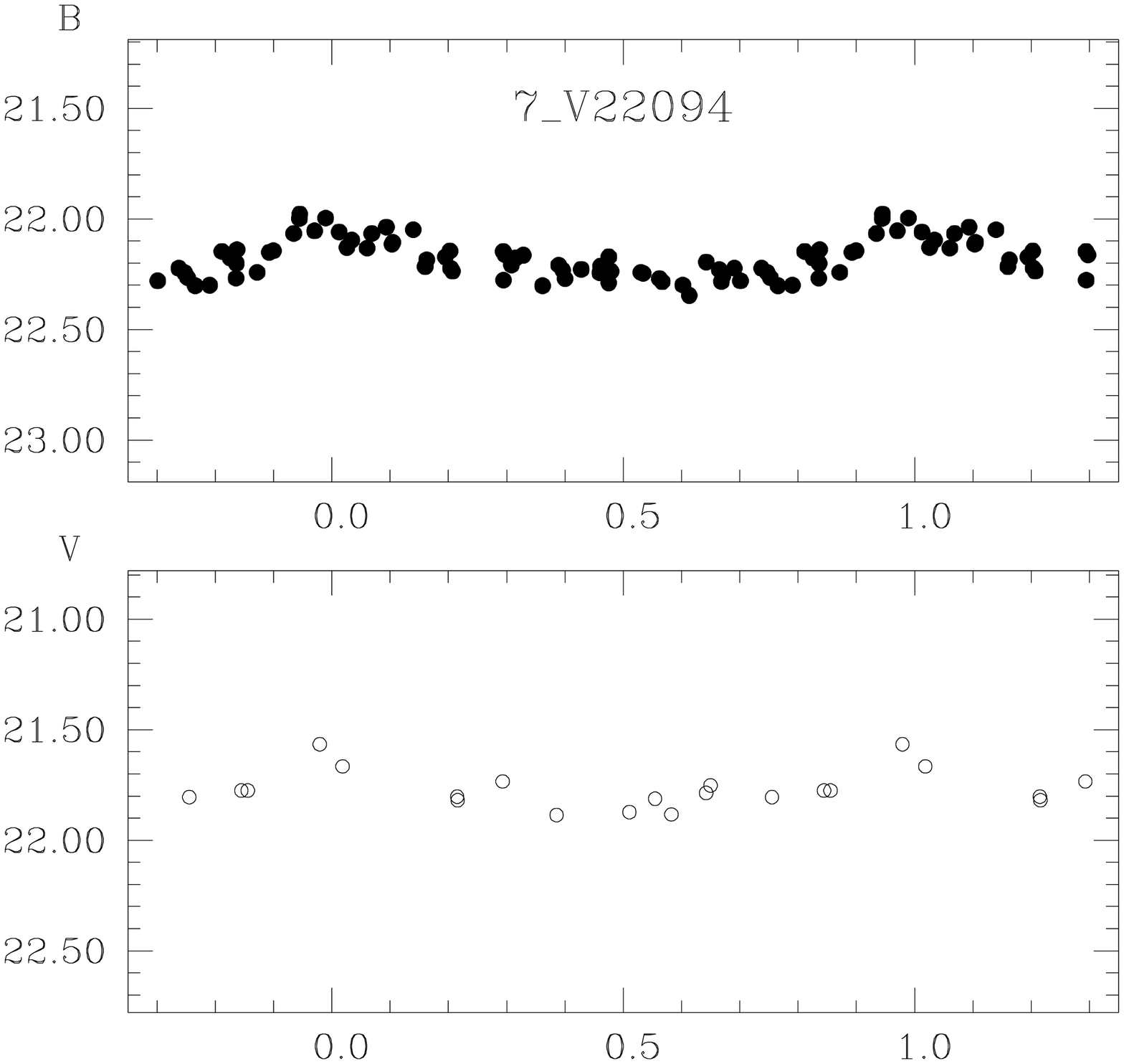}
\includegraphics[width=0.329\columnwidth,height=0.27\columnwidth]{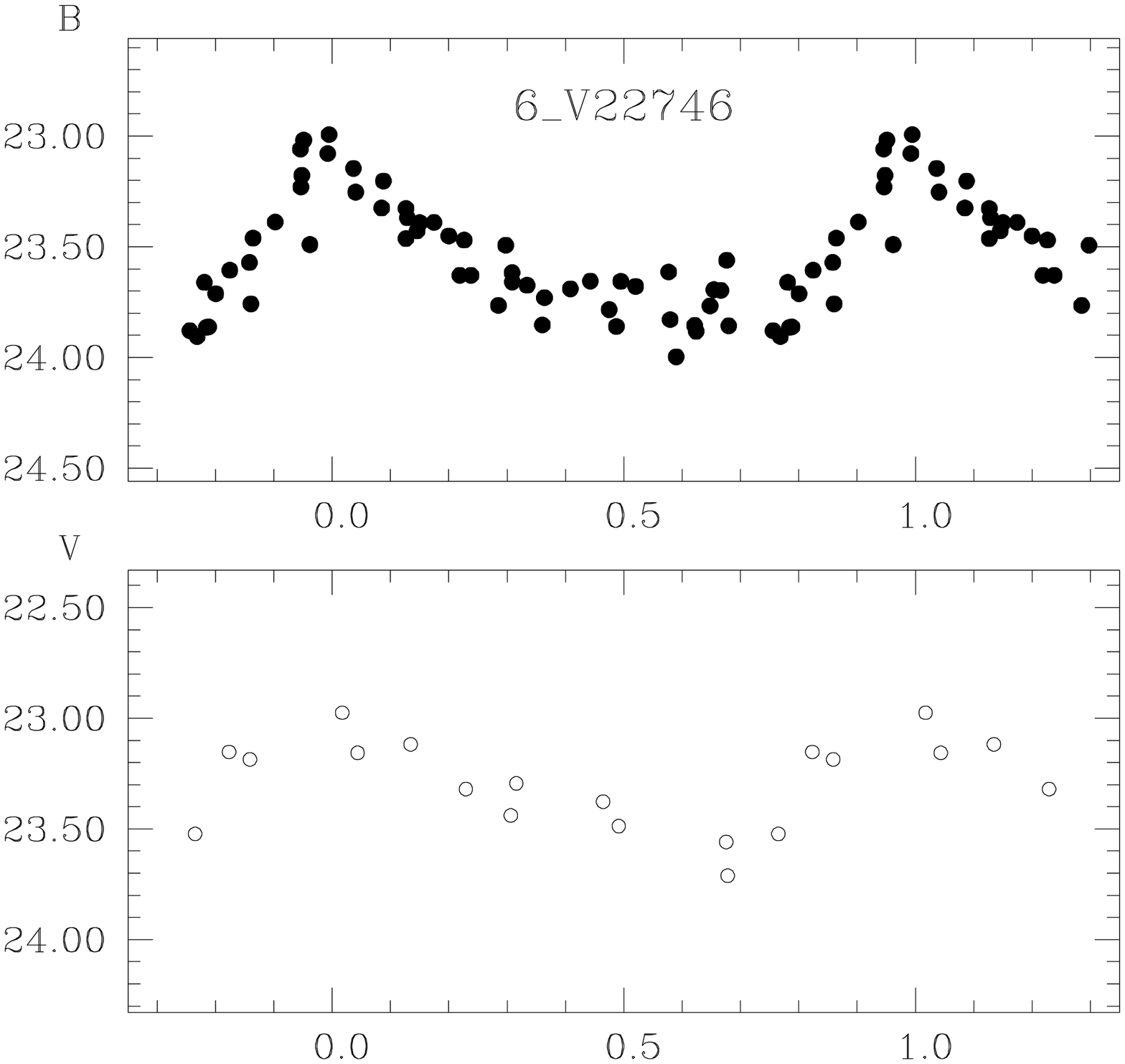}
\includegraphics[width=0.329\columnwidth,height=0.27\columnwidth]{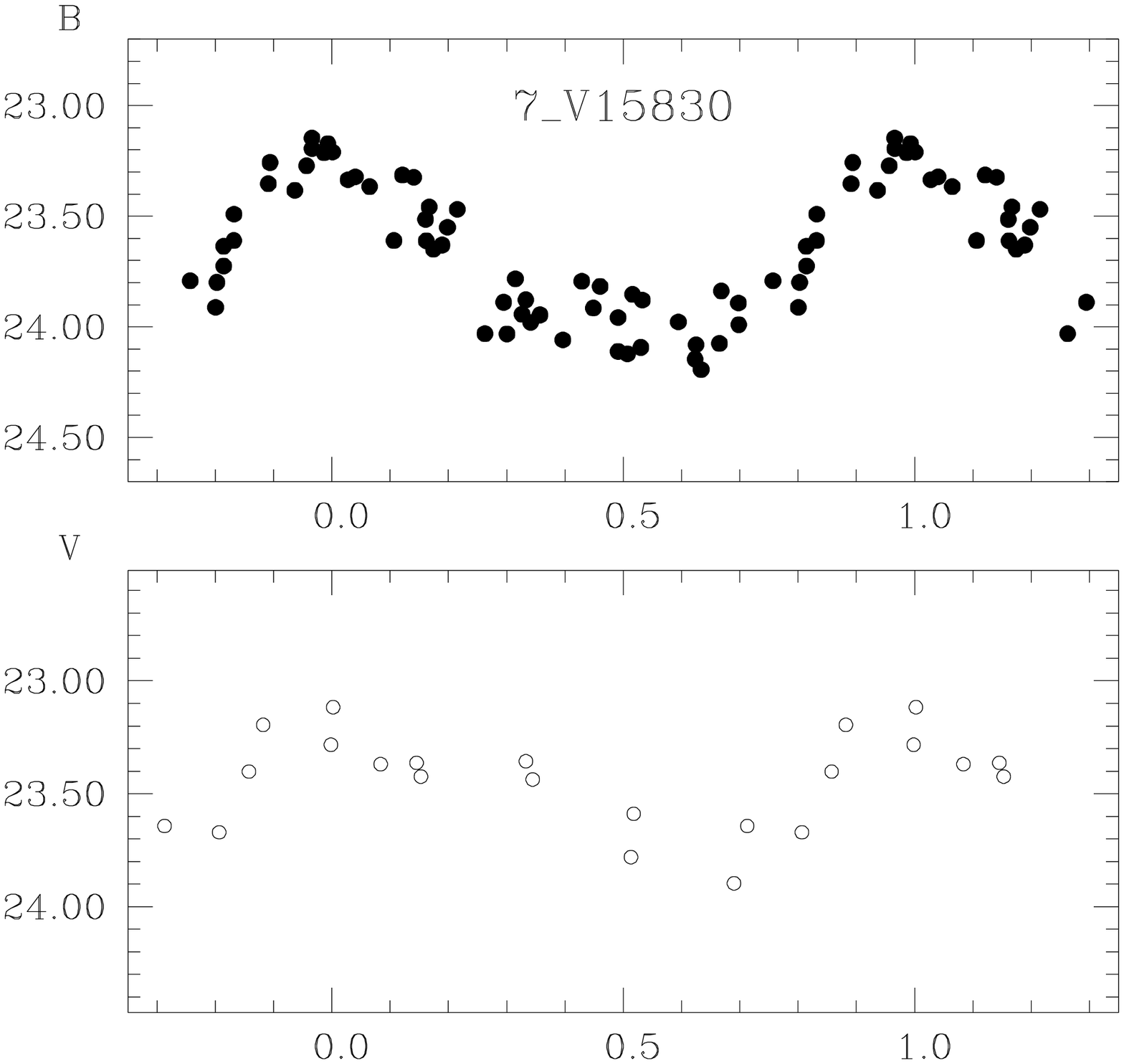}
\includegraphics[width=0.329\columnwidth,height=0.27\columnwidth]{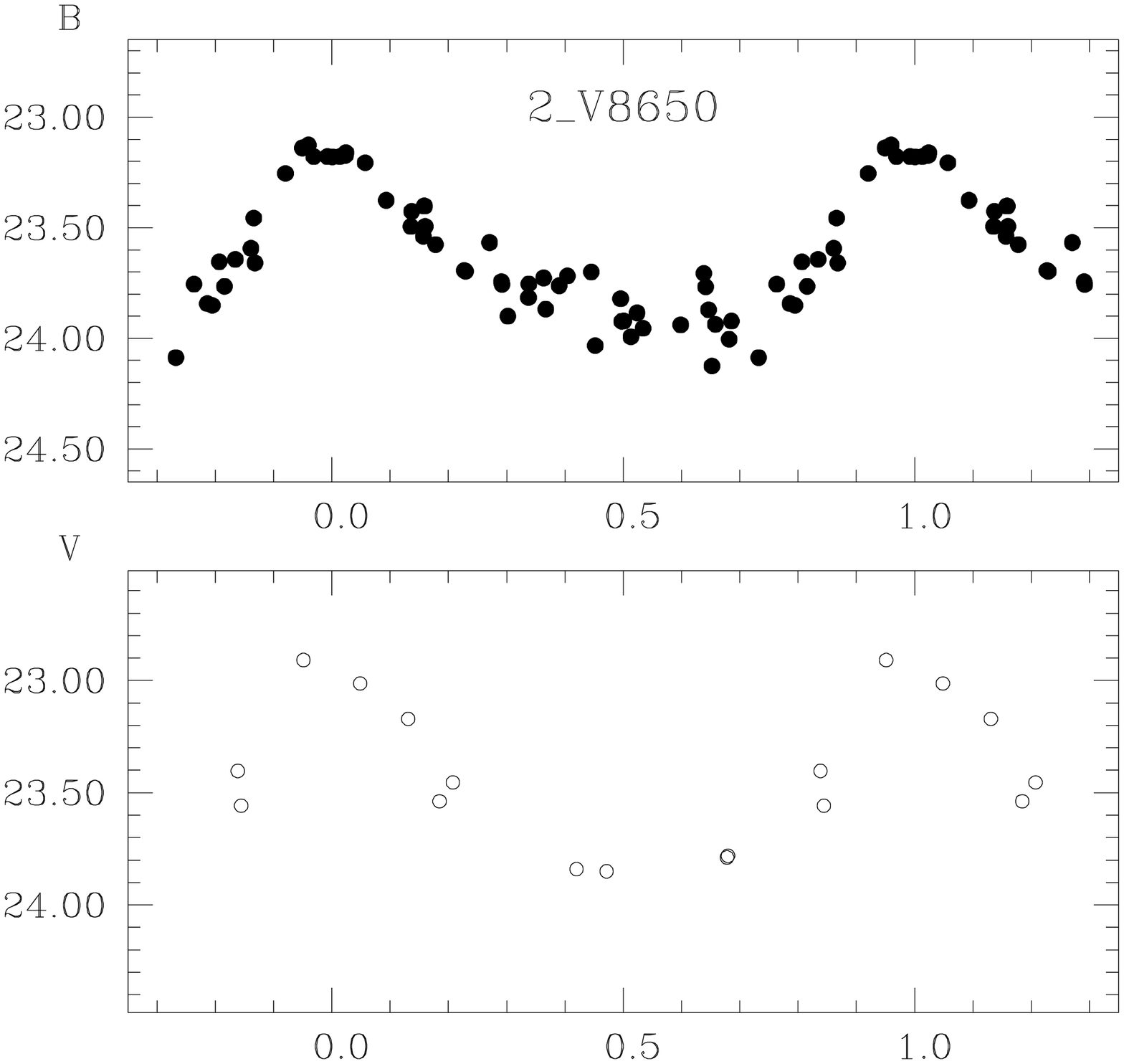}
\includegraphics[width=0.329\columnwidth,height=0.27\columnwidth]{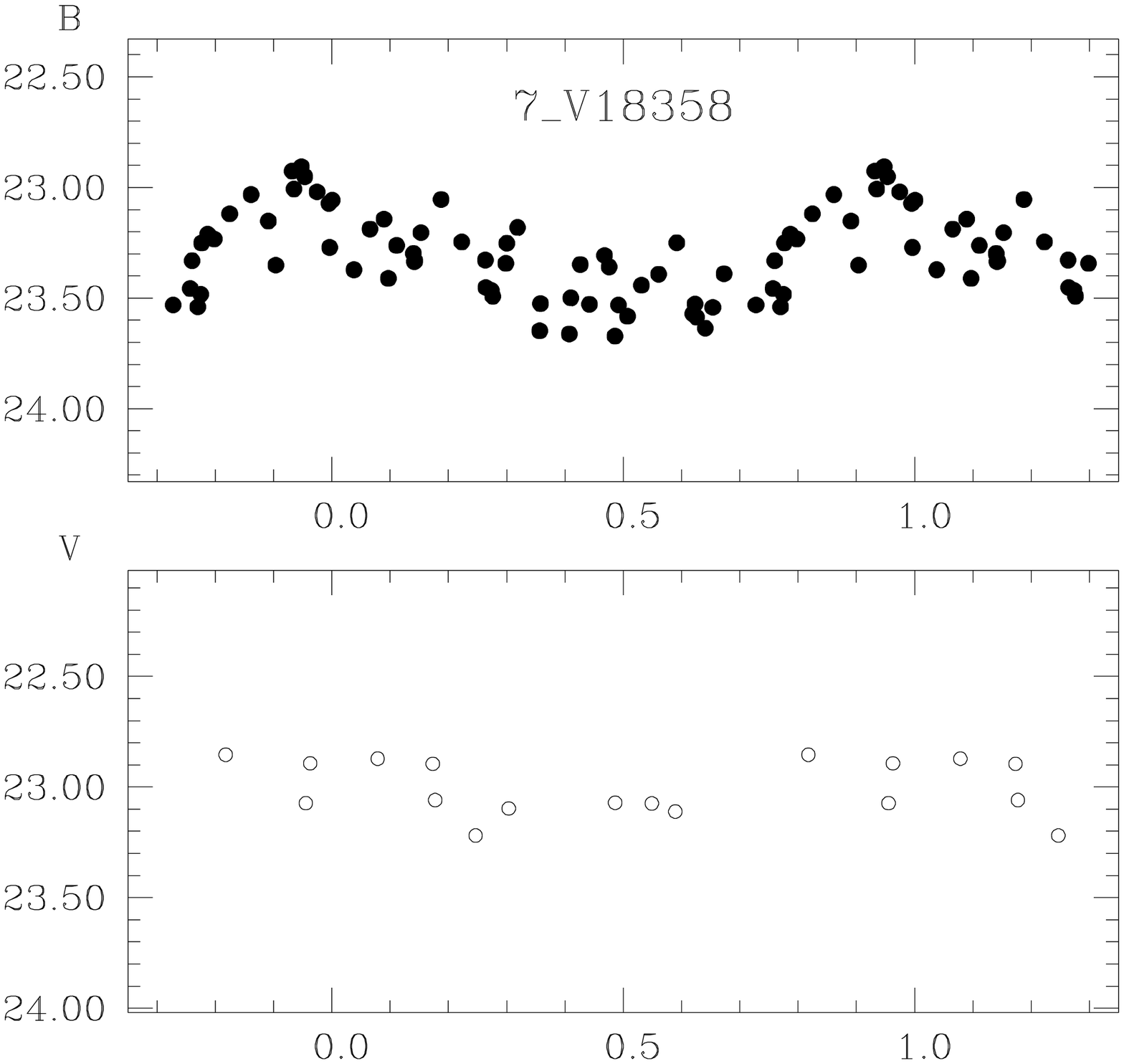}
\end{figure*}
\begin{figure*}
\includegraphics[width=0.329\columnwidth,height=0.27\columnwidth]{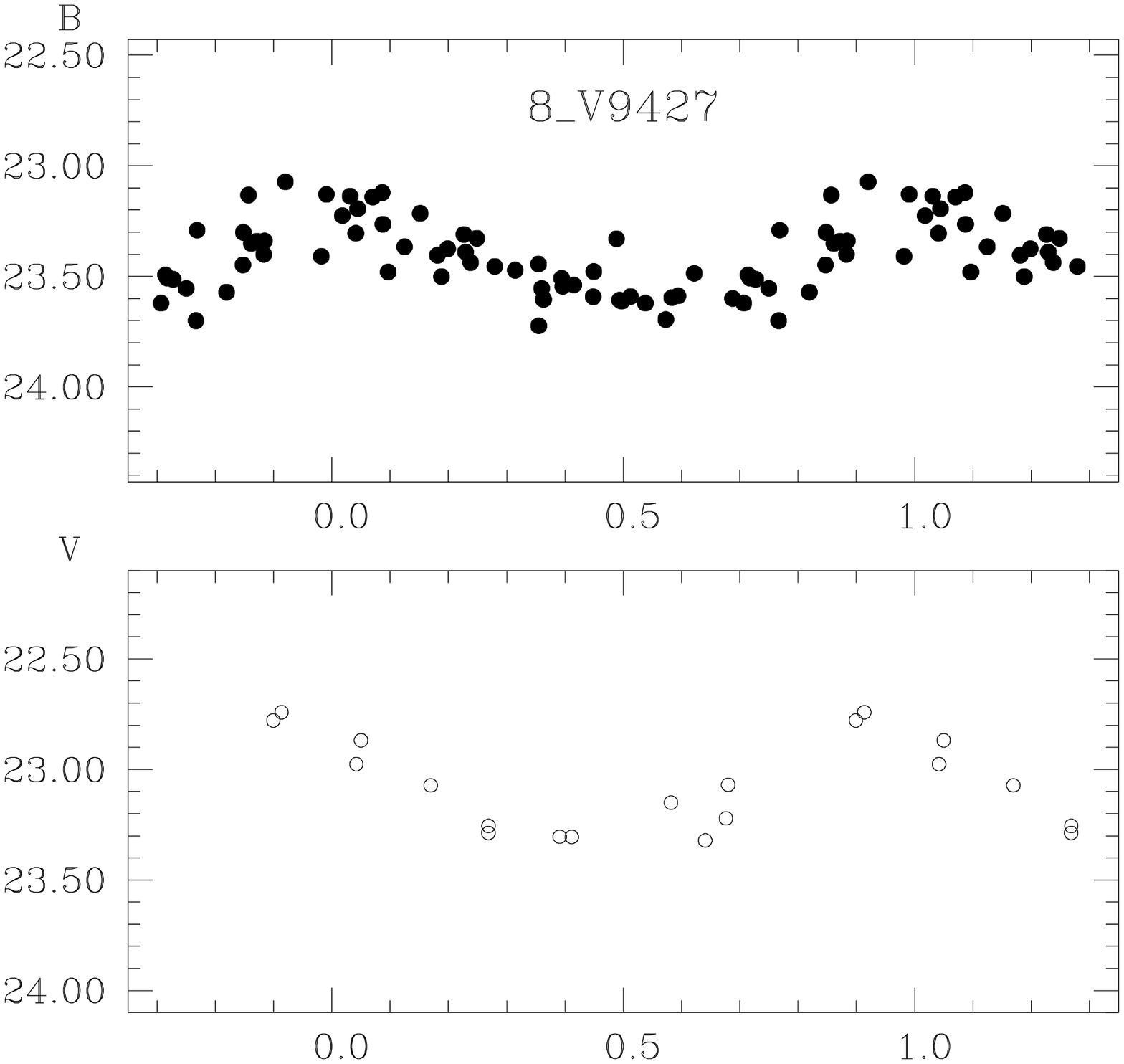}
\includegraphics[width=0.329\columnwidth,height=0.27\columnwidth]{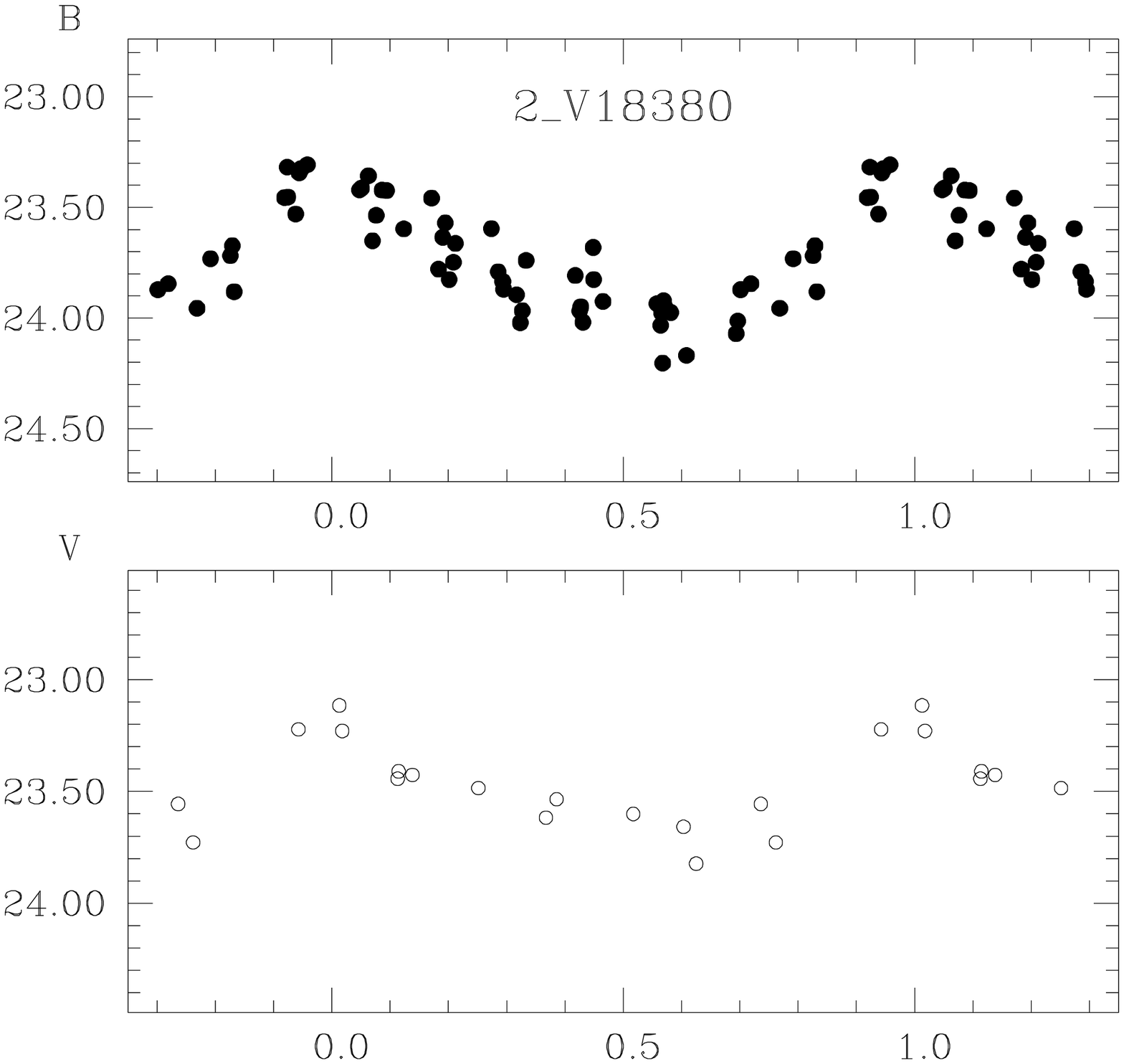}
\includegraphics[width=0.329\columnwidth,height=0.27\columnwidth]{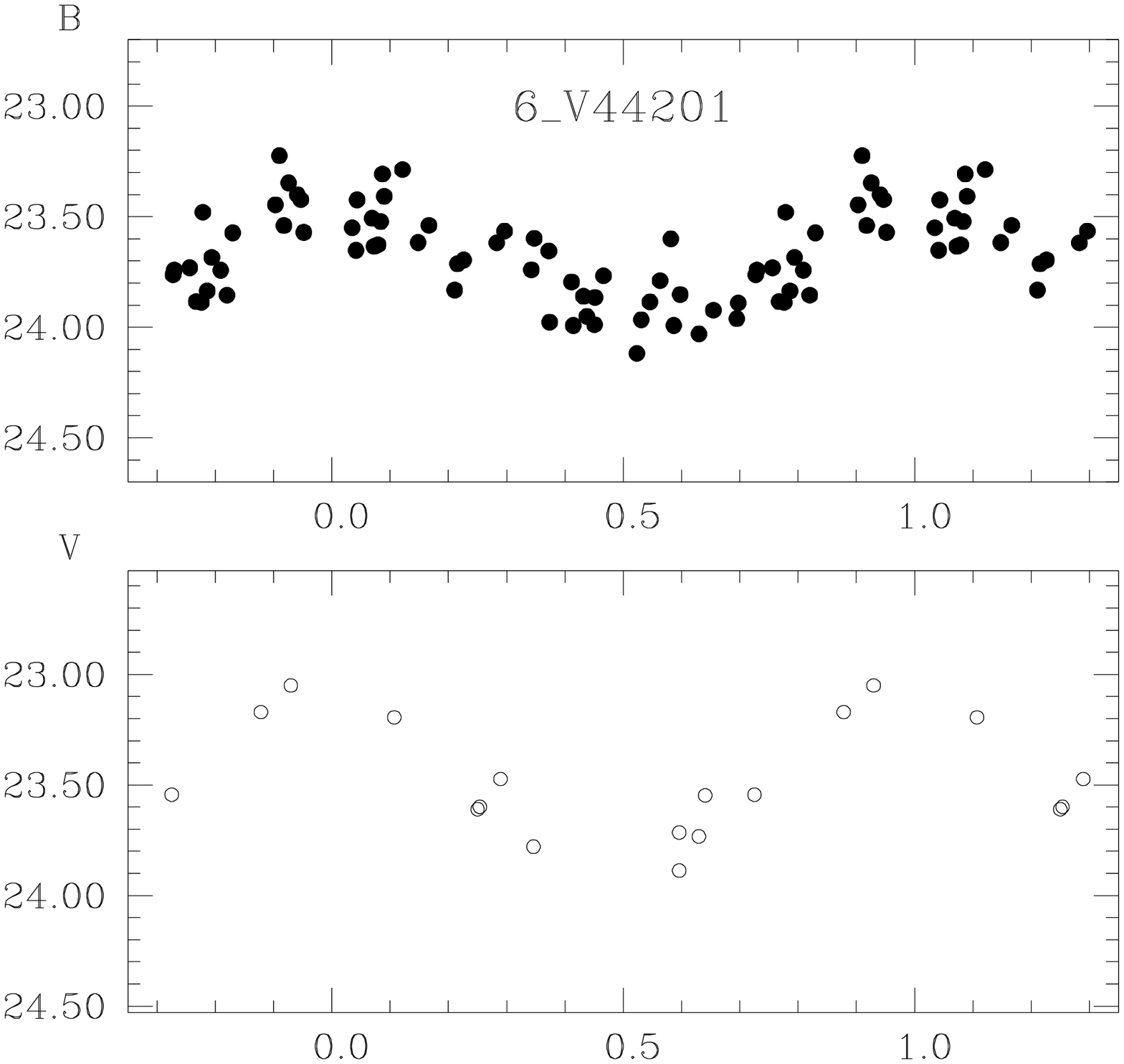}
\includegraphics[width=0.329\columnwidth,height=0.27\columnwidth]{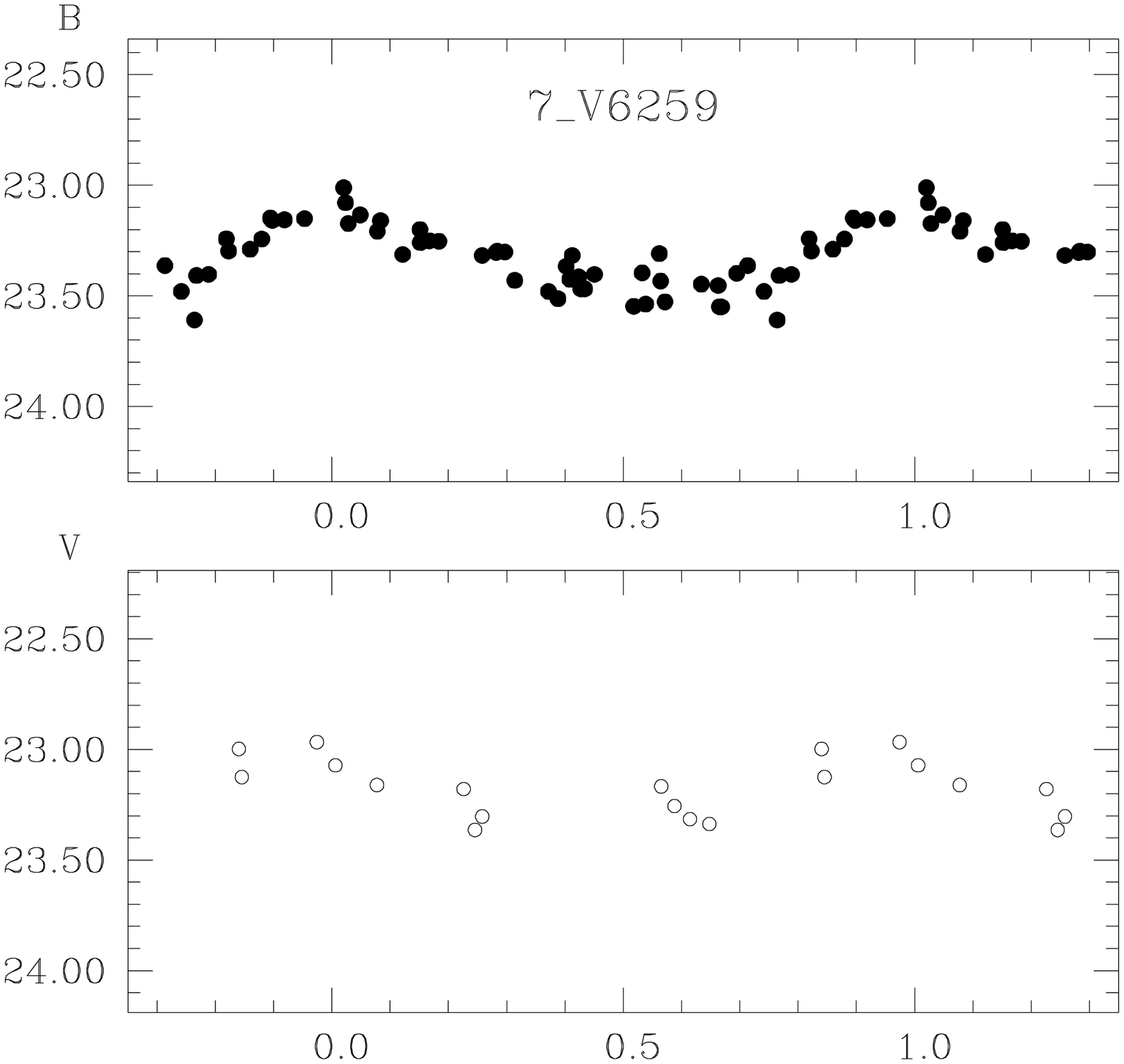}
\includegraphics[width=0.329\columnwidth,height=0.27\columnwidth]{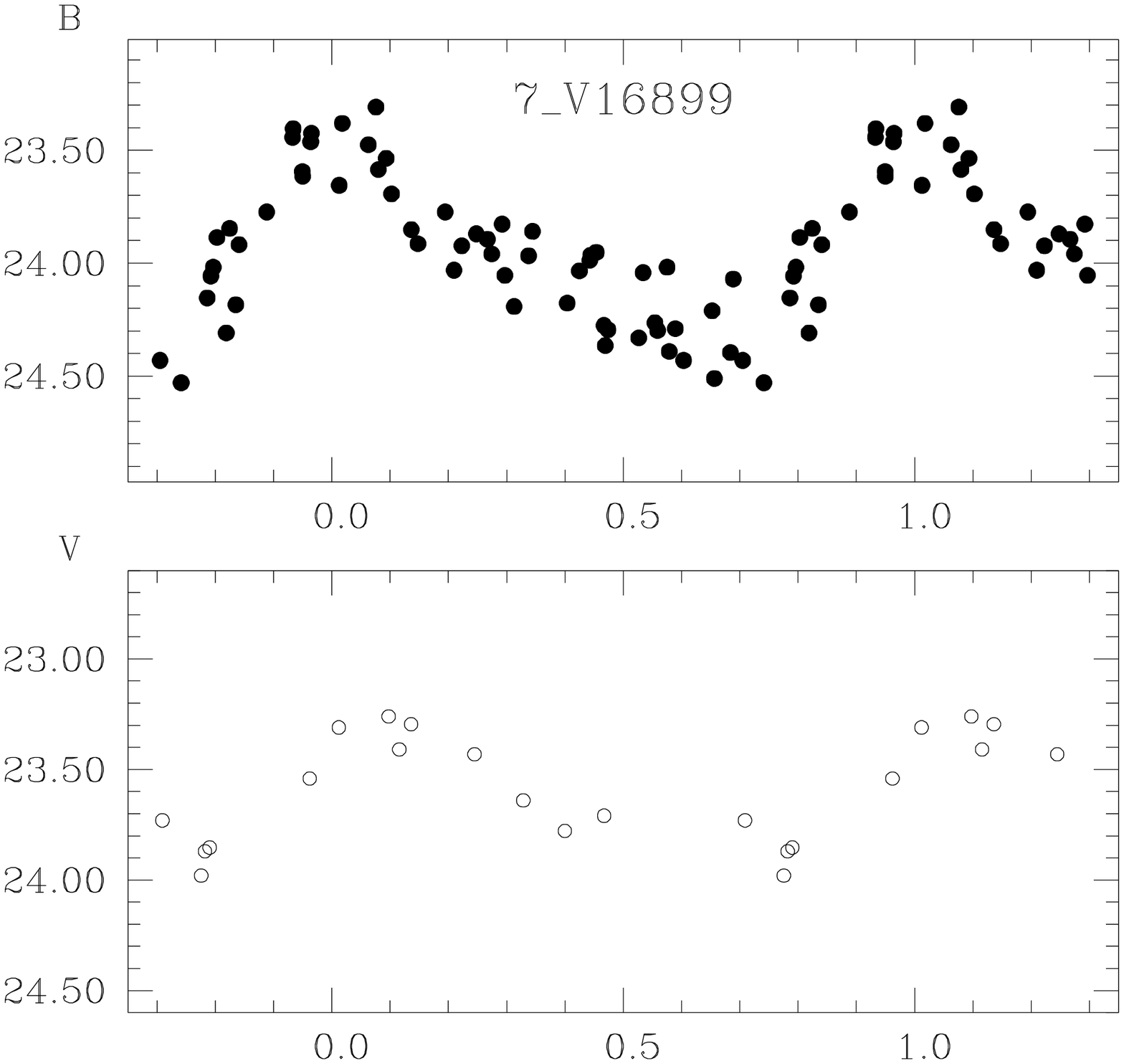}
\includegraphics[width=0.329\columnwidth,height=0.27\columnwidth]{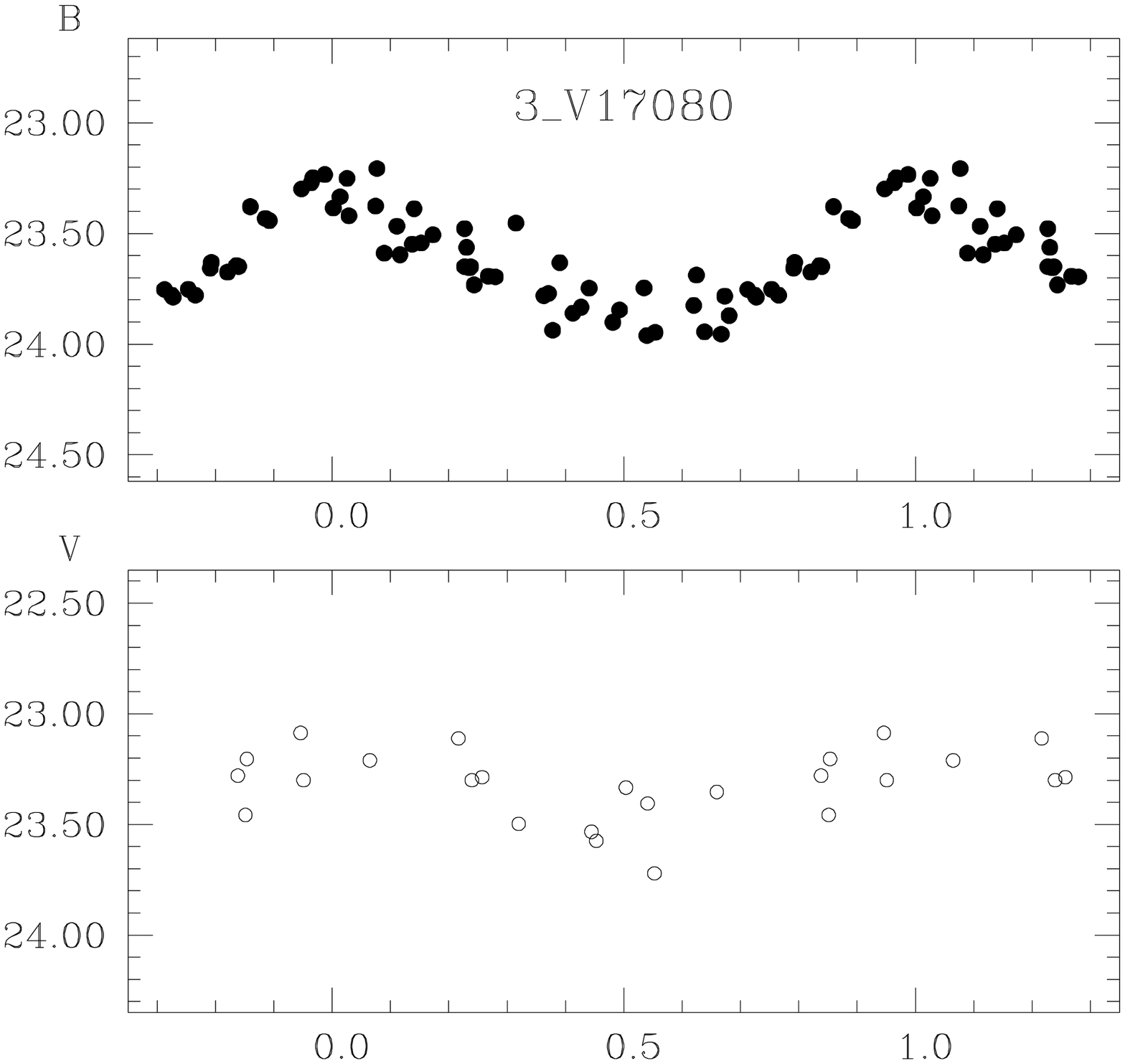}
\includegraphics[width=0.329\columnwidth,height=0.27\columnwidth]{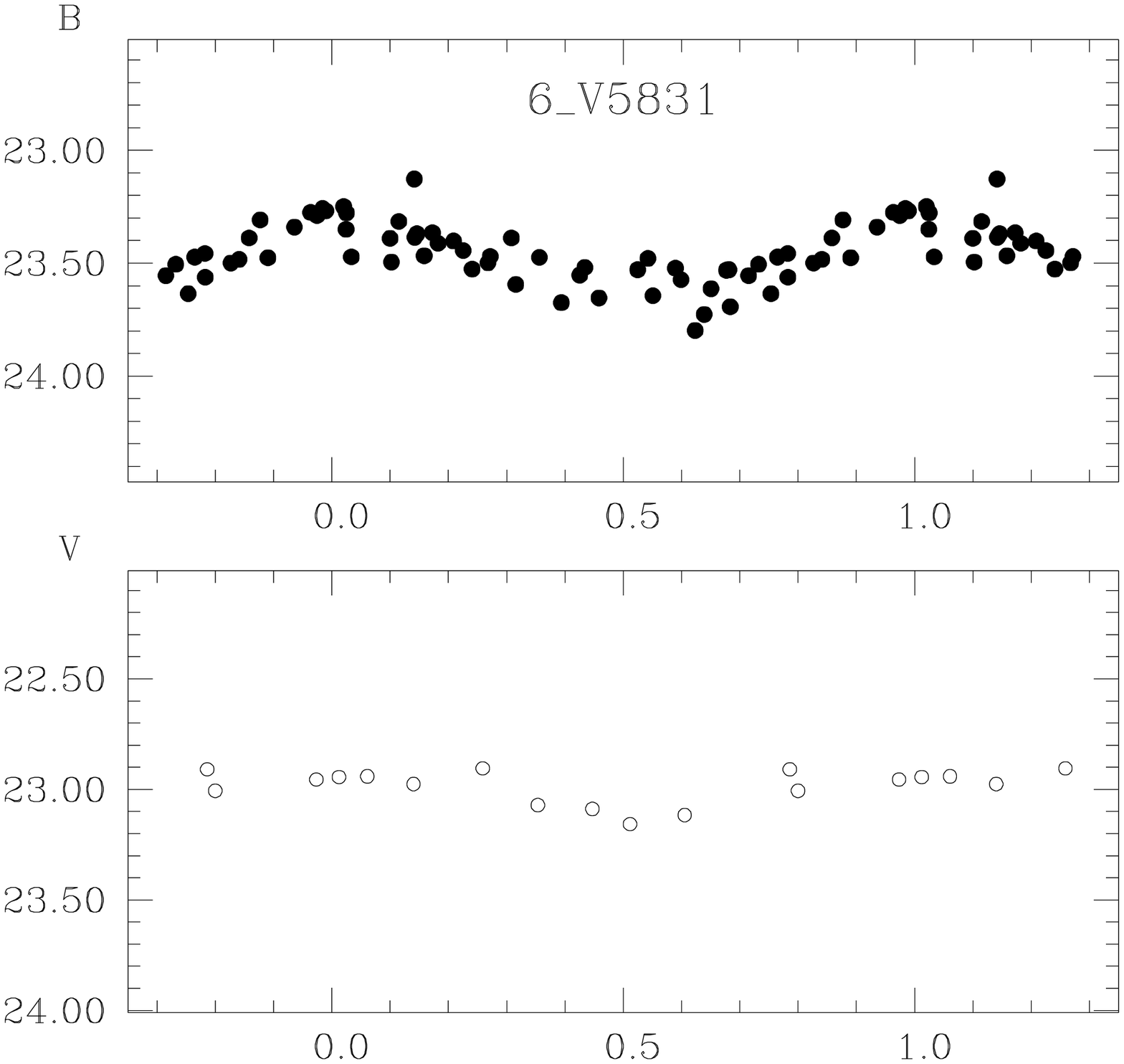}
\includegraphics[width=0.329\columnwidth,height=0.27\columnwidth]{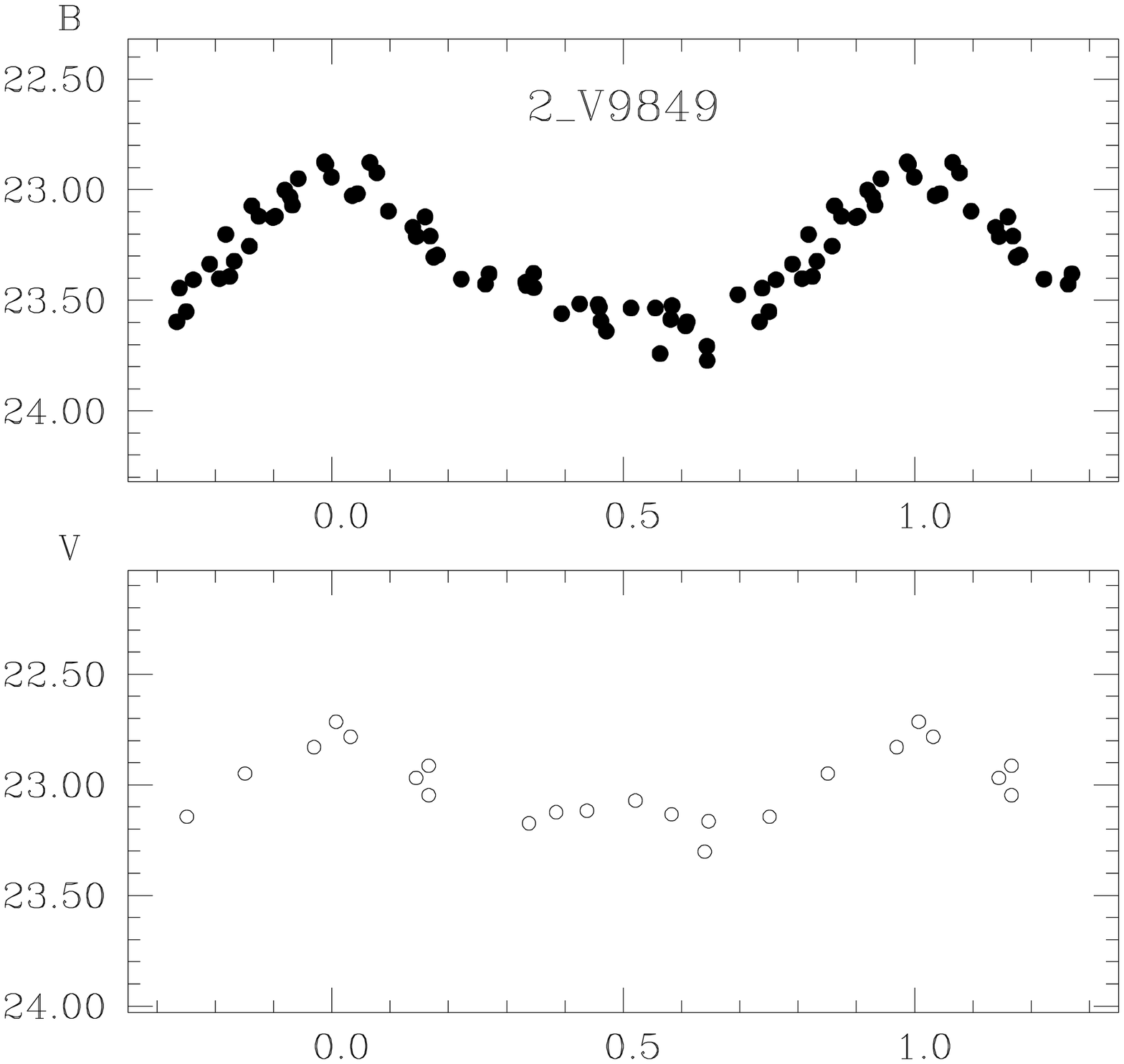}
\includegraphics[width=0.329\columnwidth,height=0.27\columnwidth]{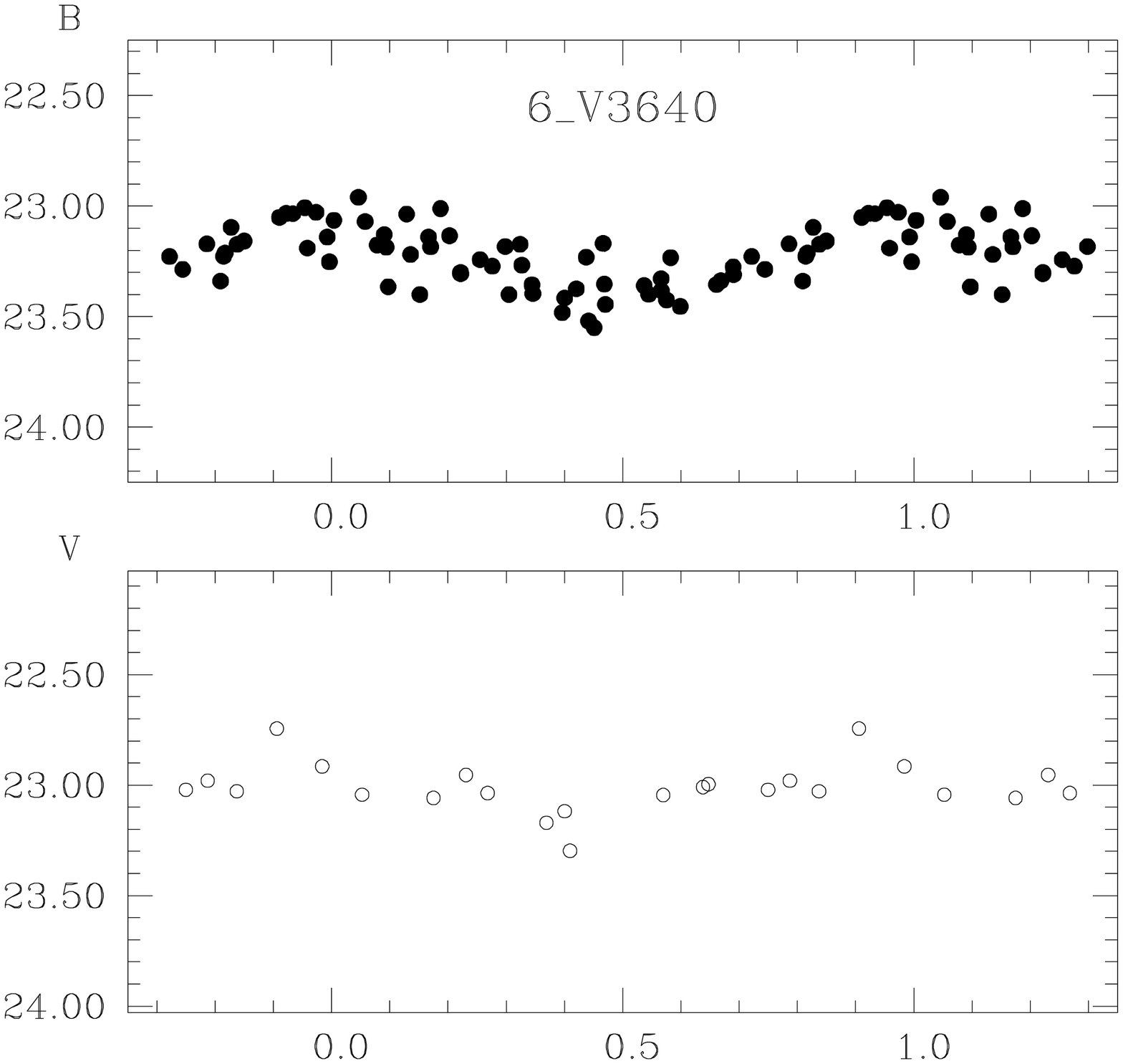}
\includegraphics[width=0.329\columnwidth,height=0.27\columnwidth]{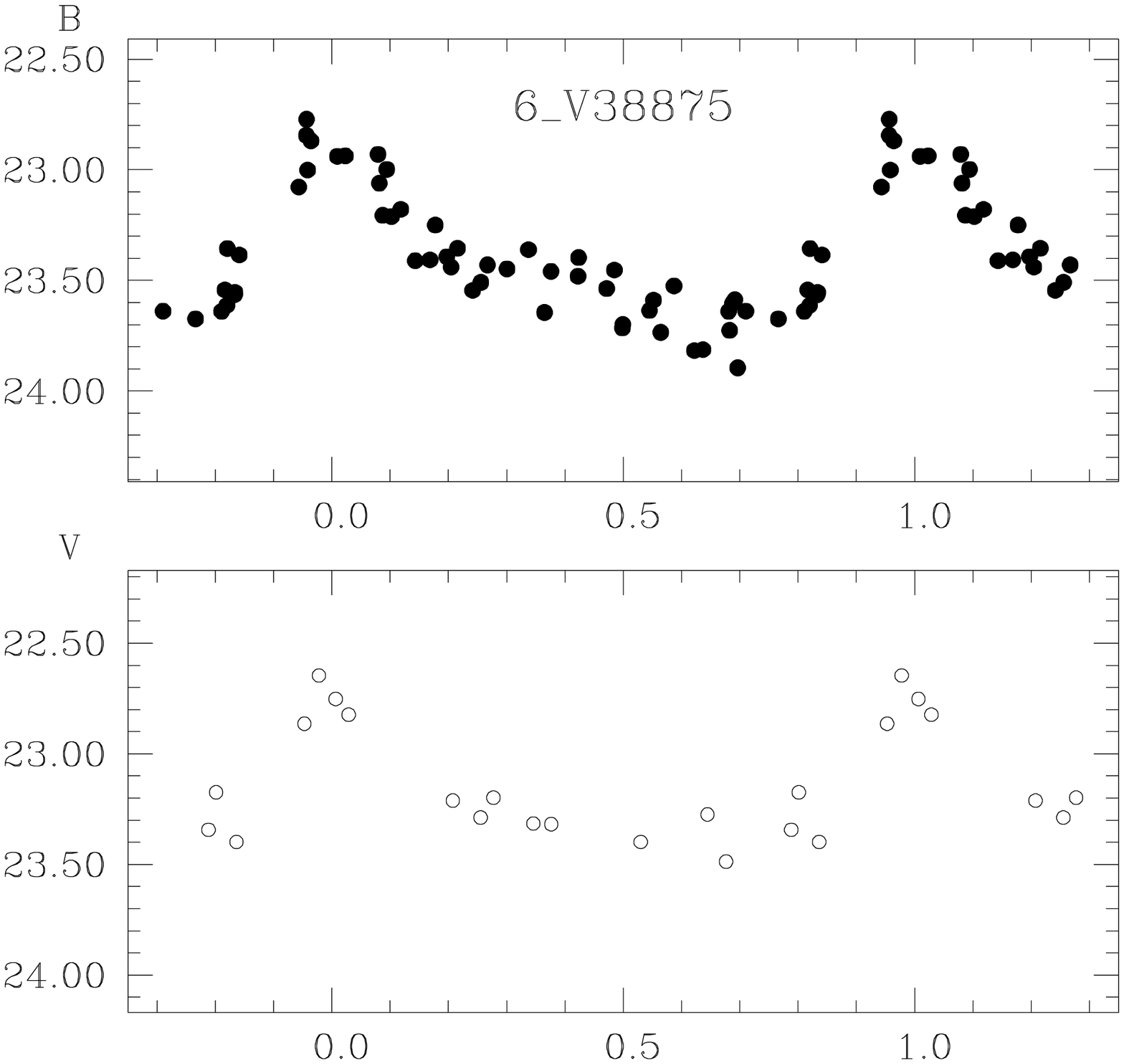}
\includegraphics[width=0.329\columnwidth,height=0.27\columnwidth]{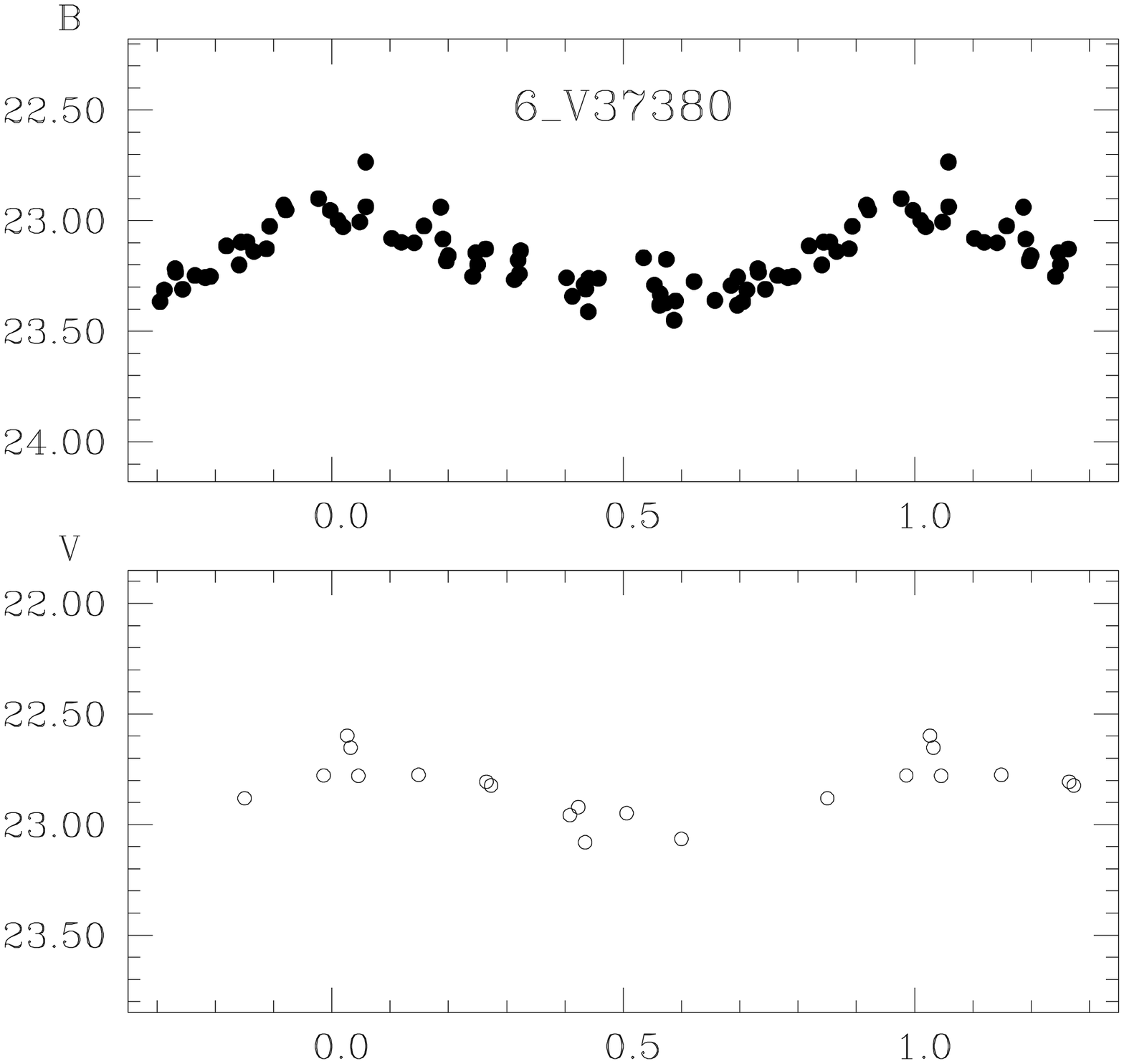}
\includegraphics[width=0.329\columnwidth,height=0.27\columnwidth]{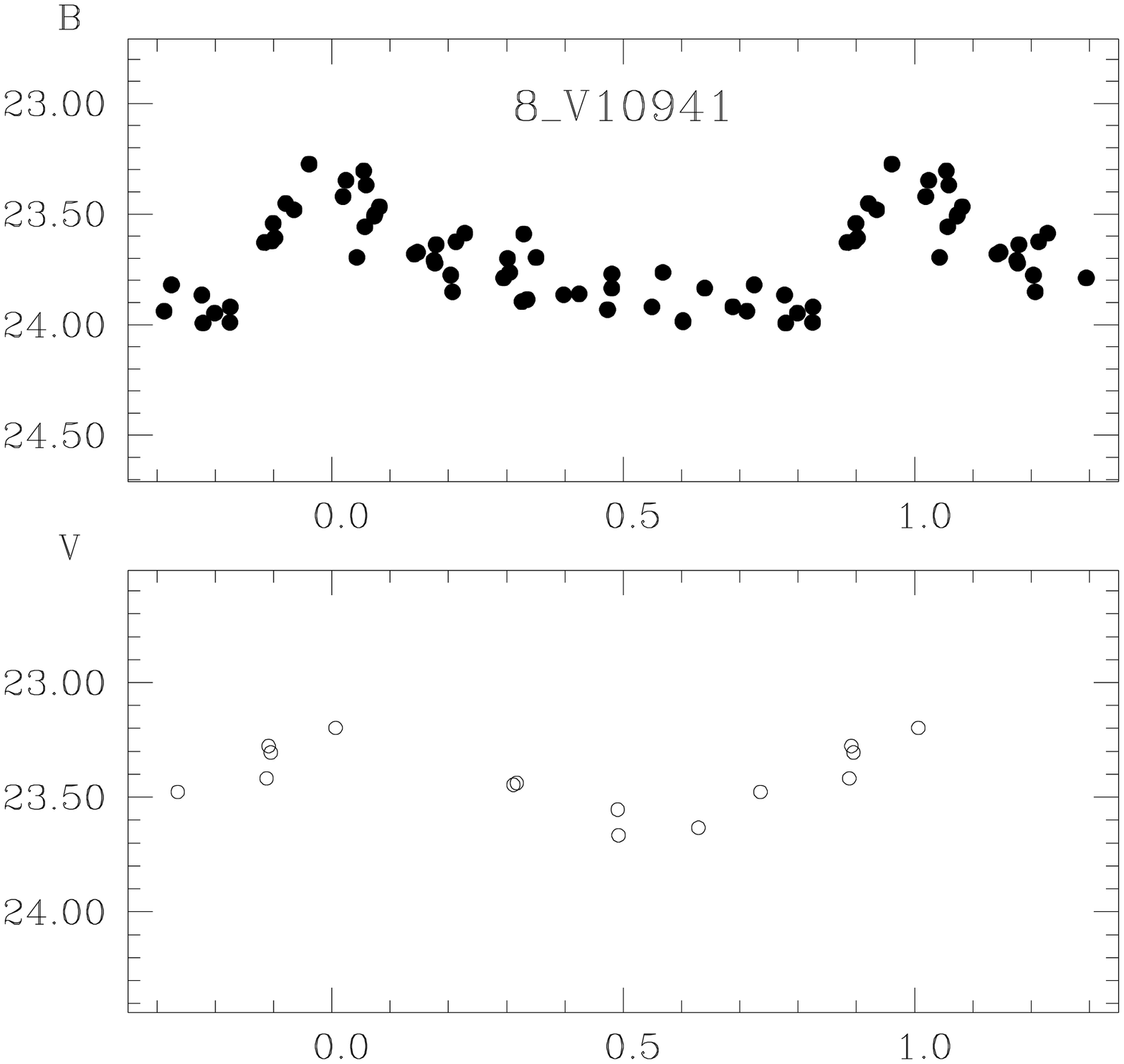}
\includegraphics[width=0.329\columnwidth,height=0.27\columnwidth]{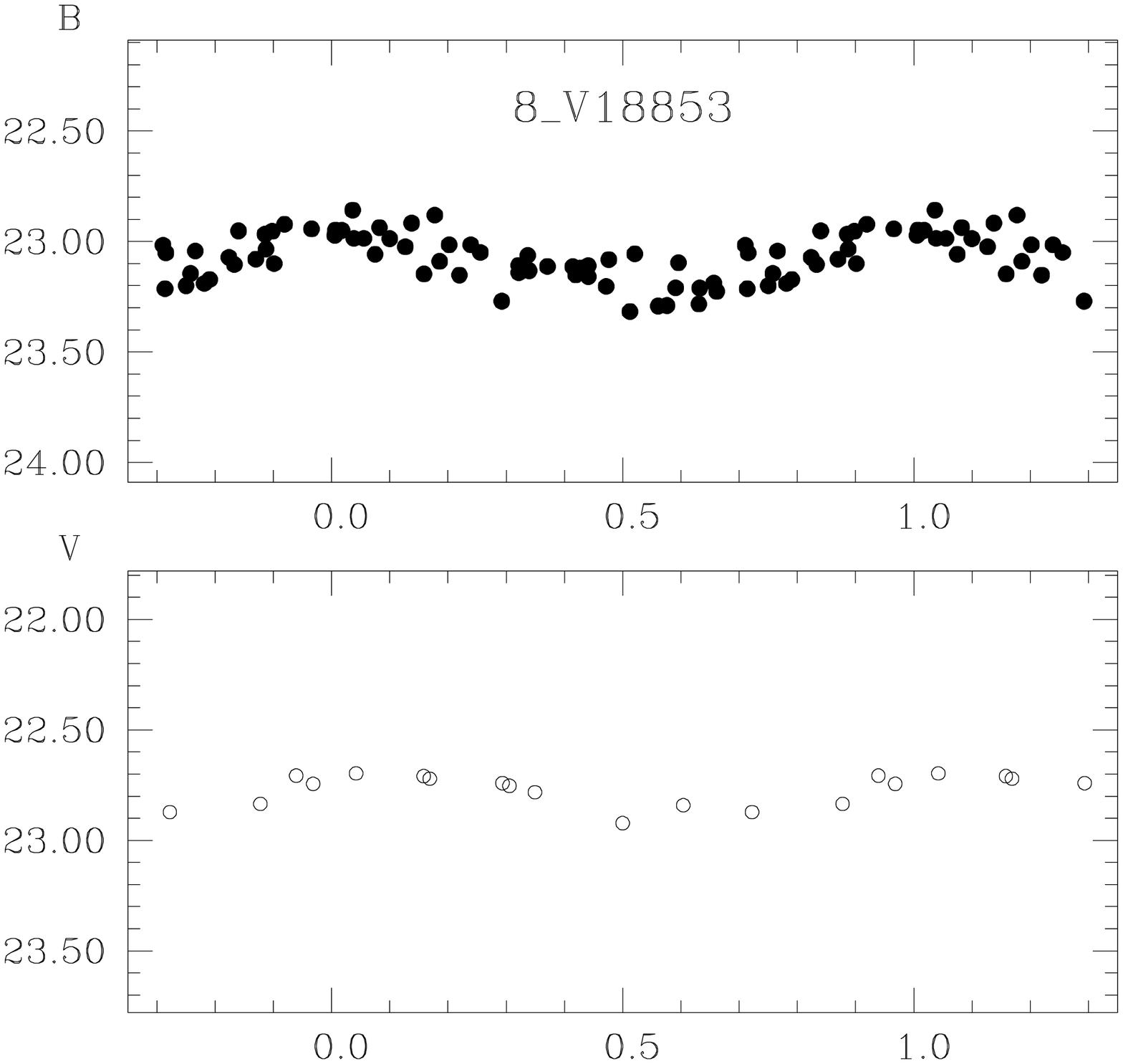}
\includegraphics[width=0.329\columnwidth,height=0.27\columnwidth]{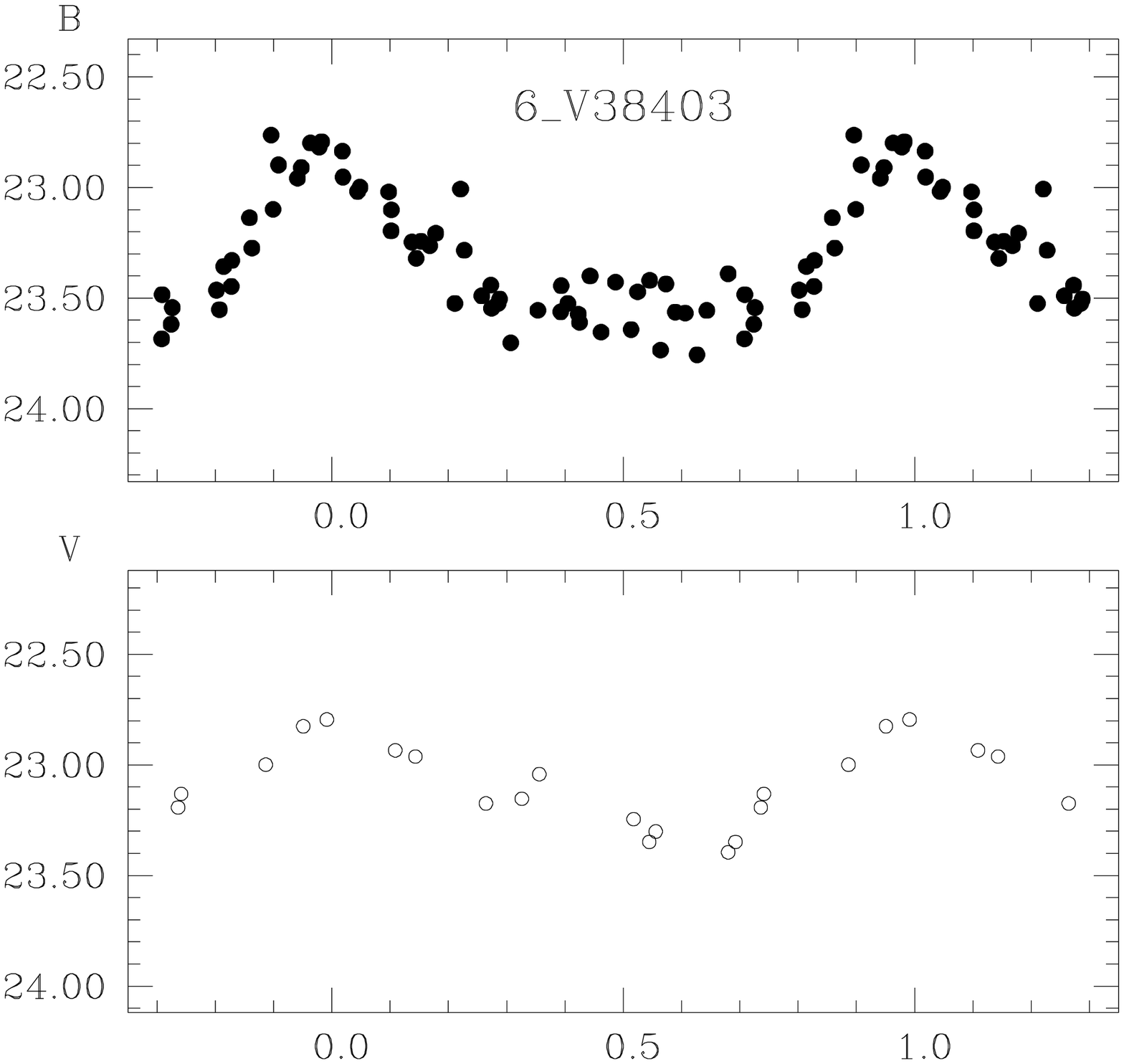}
\includegraphics[width=0.329\columnwidth,height=0.27\columnwidth]{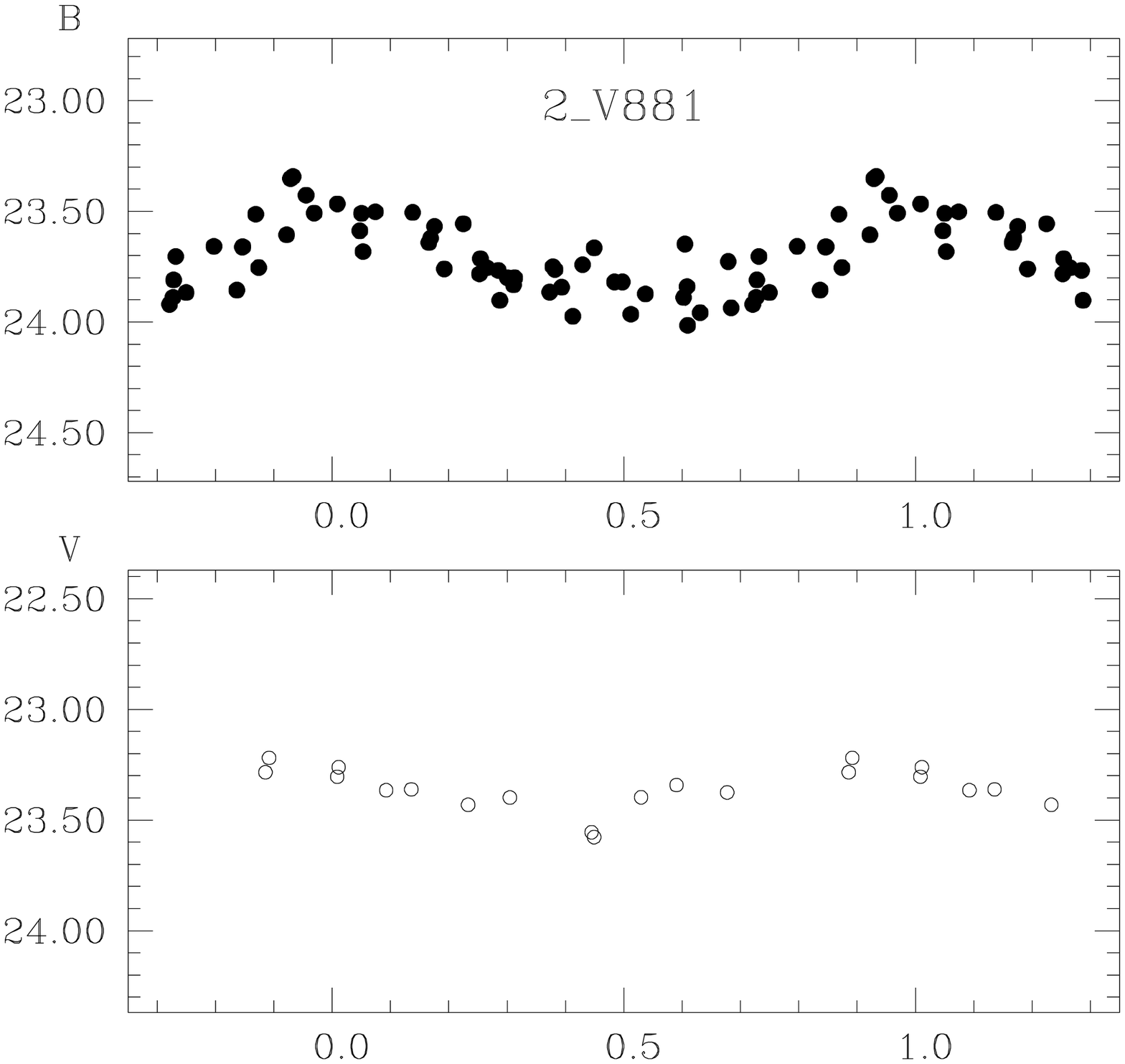}
\end{figure*}
\begin{figure*}
\includegraphics[width=0.329\columnwidth,height=0.27\columnwidth]{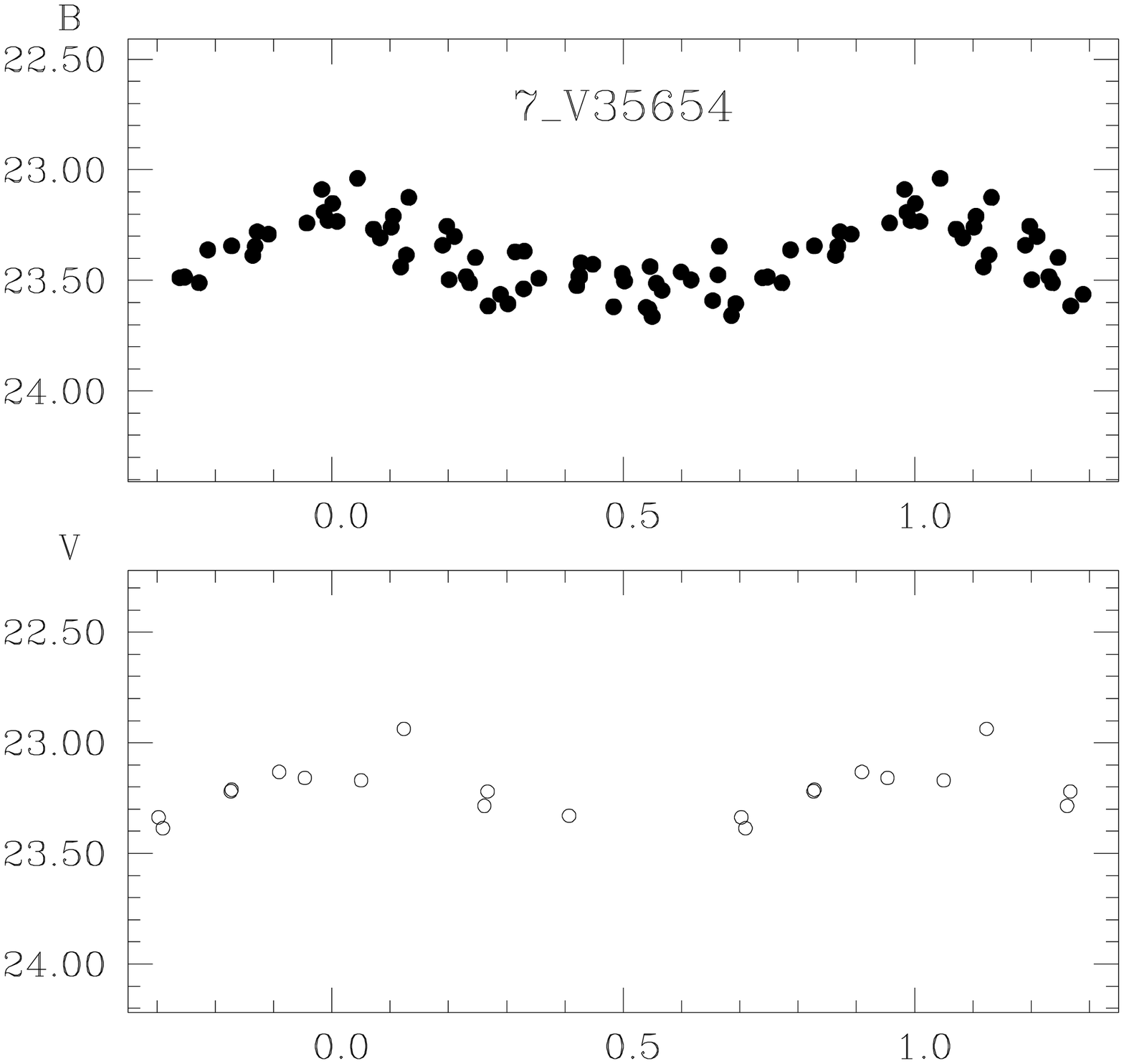}
\includegraphics[width=0.329\columnwidth,height=0.27\columnwidth]{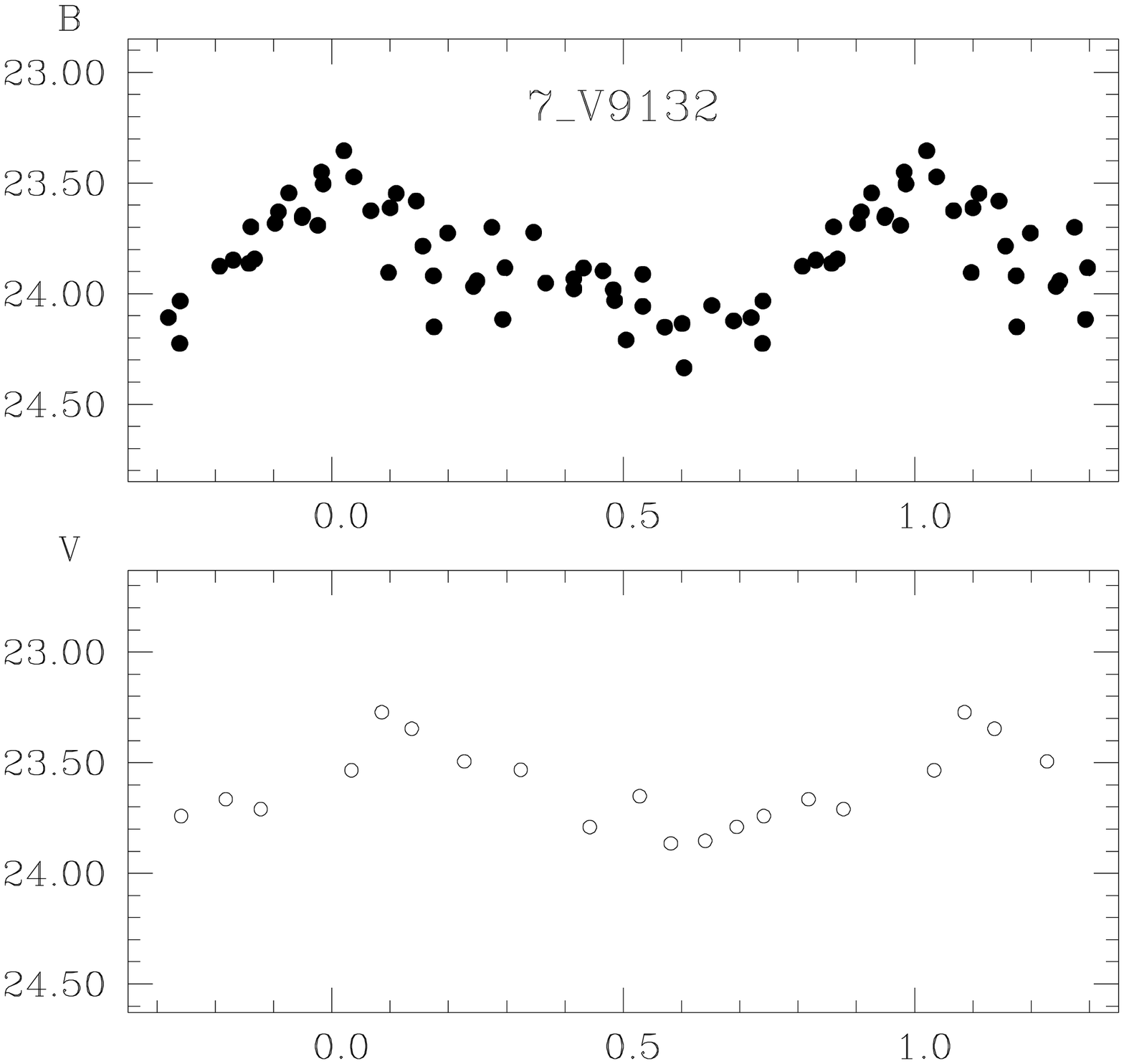}
\includegraphics[width=0.329\columnwidth,height=0.27\columnwidth]{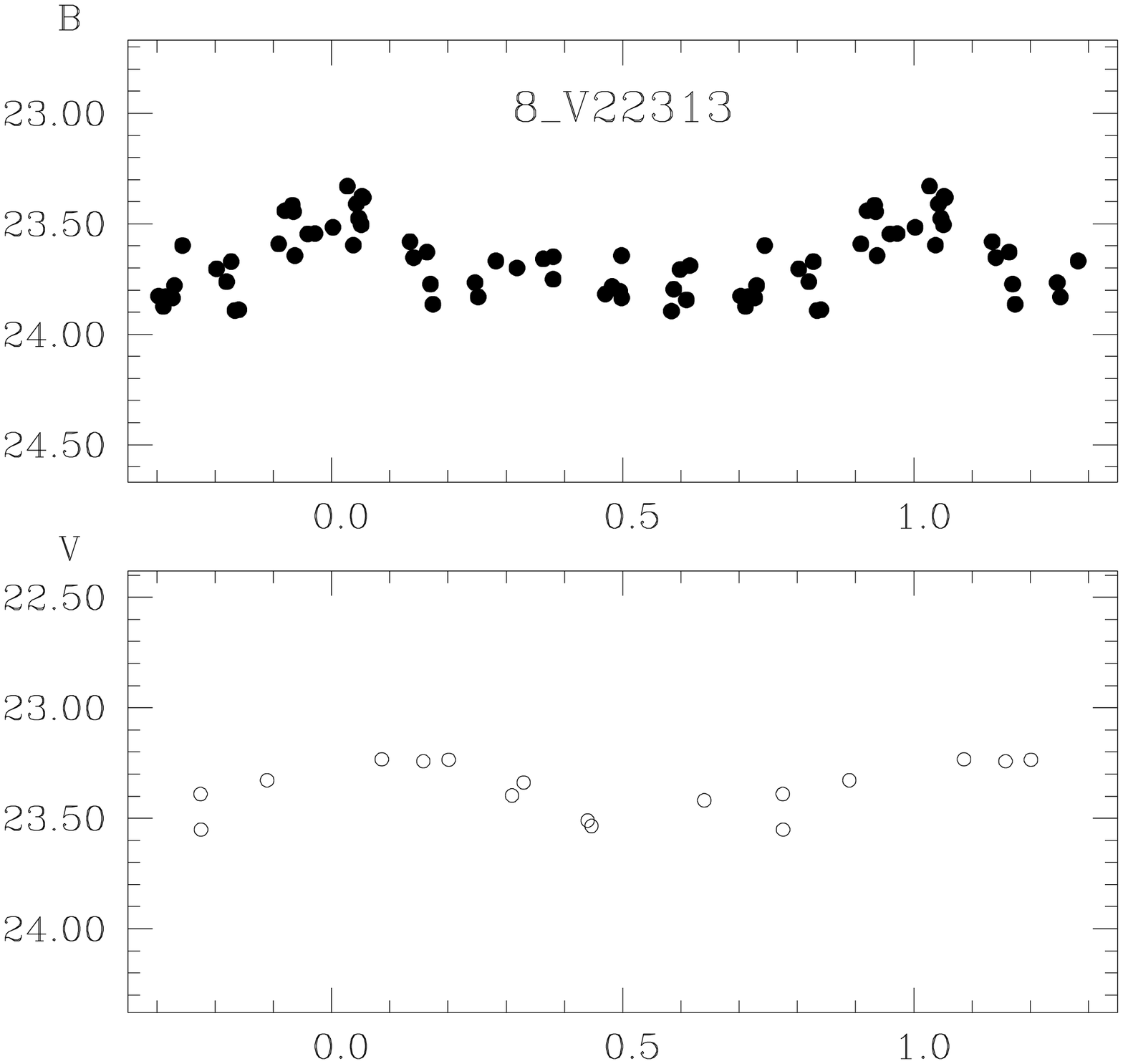}
\includegraphics[width=0.329\columnwidth,height=0.27\columnwidth]{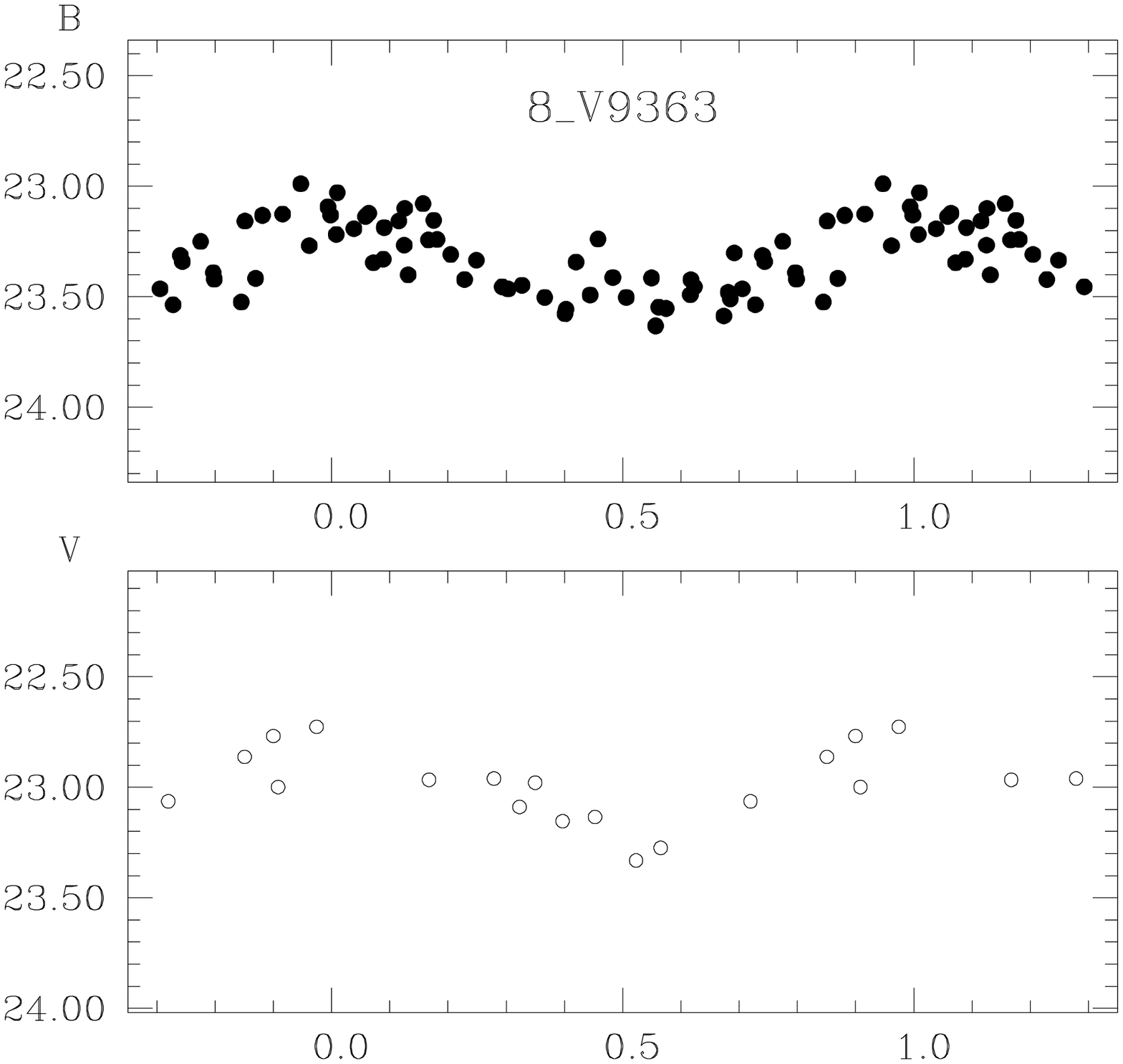}
\includegraphics[width=0.329\columnwidth,height=0.27\columnwidth]{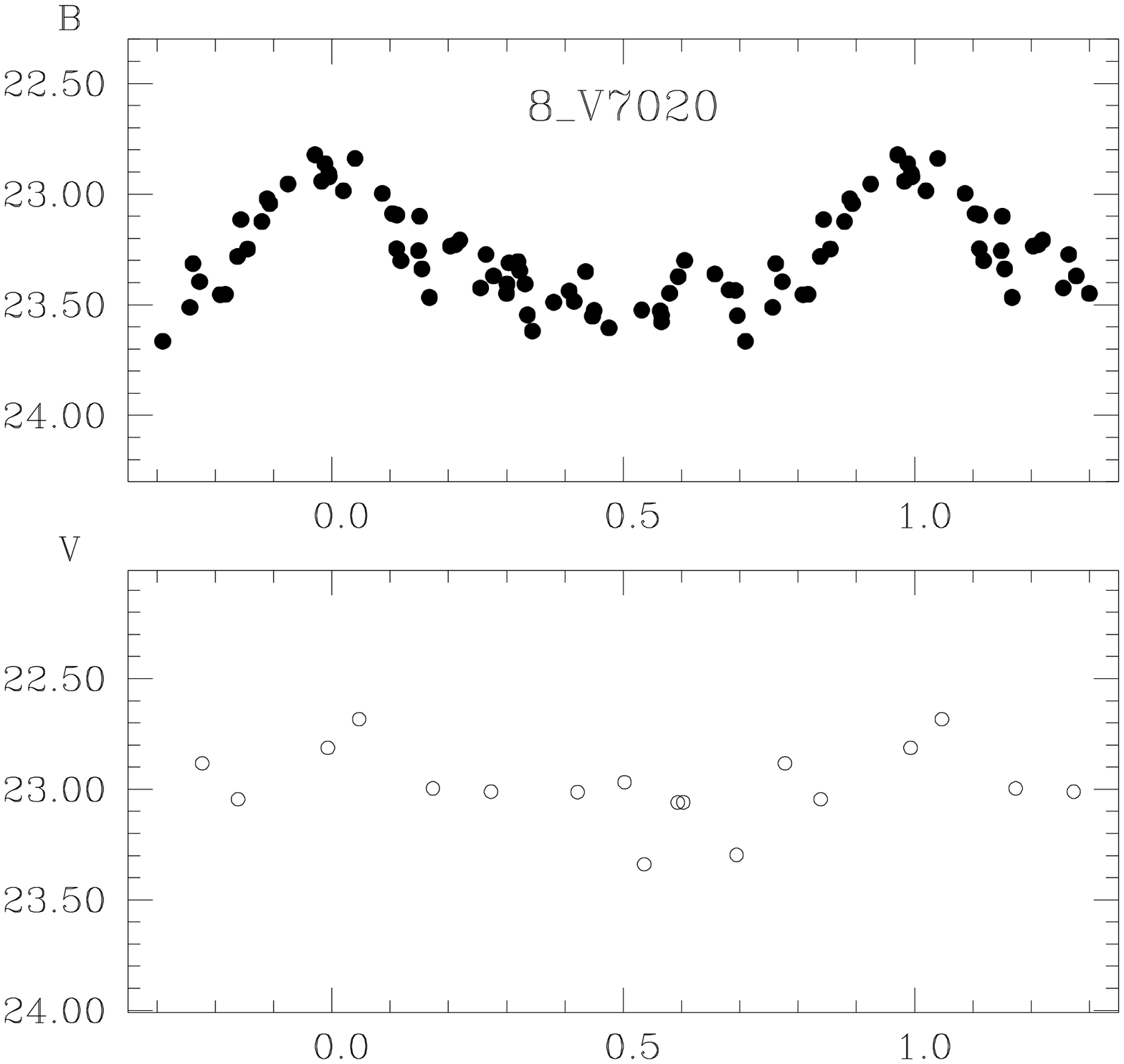}
\includegraphics[width=0.329\columnwidth,height=0.27\columnwidth]{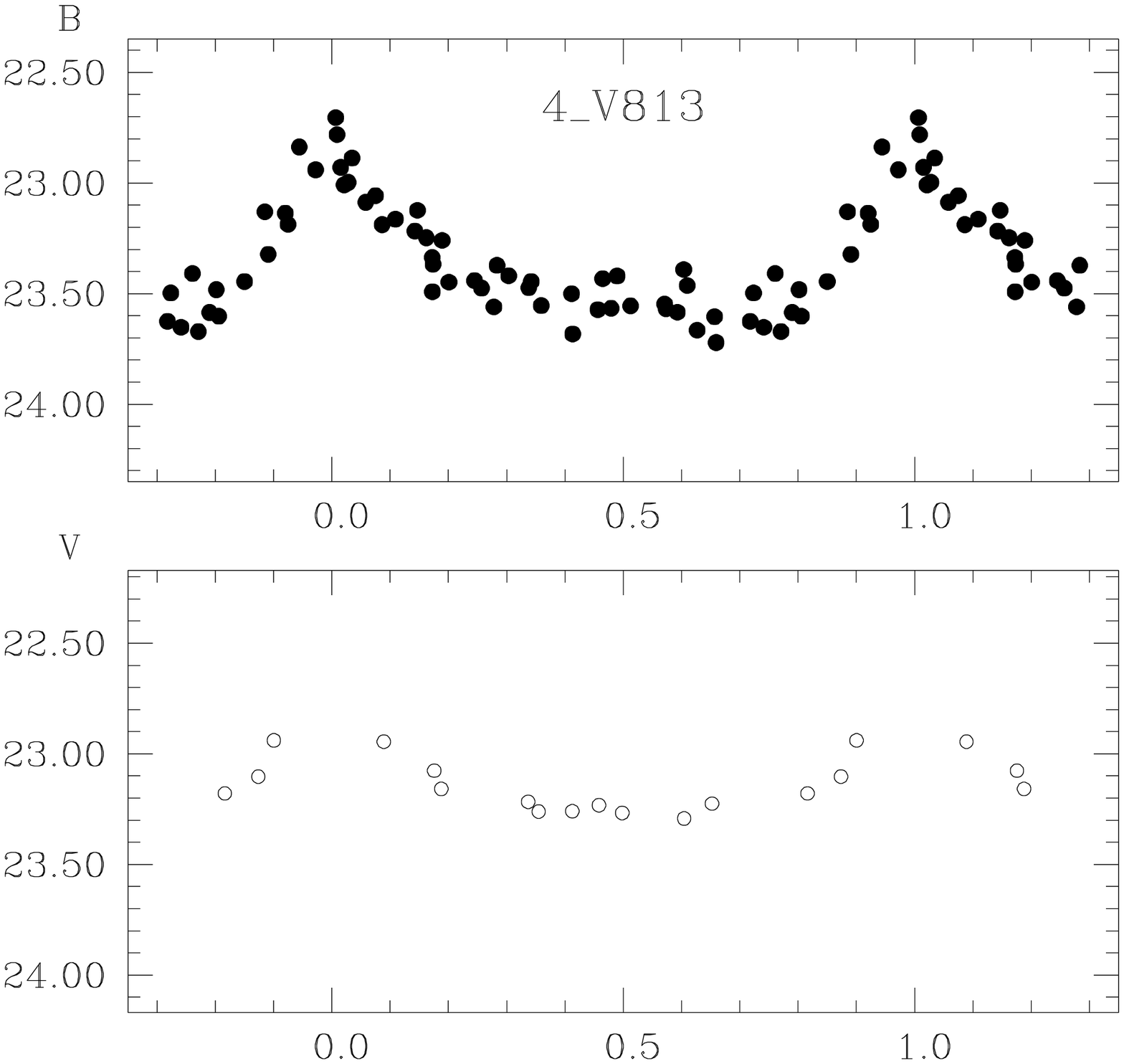}
\includegraphics[width=0.329\columnwidth,height=0.27\columnwidth]{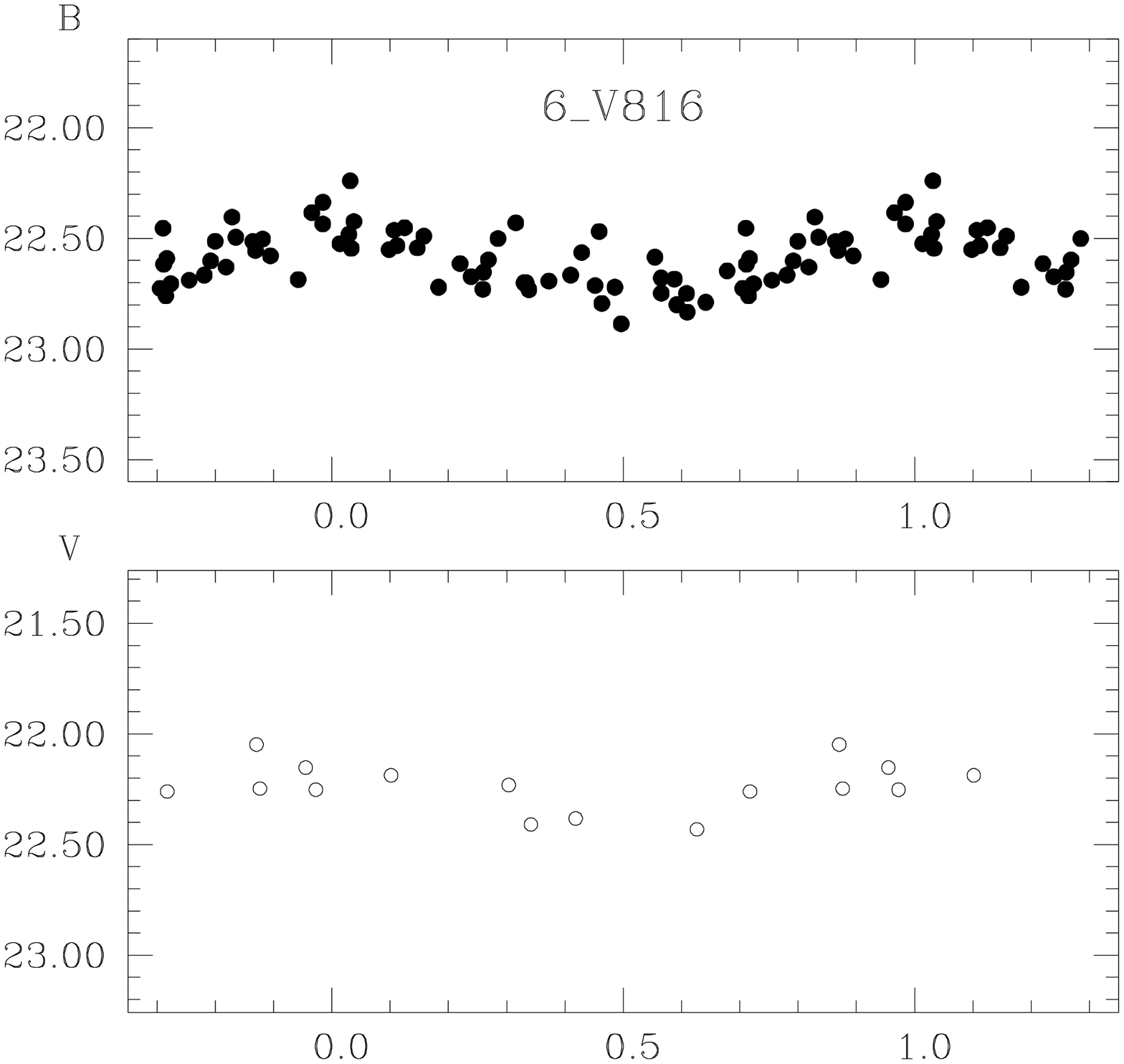}
\includegraphics[width=0.329\columnwidth,height=0.27\columnwidth]{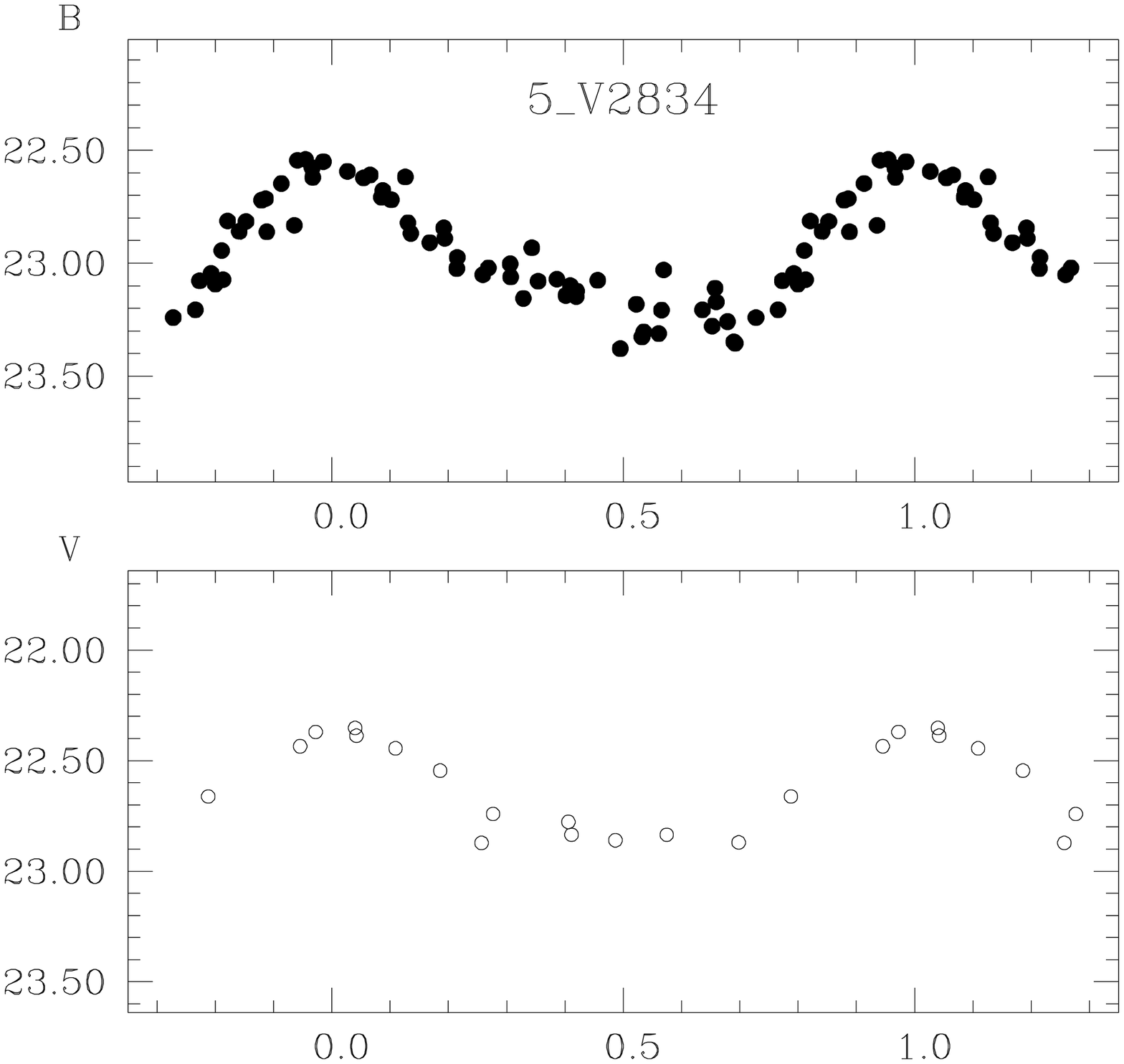}
\includegraphics[width=0.329\columnwidth,height=0.27\columnwidth]{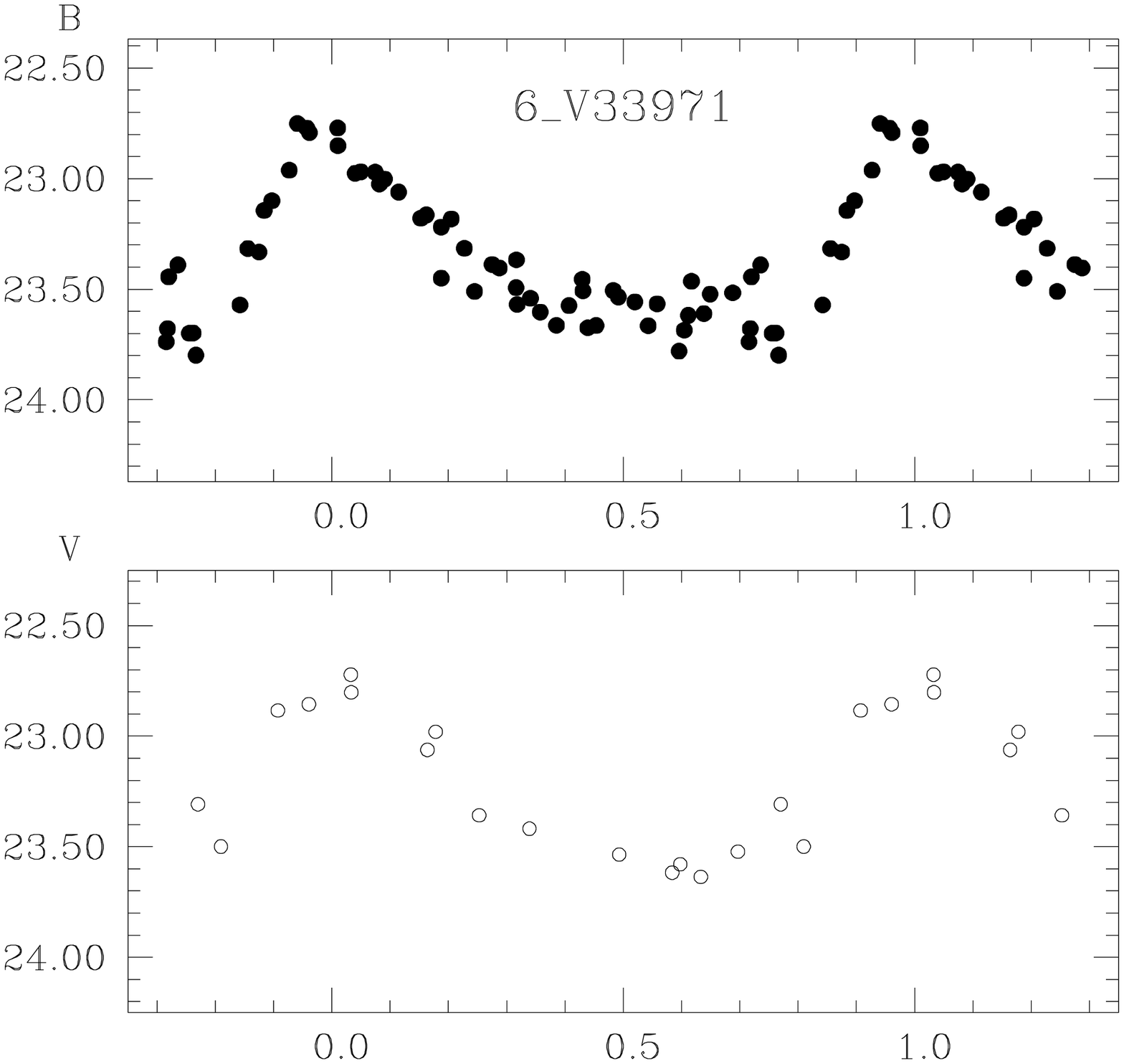}
\includegraphics[width=0.329\columnwidth,height=0.27\columnwidth]{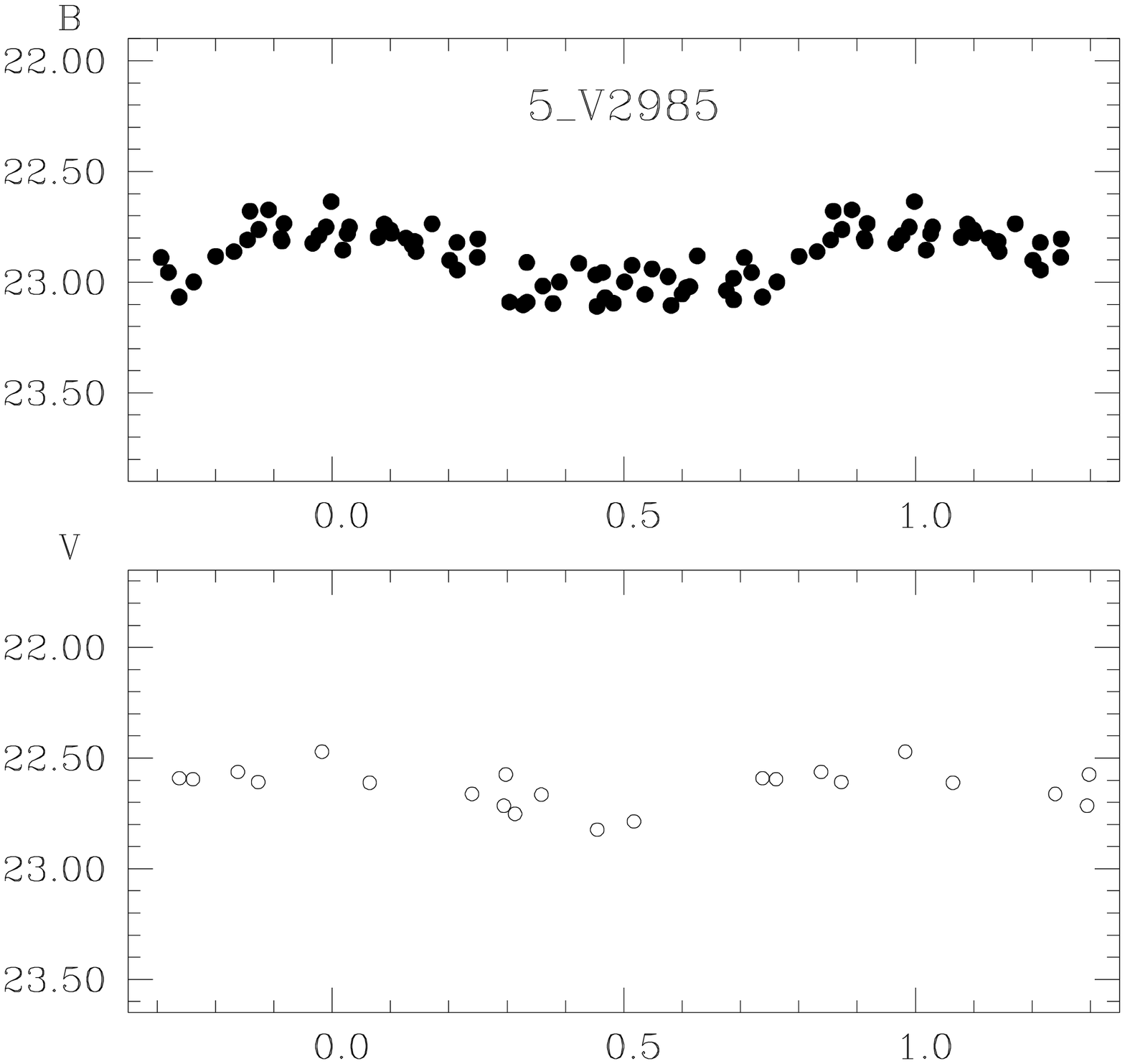}
\includegraphics[width=0.329\columnwidth,height=0.27\columnwidth]{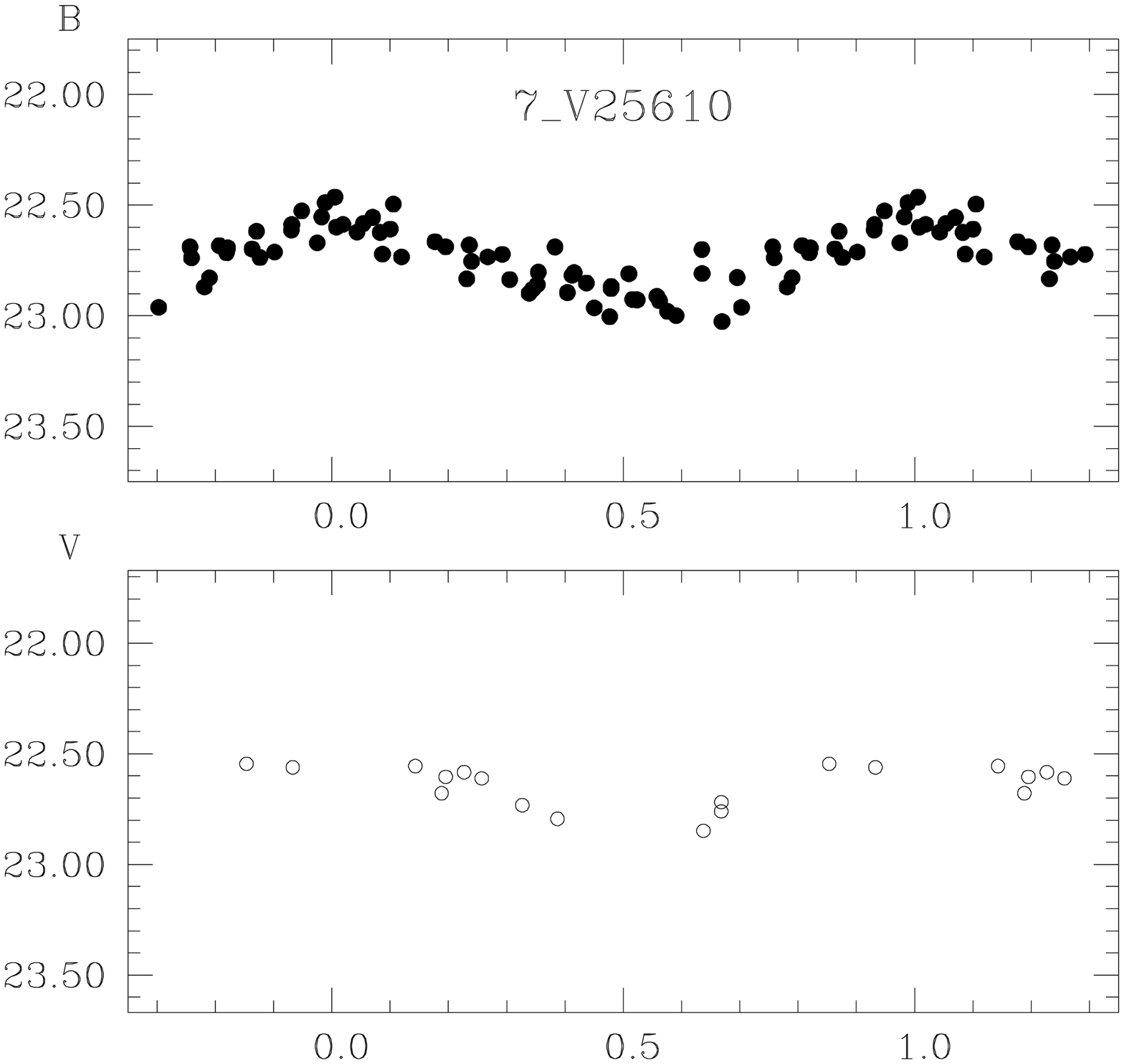}
\includegraphics[width=0.329\columnwidth,height=0.27\columnwidth]{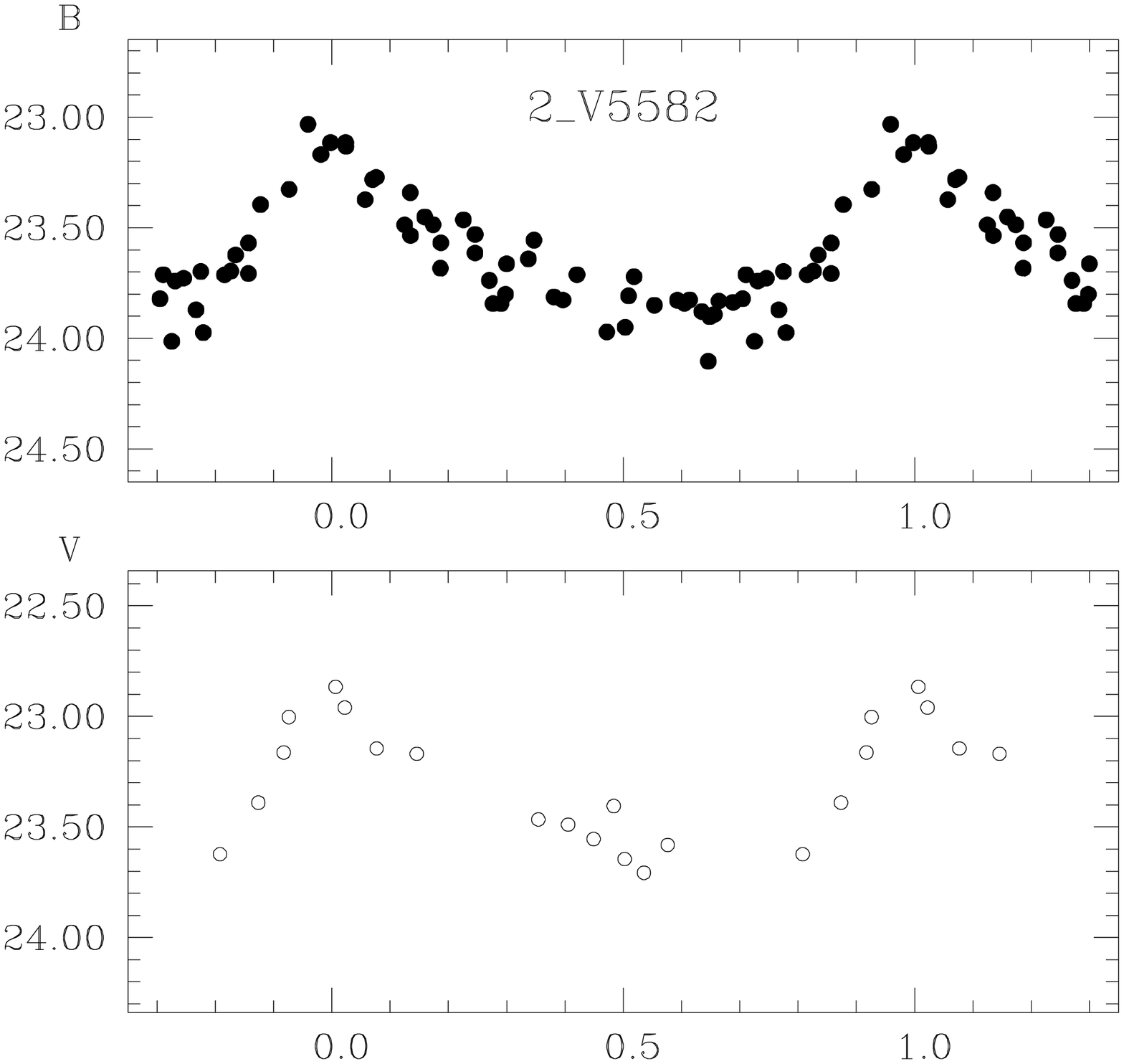}
\includegraphics[width=0.329\columnwidth,height=0.27\columnwidth]{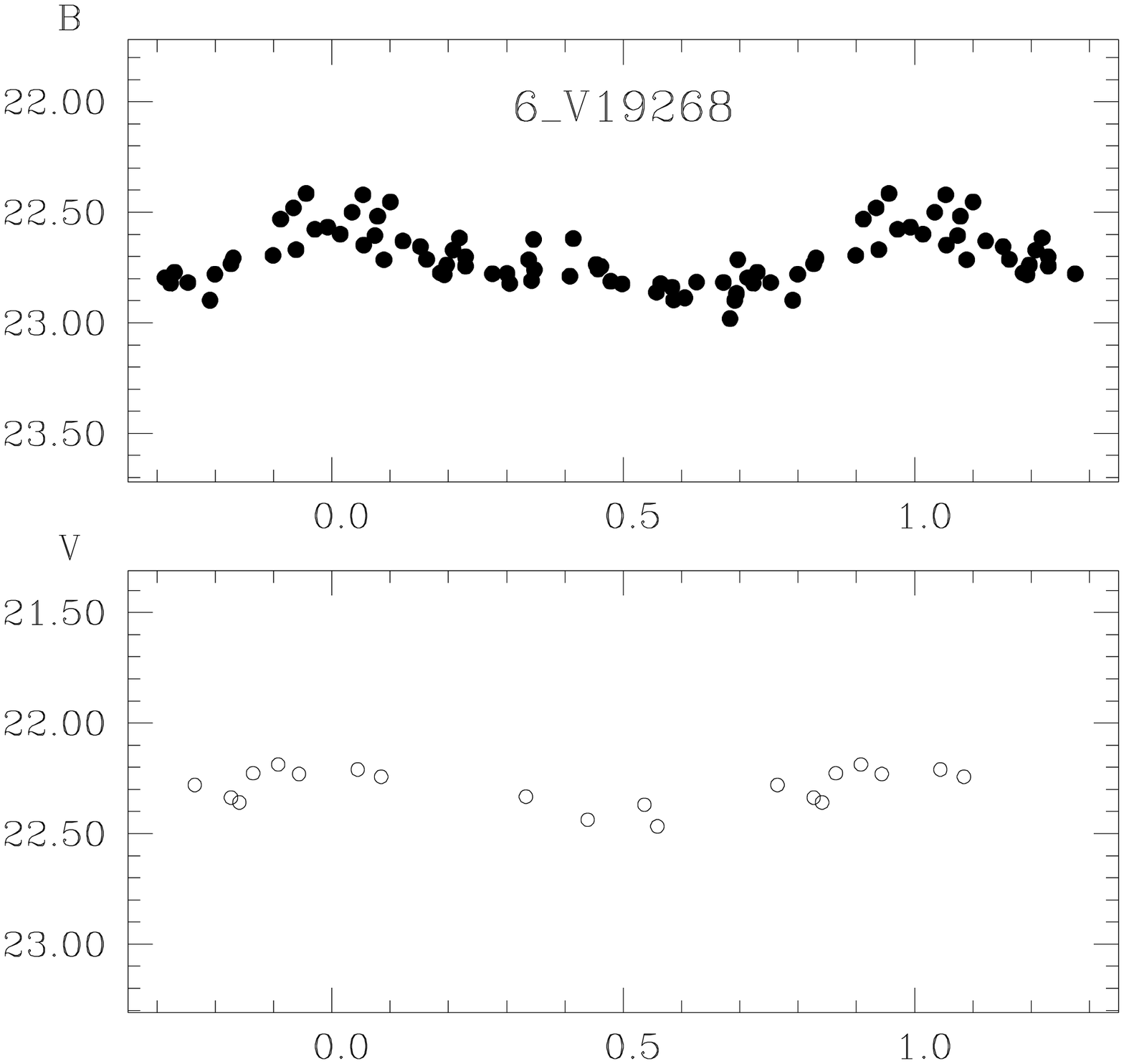}
\includegraphics[width=0.329\columnwidth,height=0.27\columnwidth]{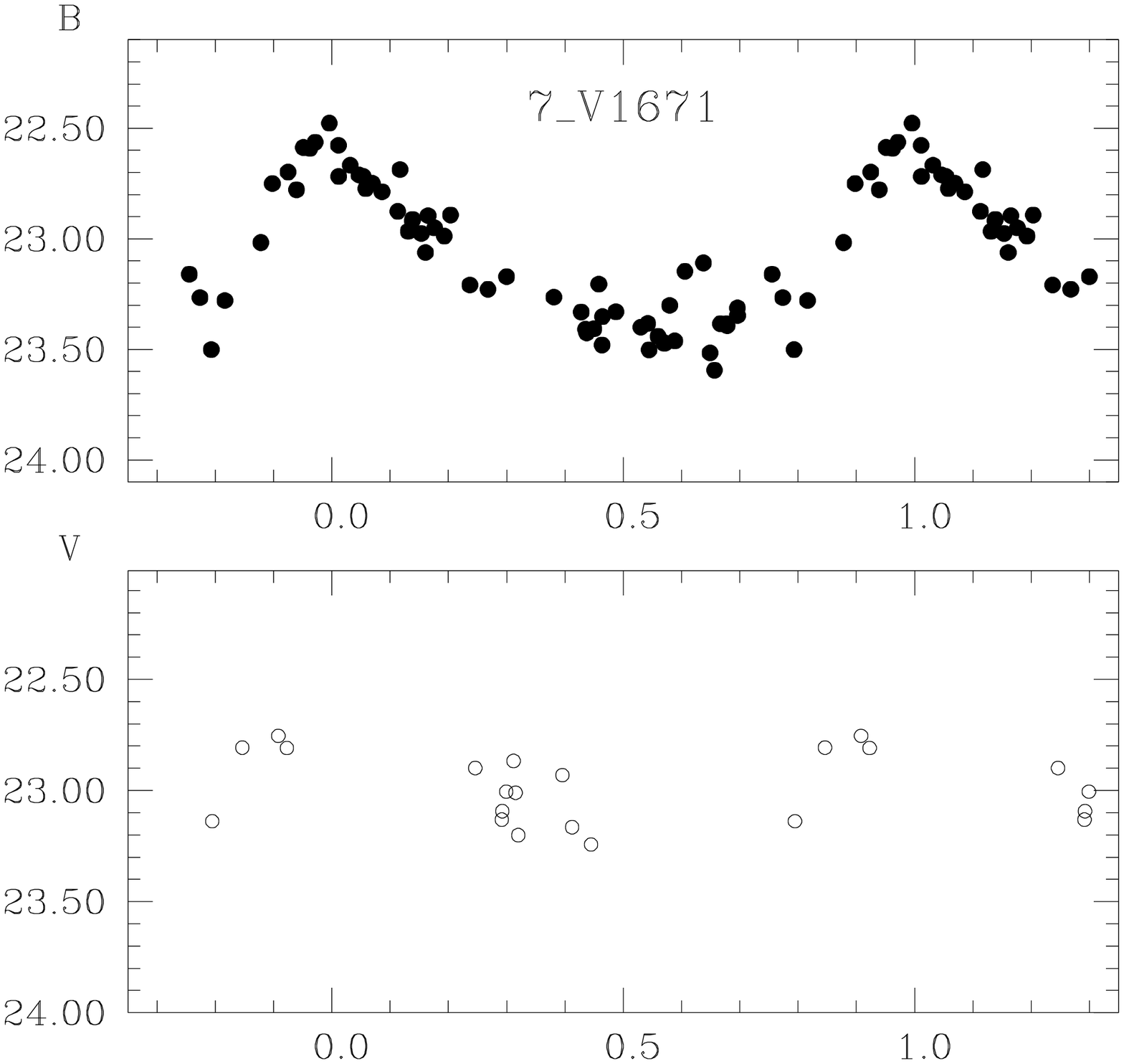}
\includegraphics[width=0.329\columnwidth,height=0.27\columnwidth]{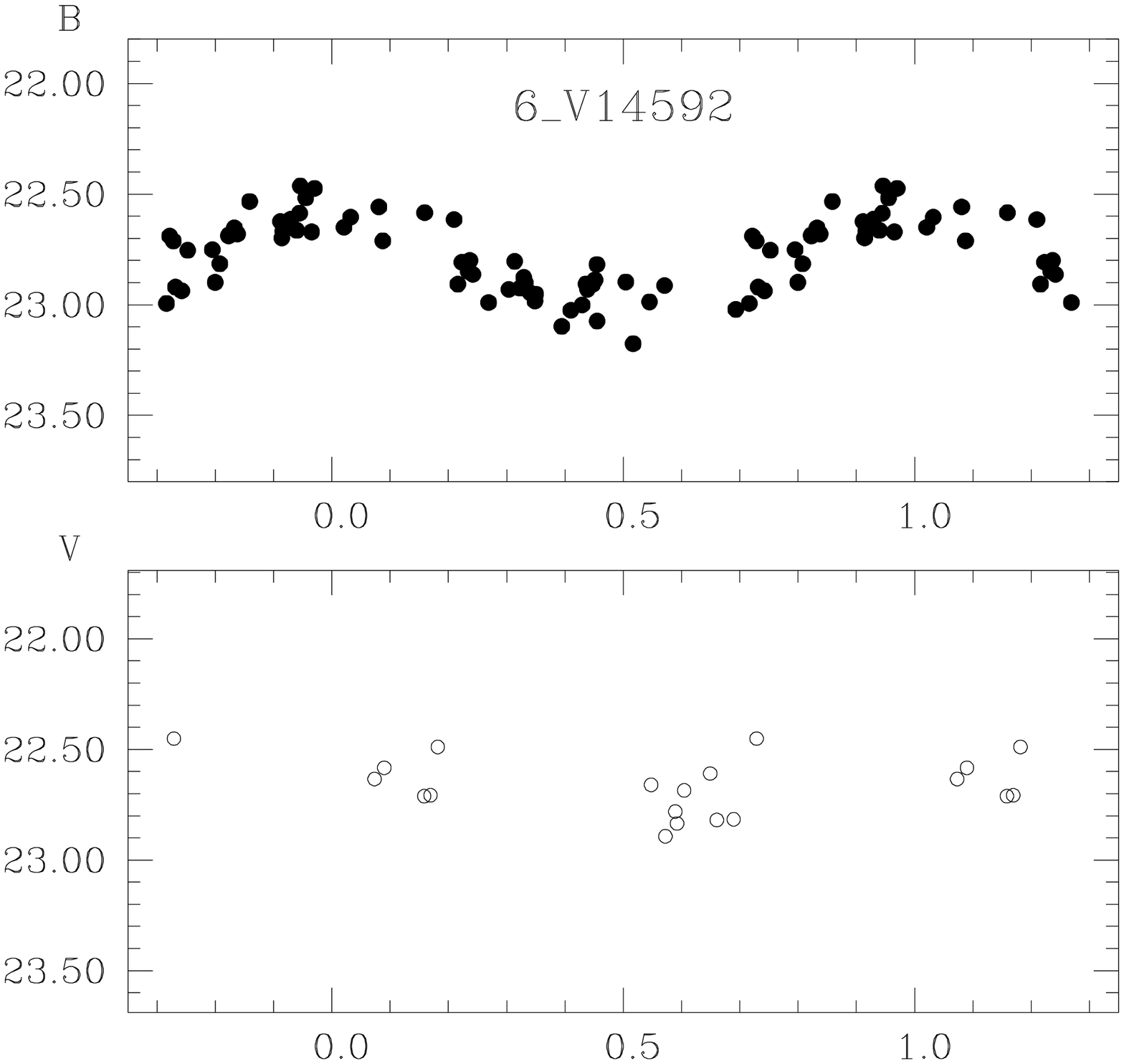}
\end{figure*}
\begin{figure*}
\includegraphics[width=0.329\columnwidth,height=0.27\columnwidth]{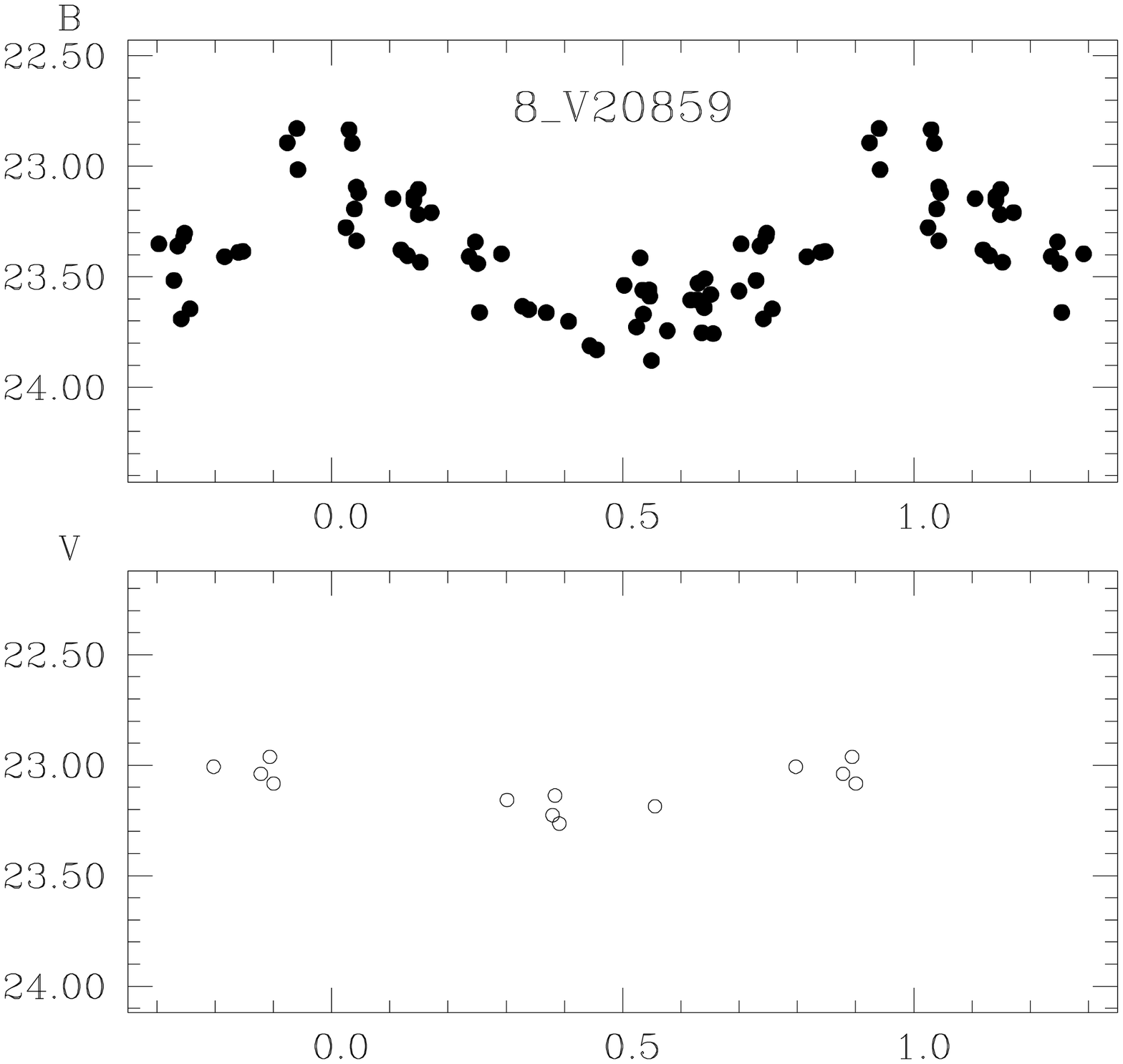}
\includegraphics[width=0.329\columnwidth,height=0.27\columnwidth]{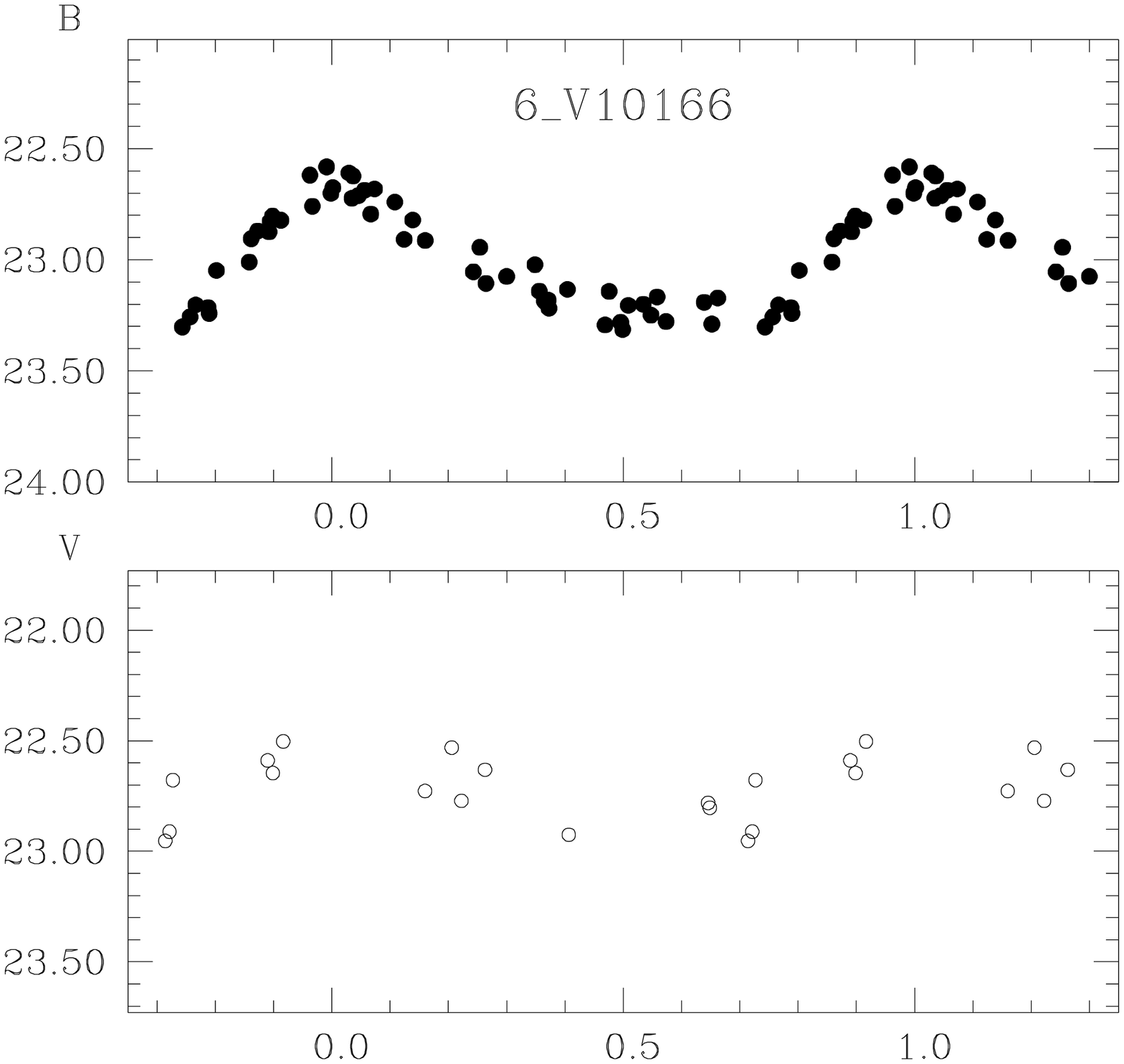}
\includegraphics[width=0.329\columnwidth,height=0.27\columnwidth]{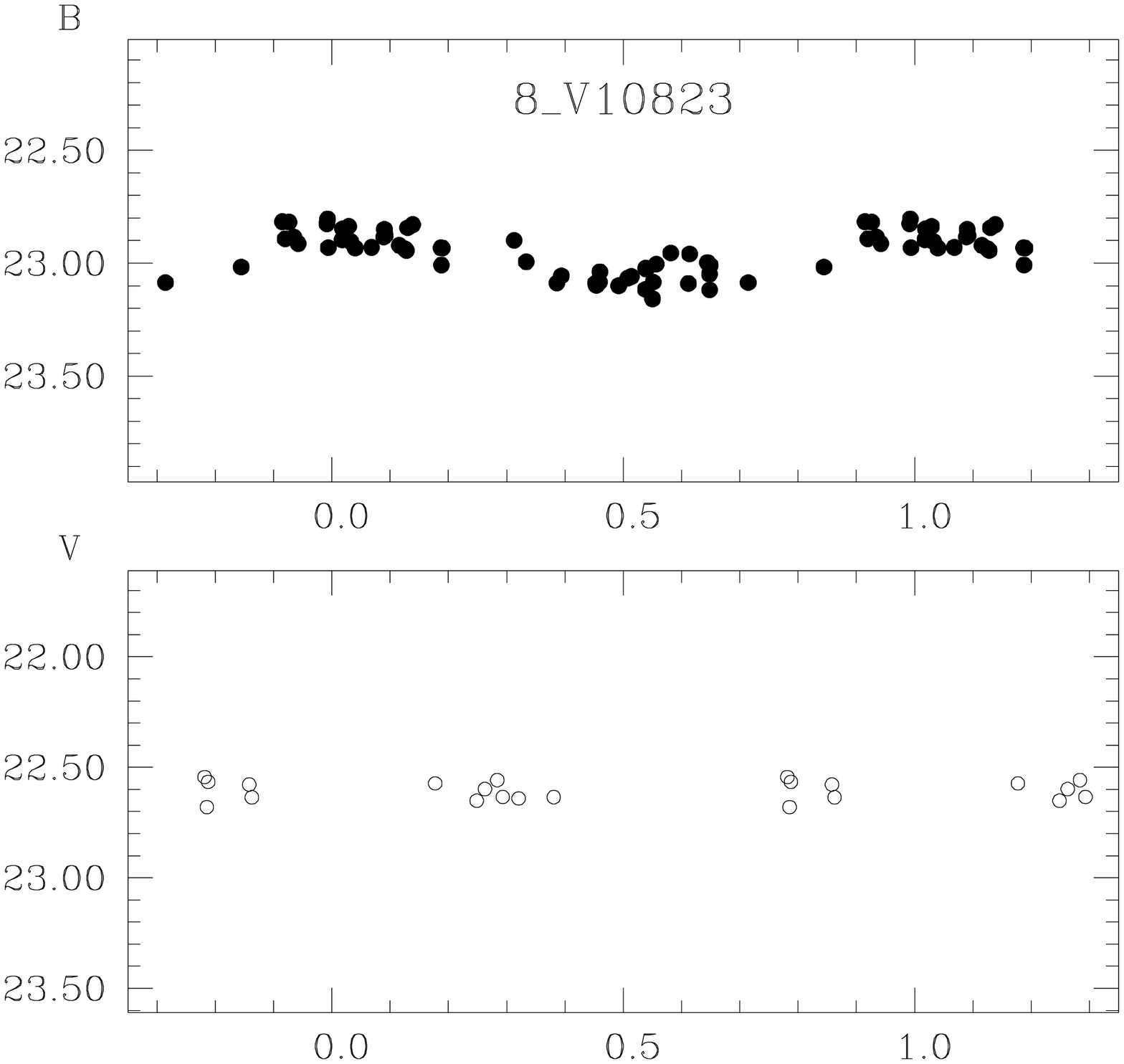}
\includegraphics[width=0.329\columnwidth,height=0.27\columnwidth]{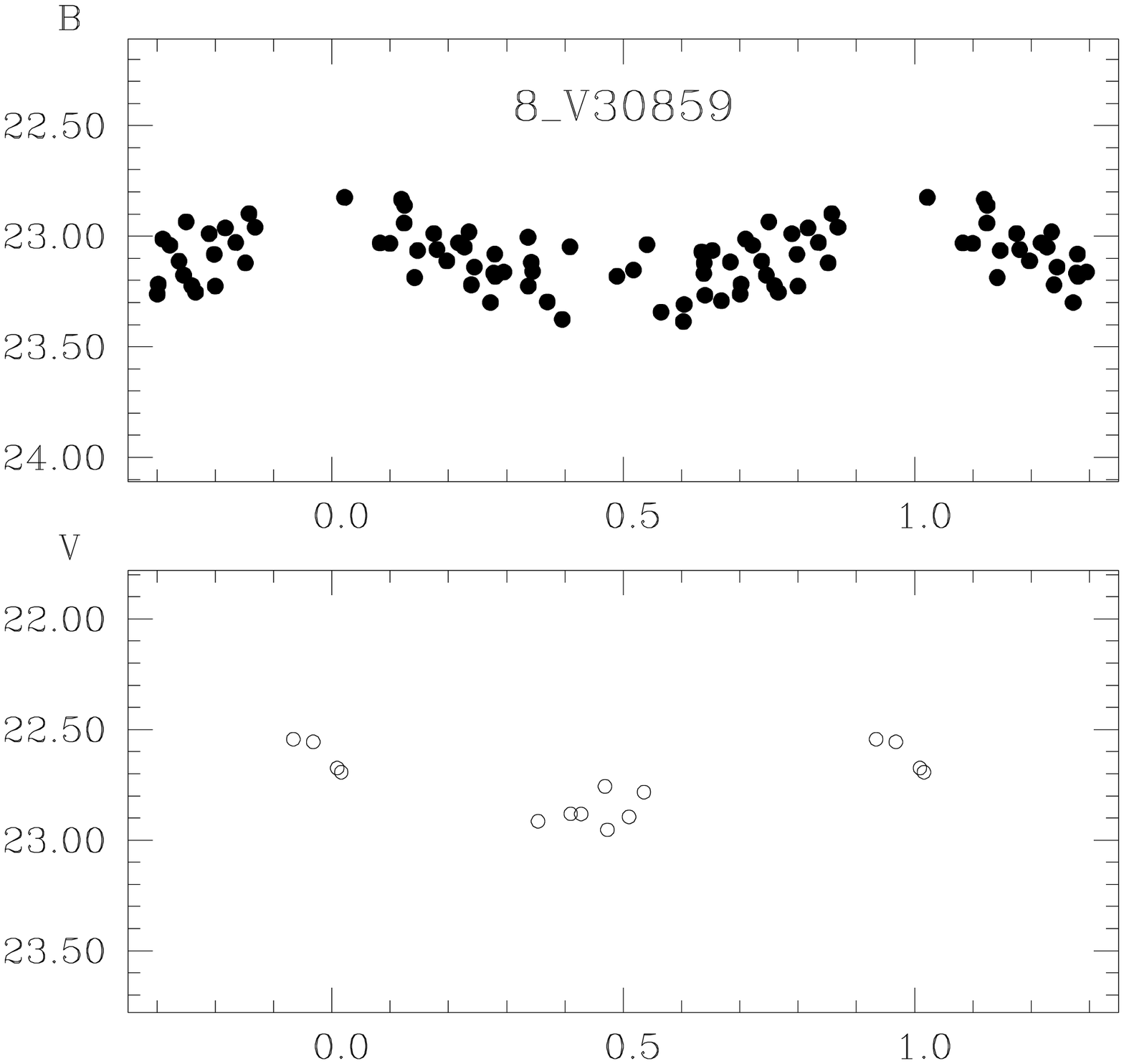}
\includegraphics[width=0.329\columnwidth,height=0.27\columnwidth]{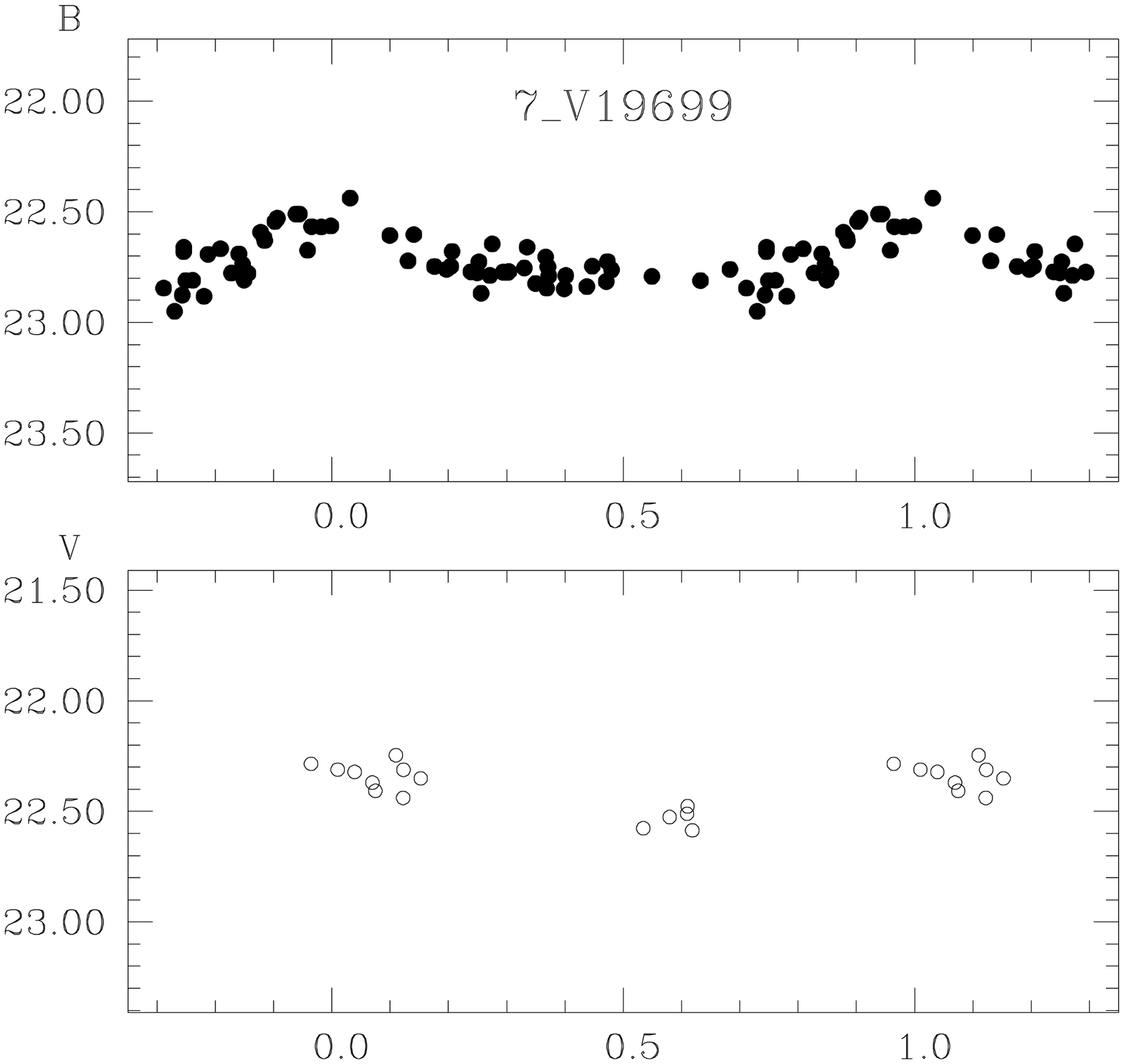}
\includegraphics[width=0.329\columnwidth,height=0.27\columnwidth]{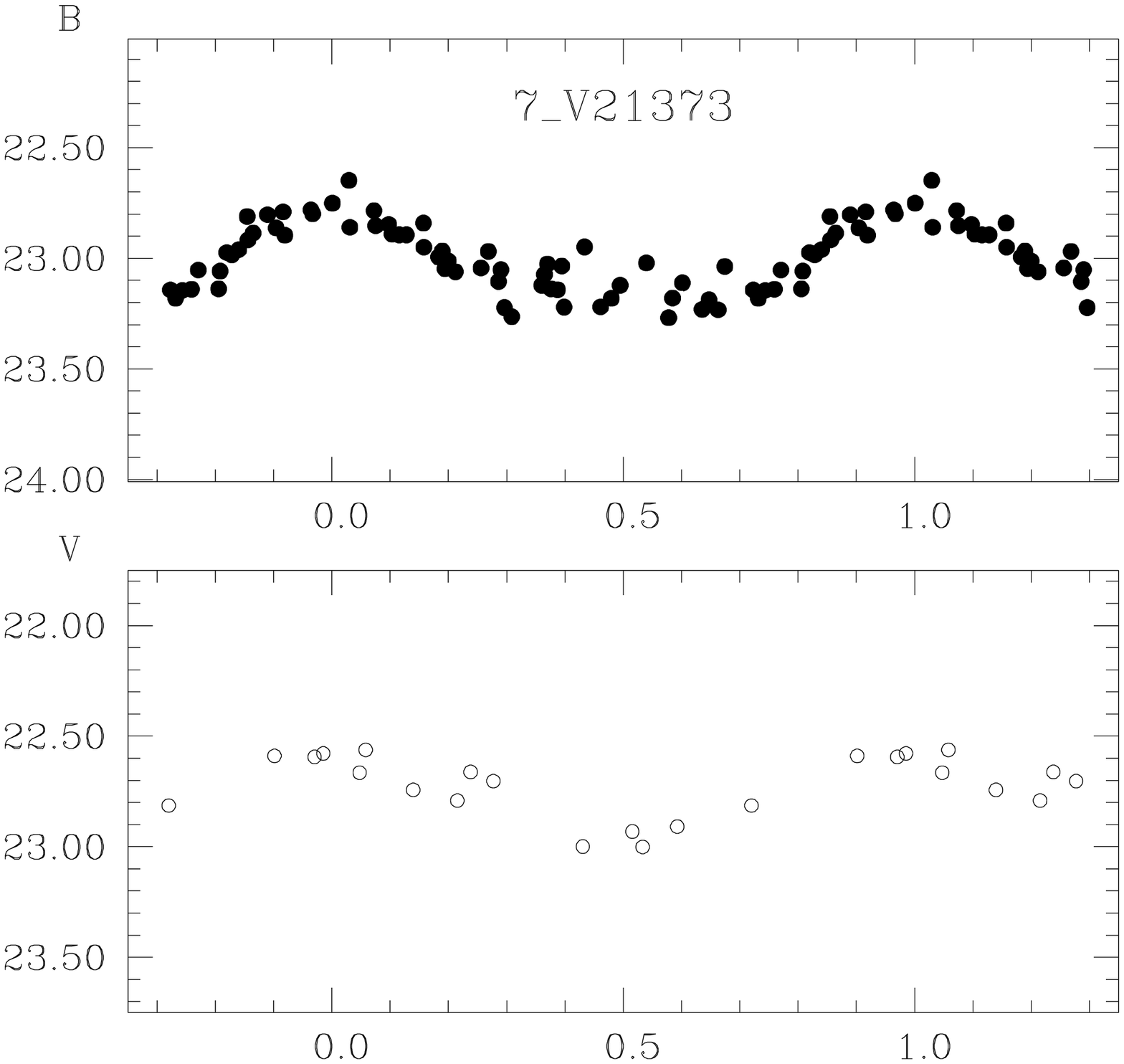}
\includegraphics[width=0.329\columnwidth,height=0.27\columnwidth]{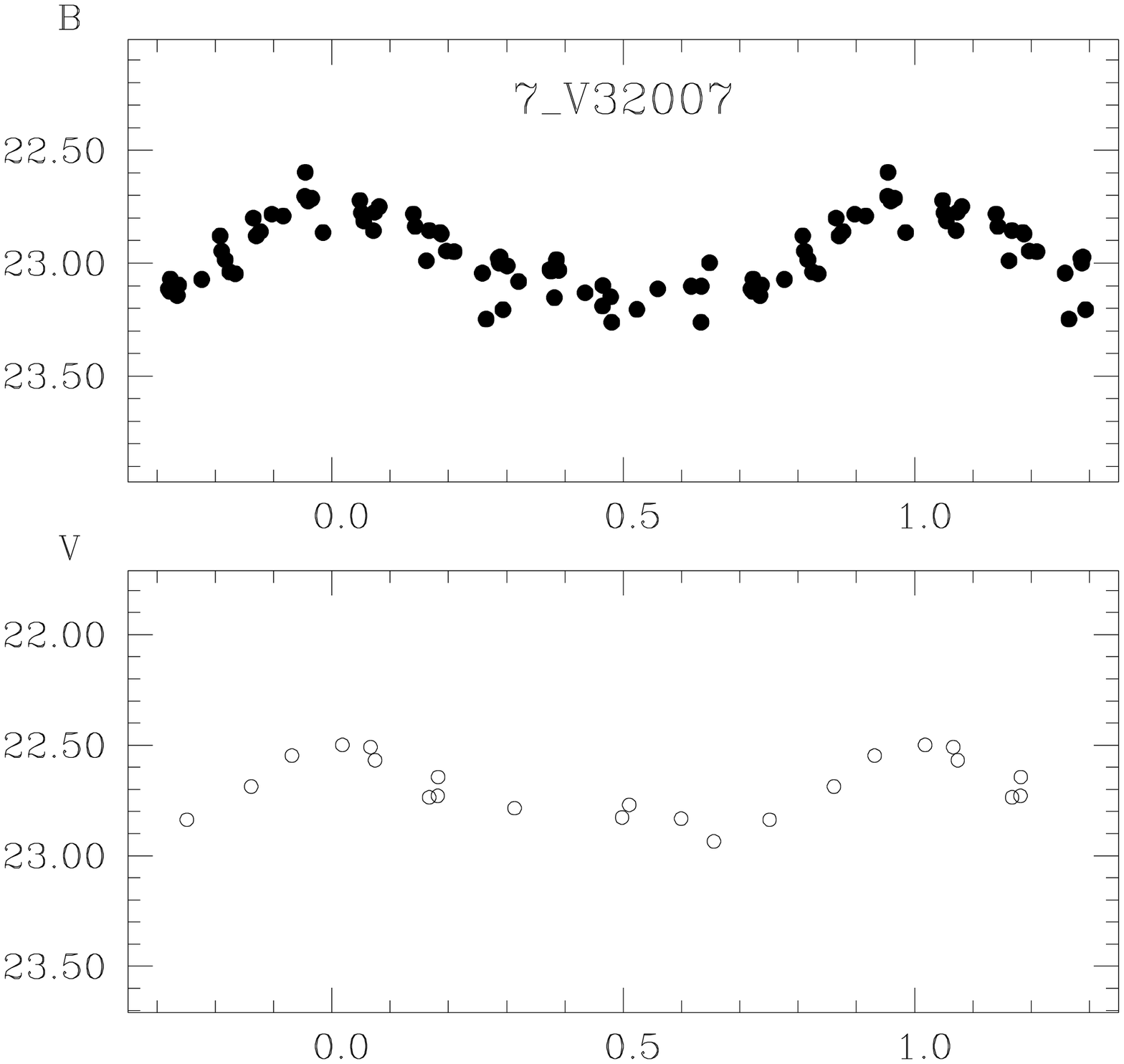}
\includegraphics[width=0.329\columnwidth,height=0.27\columnwidth]{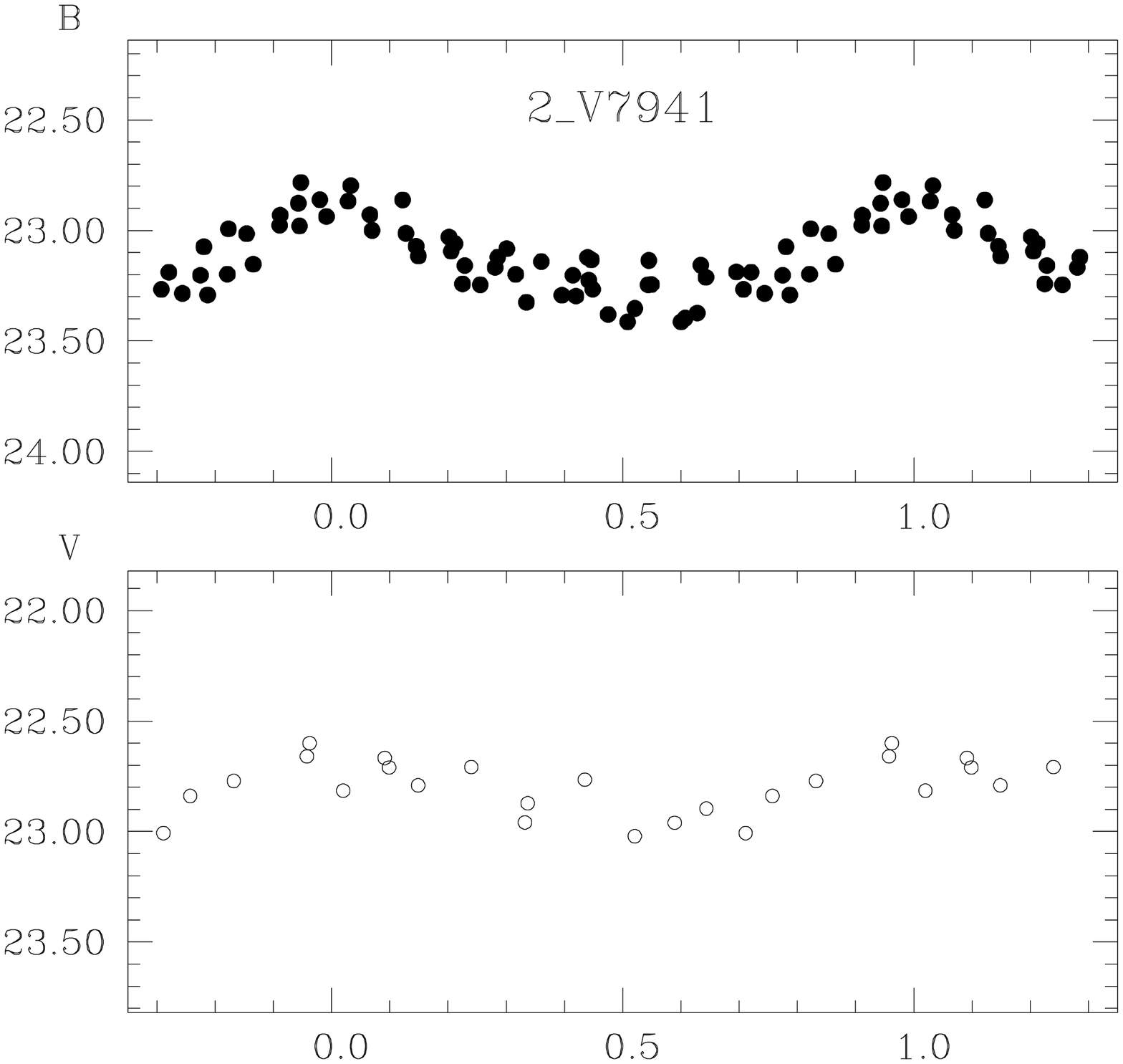}
\includegraphics[width=0.329\columnwidth,height=0.27\columnwidth]{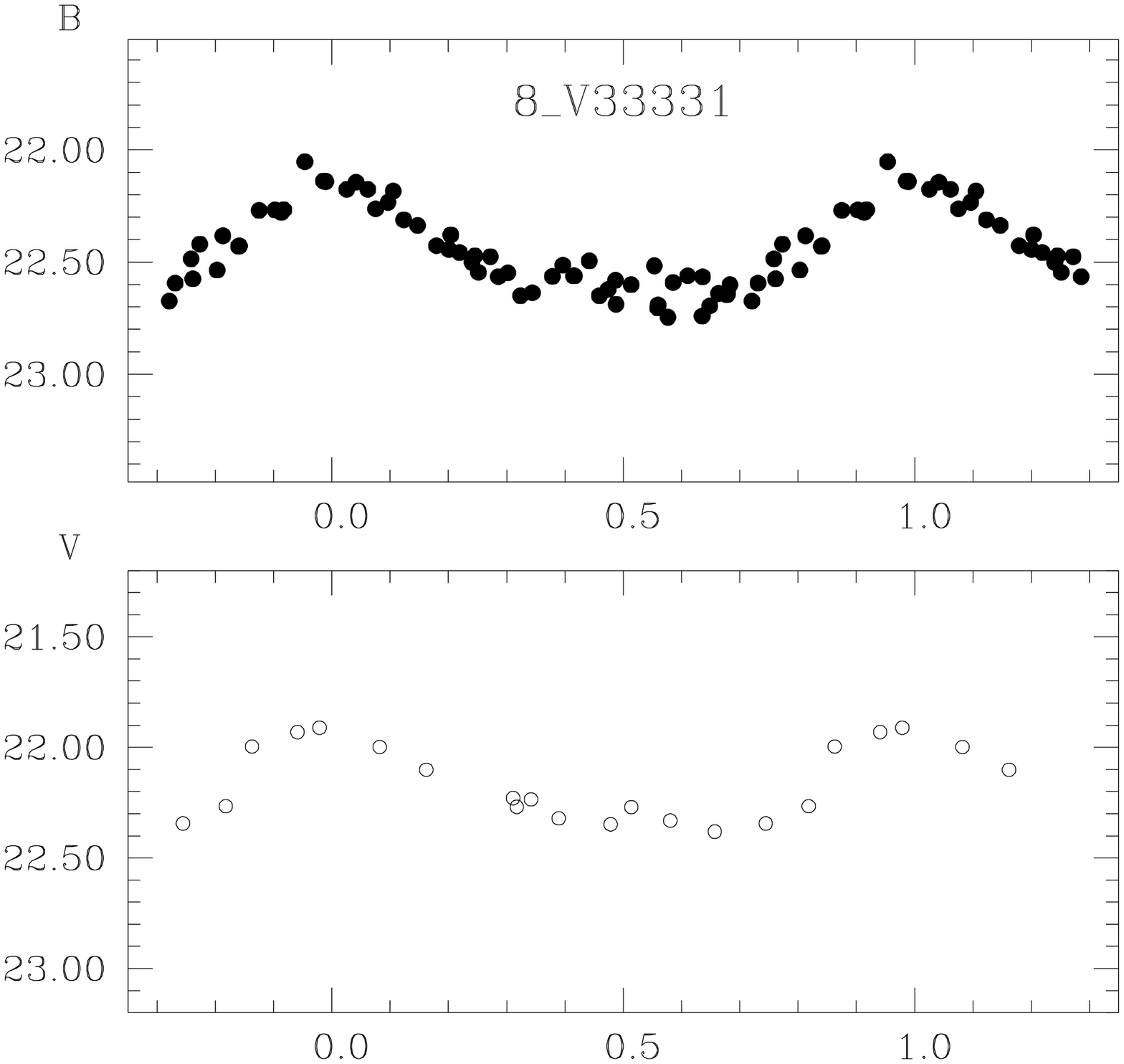}
\includegraphics[width=0.329\columnwidth,height=0.27\columnwidth]{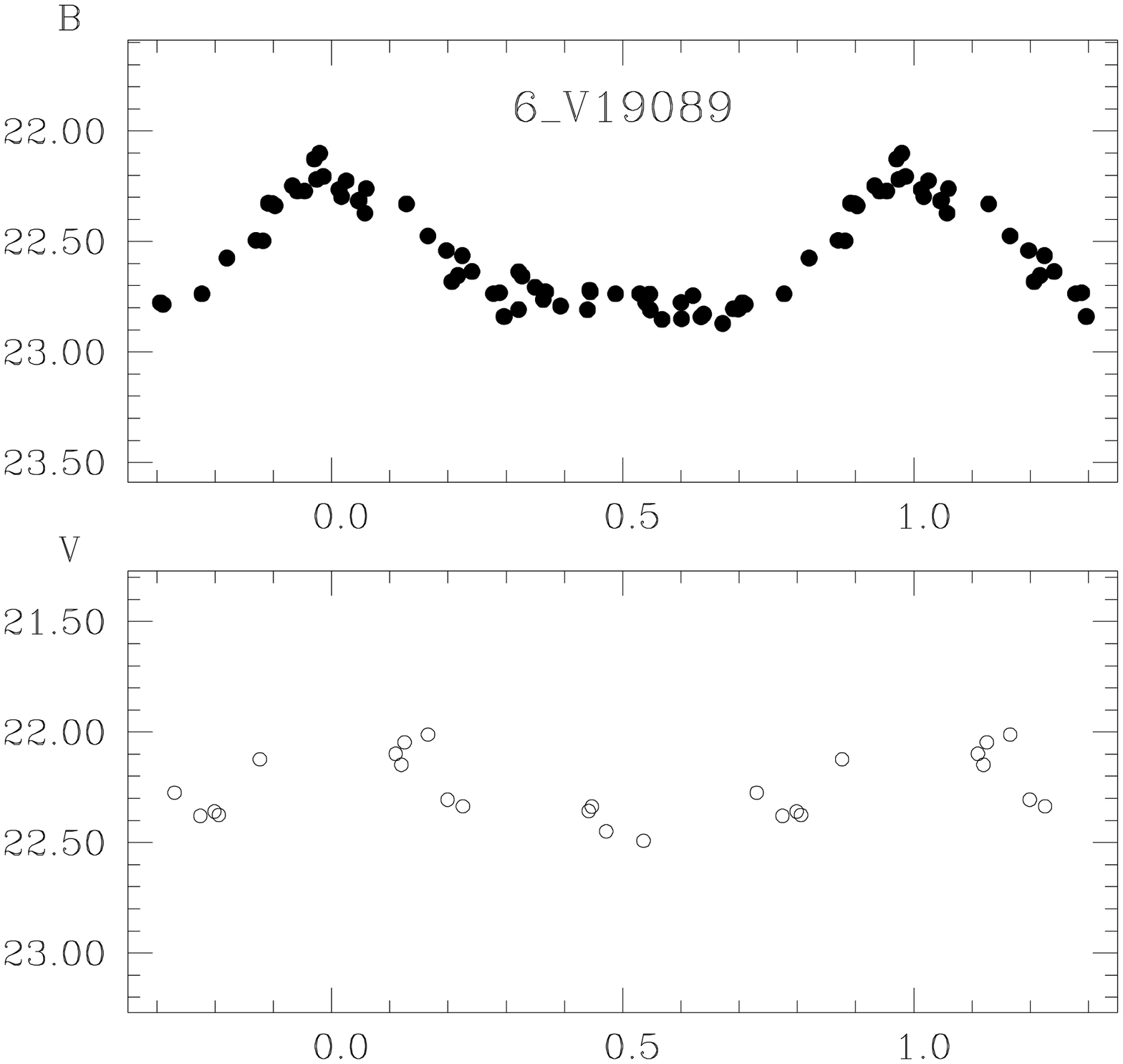}
\label{lcapp}
\end{figure*}

\end{document}